\documentclass[twoside,11pt]{article}

\usepackage{jmlr2e}

\usepackage{amsmath}
\usepackage{bm}
\usepackage{dsfont}
\usepackage{enumitem}
\usepackage{comment}
\usepackage{xcolor}
\usepackage{subcaption}
\usepackage{setspace}

\hypersetup{hidelinks}

\makeatletter

\let\c@remark\relax
\makeatother
\newtheorem{remark}{Remark}
\newtheorem{assumption}{Assumption}

\numberwithin{equation}{section}

\newcommand{\obs}{\mathrm{obs}}
\newcommand{\zvec}{\mathbf z}
\newcommand{\Zvec}{\mathbf Z}
\newcommand{\tespace}{\mathcal X}
\newcommand{\Dcond}{D}
\newcommand{\Qpol}{\mathcal Q}
\newcommand{\allocspace}{\Omega}

\usepackage{lastpage}
\jmlrheading{}{}{}{}{}{}{}
\makeatletter
\def\ps@jmlrtps{\let\@mkboth\@gobbletwo%
  \def\@oddhead{}\def\@evenhead{}%
  \def\@oddfoot{\hfil\thepage\hfil}\def\@evenfoot{\hfil\thepage\hfil}}
\def\@starteditor{}\def\@endeditor{}
\def\editor#1{\gdef\@editor{}}
\makeatother
\ShortHeadings{Average versus expected causal effects under interference}{Karwa, Airoldi and Rubin}
\firstpageno{1}

\begin{document}

\title{Average versus expected causal effects under interference: Modeling assumptions and estimation bottlenecks}

\author{\name Vishesh Karwa$^{*\dagger}$ \email vishesh.karwa@temple.edu\\
\name Edoardo M.\ Airoldi$^{*\dagger}$ \email edoardo.airoldi@temple.edu\\
\name Donald B.\ Rubin$^{*\dagger\ddagger}$ \email rubin@stat.harvard.edu\\[0.25em]
\addr $^{*}$Department of Statistics, Operations, and Data Science, Temple University\\ %
\addr $^{\dagger}$Data Science Institute, Temple University, Philadelphia, PA 19122, USA\\
\addr $^{\ddagger}$Department of Statistics, Harvard University, Cambridge, MA 02138, USA}

\editor{AE}

\maketitle
\begin{abstract}
Causal inference under treatment interference requires assumptions to reduce the exponentially large number of potential outcomes. We develop hierarchical finite-population models for potential outcomes by separating these assumptions into three layers: \emph{interference neighborhoods}, \emph{exposure mappings}, and \emph{outcome models}. Given an interference neighborhood and exposure mapping, we show that potential outcomes admit a nonparametric decomposition into direct, spillover, and treatment--spillover interaction components. This decomposition provides a common coordinate system for nonparametric, semiparametric, and parametric models of finite-population potential outcomes. For a given potential-outcome model, we define two classes of estimands. Average causal effects (ACEs), such as direct and total treatment effects, contrast potential outcomes under fixed treatment--exposure conditions. Expected causal effects (ECEs), such as expected direct and indirect effects, contrast expectations of potential outcomes under a specified allocation policy. For both classes, we bound the variance of the canonical Horvitz--Thompson estimator by three terms: a diagonal weight term, a term from overlapping interference neighborhoods, and a residual design-dependence term. Our upper and lower bounds show that under \emph{graph-unaware} designs, nonparametric design-unbiased estimation of ACEs can require sample sizes growing exponentially in the interference degree, whereas polynomial growth can suffice for design-aligned ECEs. Under neighborhood interference, we construct \emph{graph-aware} independent-set designs with variance bounds polynomial in the maximum degree for direct and total treatment effects on every graph, whereas graph cluster randomization can incur worst-case variance exponential in the degree.

\end{abstract}

\begin{keywords}
  causal inference, interference, potential outcomes, network experiments, randomization inference
\end{keywords}

\tableofcontents

\section{Introduction}
\label{sec:intro}
Under treatment interference, the potential outcome of one unit may depend on the
treatment assignments of other units in the finite population. Consequently, the table of science is exponentially large. Moreover, unlike the classical case of no interference, there is no unique canonical causal estimand like the average treatment effect.
Thus causal inference under interference requires assumptions on the structure of
the table of science, together with a specification of a causal estimand and a design and estimator used to learn the estimand.

A large literature has emerged for causal inference under interference across settings
such as peer effects
\citep{manski1993identification,bramoulle2009identification,graham2008identifying},
vaccination studies \citep{hudgens2012toward}, spatial
environments \citep{zigler2012,giffin2023generalized}, and online and network
experiments \citep{aral2011}.
These approaches use a variety of assumptions to reduce the number of potential outcomes per unit
and define different causal estimands along with designs and estimators to estimate them.
However, two layers of the causal inference problem are often specified jointly, sometimes even implicitly.  The first concerns the modeling assumptions on potential outcomes.  Assumptions under interference involve three distinct choices:
which units' treatment assignments in the finite population may affect a given unit's potential outcome, how those assignments
are summarized or aggregated, and what functional restrictions, if any, are imposed on the resulting potential outcomes. These choices correspond to \emph{interference neighborhoods}, \emph{exposure mappings}, and \emph{outcome models} respectively.  These three modeling assumptions are often conflated in the literature, but they play different mathematical roles.

The second layer concerns the definition of causal effects and its interaction with
design and estimation.  Some causal estimands under interference are fixed
potential-outcome contrasts: they compare outcomes under two specified treatment allocations or treatment--exposure conditions.  Other estimands are allocation-averaged contrasts: they average potential outcomes under a specified allocation distribution, often corresponding to a policy implementation.  We treat the allocation distribution that is part of the definition of an estimand as distinct from the design used to estimate it.

Without separating these layers, it can be difficult to determine whether differences across methods arise from different assumptions on the potential-outcome table, different causal estimands, different allocation policies, different experimental designs, or different estimators. In this paper, we develop a unified framework for finite population causal inference under interference
that makes these distinctions explicit.  We consider the Table of Science as the primitive
finite-population object.  We develop a set of hierarchical modeling assumptions on the Table of Science in the form of interference neighborhoods, exposure mappings, and outcome models. The first two are exclusion restrictions: they reduce the Table of Science by declaring which assignment coordinates can matter and which assignments are equivalent. On the other hand, an outcome model is a functional restriction on the potential outcomes. Once an interference neighborhood and exposure mapping are
specified, the full-assignment potential outcome \(Y_i(\mathbf z)\) is represented by a lower-dimensional \emph{reduced potential outcome}
\(Y_i(z,e)\), where \(\mathbf z\) is the vector of treatment allocations, \(z_i\) is unit \(i\)'s treatment, and \(e_i\) is its \emph{exposure}.  A central consequence of this representation is a nonparametric linear decomposition of the reduced potential outcome \(Y_i(z,e)\) given by
\[
Y_i(z,e)=A_i(z)+B_i(e)+zC_i(e),
\]
where \(A_i\) captures the direct treatment component, \(B_i\) the spillover component,
and \(C_i\) the treatment--spillover interaction.  This decomposition is a
reparameterization of the potential outcomes.  It gives a common language for specifying nonparametric, semiparametric,
and parametric finite population models of potential outcomes under interference.  For example, a nonparametric model can leave the exposure
as the full interference neighborhood and leave \(A_i,B_i,C_i\) unrestricted.
Semiparametric and parametric outcome models impose additional restrictions such as additivity, homogeneity, linearity in exposure, or parametric forms
for the spillover and interaction components.  All potential-outcome models used in the
interference literature, including arbitrary interference, linear-in-means models, stratified interference, binary exposure, spatial decay, and threshold or contagion models, can be represented within this hierarchy by appropriate choices of the interference neighborhood, exposure mapping, and outcome-model restrictions.

The framework also separates two natural classes of causal estimands under interference, assuming that the interference neighborhood and exposure mappings are correctly specified. Average causal effects (ACEs), such as the direct and total treatment effects, are contrasts of fixed potential outcomes under specific treatment allocations or treatment--exposure combinations. They are properties of the finite population and do not involve an allocation distribution. In contrast, expected causal effects (ECEs), such as the expected direct effect (EDE) and expected indirect effect (EIE), are contrasts of expectations of potential outcomes under allocation distributions--sometimes also referred to as policies. The distributions entering the two arms of the contrast may be the same, different, or induced by a common allocation rule; in all cases, the allocation distribution used to average potential outcomes is part of the causal question itself. The experimental design remains distinct from this allocation distribution, though it is usually informed by it.

We show that there is a fundamental separation between the two classes for nonparametric estimation. For ACEs, the design must reveal the particular treatment--exposure potential outcomes that define the target estimand.  Potential outcomes revealed by the design but outside this set are nuisance for that estimand: without further outcome-model restrictions on the causal response surface, observing them gives no information about the fixed target potential outcomes.  For ECEs, by contrast, the exposure states generated by the target allocation distribution enter the estimand.  The design does not need to isolate a set of fixed treatment-exposure conditions; it must reproduce the allocation distribution's averaging weights.

We illustrate this through design-based variance bounds for canonical weighted unbiased Horvitz--Thompson estimators. We show that, for both classes of estimands, the variance of the canonical estimator decomposes into a diagonal weight term, an interference-dependence term from overlapping neighborhoods, and a residual design-dependence term induced by the randomization design.  The difference lies in the weight term.  ACE weights involve design-induced inverse propensity scores for fixed relevant treatment--exposure conditions.  Under classical graph-unaware designs, these propensity scores can be exponentially small in the degree of the interference graph, and matching lower bounds show that this obstruction applies to all unbiased estimators. By contrast, ECE weights are allocation policy-to-design importance weights; when the design is aligned with the allocation policy, the diagonal term is mild and the dominant cost is the dependence among observed weighted outcomes.  Consequently we show that, under the same classical designs, fixed treatment-exposure ACE estimation can require sample sizes growing exponentially in the degree of the interference graph, whereas ECE estimation is limited by polynomial dependence on the degree of an appropriate dependence graph.

These results have direct practical implications for design and estimation. Designs for nonparametric ACE estimation must be aligned with the treatment--exposure combinations that define the target estimand. Potential outcomes that do not appear in the definition of the estimand are \emph{nuisance} potential outcomes for nonparametric estimation. Thus, nonparametric estimation of ACEs requires specialized \emph{graph-aware} designs that place sufficient probability on revealing the relevant potential outcomes. We construct such designs from independent sets of the interference graph: a distance-one independent-set design targets the direct treatment effect and a distance-two independent-set design targets the total treatment effect. We show that in each case, these graph aware designs replace the exponentially small propensity of the relevant potential outcome by a focal probability that is polynomial in the relevant local graph size. Sparse seeding of the distance-one independent set additionally estimates the influencer spillover effect---the average effect of treated influencers on untreated peers under the any-treated exposure map---at constant-order focal propensities on locally tree-like regular graphs (Section~\ref{sec:is-indirect}). The resulting estimators have mean squared error polynomial in the graph degree scale for broad graph classes. For the total effect, cluster randomized designs can be sharper than the independent-set construction when closed neighborhoods are contained in few clusters, though this requires the clustering to be aligned with the interference graph. Outcome-model restrictions provide another route for ACE estimation by relating nuisance potential outcomes to target potential outcomes, but this route is no longer fully nonparametric and is subject to misspecification of the outcome model. By contrast, for ECEs, the allocation policy is part of the estimand itself, and all potential outcomes with positive probability under the policy enter the estimand. Thus, ECEs can be estimated nonparametrically under classical Bernoulli designs when the target allocation policy is Bernoulli.

\noindent
The paper is organized as follows.  Section~\ref{sec:POmodels} introduces the Table of
Science and the modeling hierarchy.  Section~\ref{sec:decomposition} gives the non-parametric
decomposition of Potential Outcomes and shows how common model classes arise as special cases.  Section~\ref{sec:estimands}
defines ACEs and ECEs.  Section~\ref{sec:nonparametric-estimation} develops the generic
estimation bounds and the ACE/ECE bottleneck comparison.  %
Section~\ref{sec:implications} draws implications
for classical estimators, specialized designs, and model-assisted estimation.
Section~\ref{sec:designs} develops estimand-aligned independent-set designs that recover
polynomial rates for average causal effects---including a sparse-seeding design that estimates
the influencer spillover effect at constant-order propensities on bounded-degree graphs---together with a comparison against cluster randomized designs, and
Section~\ref{sec:discussion} concludes.

\section{Hierarchical models of potential outcomes}
\label{sec:POmodels}

Consider a finite population of $n$ units, indexed by $i \in \{1,\dots,n\}$, and a binary treatment assignment $z_i \in \{0,1\}$ for each unit $i$. The vector of treatment assignments is denoted by $\zvec = (z_1,\dots,z_n)$, with $\zvec \in \{0,1\}^n$. Under arbitrary treatment interference, the potential outcome of a unit $i$ depends on the entire assignment vector. We write
\[
Y_i(\zvec), \qquad \zvec \in \{0,1\}^n.
\]
This denotes the \emph{full-assignment potential outcome} of unit $i$ under treatment allocation $\zvec$. The collection of all full-assignment potential outcomes
\[
\mathbb{T} = \{Y_i(\zvec) : i=1,\dots,n,\ \zvec\in\{0,1\}^n\}
\]
is the \emph{Table of Science} for the study. Crucially, \(\mathbb{T}\) is a fixed, full \(n \times 2^n\) table dictated by nature, existing prior to any modeling assumptions.  Let $\Zvec$ denote the random treatment assignment generated by a design $p(\Zvec)$. The observed outcome for unit $i$ is
\[
Y_i^{\mathrm{obs}} = Y_i(\Zvec),
\]
and in any realized experiment, exactly one column of $\mathbb{T}$ is observed.

Since each unit has exponentially many counterfactuals relative to what is observable, modeling assumptions are \emph{necessary} to relate unobserved potential outcomes to observed ones. We specify finite population models of potential outcomes through three distinct hierarchical components: an interference graph, an exposure mapping, and an outcome model. Such assumptions constrain the entries of $\mathbb{T}$ in two different ways. The interference graph and exposure mappings are \emph{exclusion restrictions}. The interference graph specifies which treatment coordinates can enter a unit's full-assignment potential outcome and the exposure mapping specifies how those coordinates aggregate. When combined, the interference graph and the exposure mapping partition the $2^n$ columns of $\mathbb{T}$ into equivalence classes for each row and require $\mathbb{T}$ to be constant within each class. On the other hand, \emph{Outcome models} are a different kind of constraint: they do not merge columns of $\mathbb T$, but instead impose functional constraints on the values $\mathbb{T}$ takes across distinct potential outcomes, potentially linking entries within a row or across rows (units). The classical no-interference setting is the most restrictive exclusion restriction, forcing each row to take only two distinct values across all $2^n$ columns. Formally, each exclusion restriction is specified by an equivalence relation $\sim_i$ on $\{0,1\}^n$ that $\mathbb{T}$ must respect, meaning $\zvec \sim_i \zvec' \Rightarrow Y_i(\zvec) = Y_i(\zvec')$; $\mathbb{T}$ is then fully determined by its values on the class representatives.

\subsection{Interference graph}
\label{sec:interference-graph}
The first restriction specifies the set of units in the finite popluation whose treatment assignments effect $Y_i(\mathbf{z})$.
\begin{assumption}[Interference exclusion restriction]
\label{assump:interference-exclusion}
For each unit \(i\), there exists a subset \(N_i \subseteq \{1,\ldots,n\}\setminus\{i\}\)
such that, for any two treatment allocations \(\zvec\) and \(\zvec'\),
\[
z_i=z_i'
\quad\text{and}\quad
z_{N_i}=z'_{N_i}
\quad\Longrightarrow\quad
Y_i(\zvec)=Y_i(\zvec').
\]
\end{assumption}
Formally, Assumption~\ref{assump:interference-exclusion} is the equivalence relation
\[
\zvec \sim_i^G \zvec' \;\iff\; z_i = z_i' \text{ and } \zvec_{N_i} = \zvec'_{N_i},
\]
which partitions $\{0,1\}^n$ into $2^{d_i+1}$ equivalence classes for row $i$, where $d_i=|N_i|$.  Row $i$ of $\mathbb{T}$ takes at most $2^{d_i+1}$ distinct values---one per class---and the potential outcome can be written as
\begin{equation}
Y_i(\zvec) := Y_i(z_i, \zvec_{N_i}),
\label{eq:NIA}
\end{equation}
depending only on $i$'s own treatment and the treatments of units in $N_i$.  The neighborhoods $\{N_i\}_{i=1}^n$ may overlap, be asymmetric, or vary in size across units.
For a fixed Table of Science, there is a minimal neighborhood
\[
N_i^\star =
\left\{
j\neq i:
\exists \zvec,\zvec'
\text{ differing only in } z_j
\text{ such that }
Y_i(\zvec)\neq Y_i(\zvec')
\right\}.
\]
The collection $\{N_i^\star\}$ can be encoded via a minimal functional
\emph{interference graph} $G^\star=(V,E^\star)$ on $n$ nodes.  The vertices of
the graph $V$ correspond to the $n$ units in the finite population.  A directed edge
from node $j$ to node $i$ exists if the treatment of unit $j$ can affect the potential
outcome of unit $i$.  In the remainder of the paper, we work with a specified
interference graph $G=(V,E)$ and the corresponding neighborhoods $\{N_i\}_{i=1}^n$, under Assumption~\ref{assump:interference-exclusion}. Issues arising from random or misspecified interference graphs are discussed in the remark below. We use the degree to denote the in-degree of this graph $G$. Setting $N_i = \{1,\dots,n\}\setminus\{i\}$ recovers the complete interference case in which all $2^n$ entries of each row of $\mathbb{T}$ are distinct (the restriction is vacuous), while $N_i = \emptyset$ yields the classical SUTVA case in which each row takes only two distinct values.

It is useful to distinguish the interference graph from an observed \textit{social
graph}.  The interference graph is an abstract representation of restrictions on the
potential-outcome table.  It may or may not coincide with a social graph, be an induced
subgraph of it, or be a complete graph with the strength of interference modeled
through an exposure function. In social-network settings, a natural choice is to set
$N_i$ to $i$'s neighbors in the social graph (one-hop interference), or to all units
within graph distance $k$ (multi-hop interference).  Partial interference, where units
can be partitioned into disjoint groups and interference is assumed to occur only within groups, is
another special case \citep{sobel2006randomized,hudgens2012toward}.  Spatial and
bipartite interference models can also be represented by appropriate choices of $G$;
see Appendix~\ref{app:additional-framework-examples}.

\begin{remark}[Random and misspecified interference]
The minimal functional interference graph $G^\star$ may not be known in applications, and the specified interference graph $G$ may itself be noisy or random.  The graph $G$ used in analysis should therefore be understood as an assumed interference graph.  If the assumed interference neighborhoods $N_i$ satisfies $N_i^\star\subseteq N_i$, the exclusion restriction remains
valid but the representation is nonminimal.  If, on the other hand, $N_i$ omits units in $N_i^\star$, the
interference graph is misspecified.  Another approach is to treat the interference graph as random, and model the observed graph
using a distribution; see \cite{bhattacharya2020causal},
\cite{liWagerAnnals}, and \cite{ohnishi2025degree} for examples.  For doubly robust
estimators under a misspecified $G$, see \cite{weinstein2026causal}.
\end{remark}

\paragraph{Dependence graphs}
The interference graph should not be confused with the dependence graph used in
large-sample analysis of design based estimators.  The interference graph is a
restriction on the arguments of the potential-outcomes.  A dependence graph is a
probabilistic object attached to a particular collection of random variables, such as
observed outcomes or estimator summands.  Under independent Bernoulli assignment, this
dependence may be induced by overlapping interference neighborhoods.  Under dependent
randomization designs, such as complete or cluster randomization, the assignment mechanism
can induce additional dependence even when the interference neighborhoods do not overlap.
We define the dependence graph formally in Section~\ref{sec:dependence-graphs-estimation}, where it is used for variance and consistency results.

Under Assumption~\ref{assump:interference-exclusion}, row $i$ of $\mathbb{T}$ takes at most $2^{d_i+1}$ distinct values; $\mathbb{T}$ itself remains an $n\times 2^n$ table, but is fully determined by specifying these $2^{d_i+1}$ values per row.  Even so, $2^{d_i+1}$ can still be large; the exposure mapping, introduced next, provides a further exclusion restriction.

\subsection{Exposure mapping}
\label{sec:exposure-model}
An \emph{exposure mapping} \citep{aronow2017estimating} summarizes neighborhood assignments into an exposure value,
\begin{equation}
e_i = f_i(\zvec_{N_i}) \in \mathcal{E}_i.
\end{equation}

\begin{assumption}[Exposure mapping]
\label{assump:exposure-equivalence}
For each unit \(i\), there exists an exposure mapping
\(f_i:\{0,1\}^{|N_i|}\to\mathcal E_i\) such that, for any two neighborhood
assignments \(\zvec_{N_i}\) and \(\zvec'_{N_i}\),
\[
f_i(\zvec_{N_i})=f_i(\zvec'_{N_i})
\quad\Longrightarrow\quad
Y_i(z_i,\zvec_{N_i})=Y_i(z_i,\zvec'_{N_i})
\quad\text{for }z_i\in\{0,1\}.
\]
\end{assumption}
Formally, Assumption~\ref{assump:exposure-equivalence} is the equivalence relation
\[
\zvec_{N_i} \sim_i^f \zvec'_{N_i} \;\iff\; f_i(\zvec_{N_i}) = f_i(\zvec'_{N_i}),
\]
which coarsens $\sim_i^G$ by merging neighborhood assignments that produce the same exposure value.  The function $f_i$ encodes \emph{how} the treatment assignments of unit's in $N_i$ affect unit $i$'s potential outcome.
In the main text, we focus on the case where exposure values are one-dimensional and discrete, with $|\mathcal E_i|= K_i $ exposure levels.  Vector-valued and continuous exposures are discussed in Appendix~\ref{app:additional-framework-examples}. We use the symbols $e_i=0$ to denote no exposure and $e_i=1$ to denote full exposure, with the exposure mapping specifying what these labels mean in context. 

Combined, Assumptions~\ref{assump:interference-exclusion} and~\ref{assump:exposure-equivalence} define the relation
\[
\zvec \sim_i \zvec' \;\iff\; z_i = z_i' \text{ and } f_i(\zvec_{N_i}) = f_i(\zvec'_{N_i}),
\]
with $2K_i$ equivalence classes, where $K_i = |\mathcal{E}_i|$.  Row $i$ of $\mathbb{T}$ is thus fully determined by its $2K_i$ values $Y_i(z_i, e_i)$, one for each $(z_i, e_i)$ pair.  We call each such value a \emph{reduced potential outcome}. The collection of these distinct reduced potential outcomes,
\[
\mathbb S
=
\{Y_i(z,e): z\in\{0,1\},\ e\in\mathcal E_i,\ i=1,\ldots,n\},
\]
is the \emph{causal response surface} induced by the specified interference neighborhoods and exposure mappings. Thus \(Y_i(\zvec)\) denotes a full-assignment potential outcome in \(\mathbb T\), whereas \(Y_i(z,e)\) denotes the corresponding reduced potential outcome in \(\mathbb S\). When there is no interference, \(Y_i(z)\) is the classical own-treatment potential outcome, the special case with no exposure argument. The non-parametric case corresponds to the identity mapping $e_i = \zvec_{N_i}$, giving $K_i = 2^{d_i}$ and recovering $\sim_i^G$.

Common examples of exposure mappings include symmetric or anonymous exposure, where \(f_i\) depends only on
the number or fraction of treated neighbors
\citep{hudgens2012toward,manski2013identification}; binary exposure, where
\(\mathcal E_i=\{0,1\}\); additive exposure of the form
\(f_i(\zvec_{N_i})=\sum_{j\in N_i}h_j(z_j)\); and weighted or non-anonymous exposure, where \(f_i\) incorporates neighbor identity, covariates, or graph distance.  

\subsection{A non-parametric decomposition of the causal response surface}
\label{sec:decomposition}
Assumption~\ref{assump:interference-exclusion} concerns \emph{who} can interfere with unit $i$, Assumption~\ref{assump:exposure-equivalence} concerns the \emph{form} of that interference.  Under Assumptions~\ref{assump:interference-exclusion} and~\ref{assump:exposure-equivalence}, $\mathbb{T}$ is fully determined by the causal response surface $\mathbb S$, comprising $2\sum_i K_i$ values.  This response surface admits a natural non-parametric linear representation.  This representation is an
identity, and, without loss of generality, decomposes the potential outcome of each
unit into three components: (a) a component depending only on a unit's own treatment,
(b) a component depending only on a unit's exposure, and (c) the interaction between (a) and
(b).

\begin{theorem}[Non-parametric decomposition]
\label{thm:decomposition}
Under Assumptions~\ref{assump:interference-exclusion} and
\ref{assump:exposure-equivalence}, there exist functions $A_i(z)$, $B_i(e_i)$ and
$C_i(e_i)$ such that, for each unit $i$ and each
$(z_i,e_i)\in \{0,1\}\times\mathcal{E}_i$, a potential outcome $Y_i(z,e) \in \mathbb S$ can be written as
\begin{equation}
Y_i(z_i,e_i)
= A_i(z_i) + B_i(e_i) + z_i C_i(e_i) = \alpha_i + \beta_i z_i + B_i(e_i) + z_i C_i(e_i),
\label{eq:nonparametric}
\end{equation}
with the normalization $B_i(0)=C_i(0)=0$.
\end{theorem}

The decomposition has a nice interpretation: the \(A_i\) term captures the part of the potential outcome depending only on unit \(i\)'s own treatment, \(B_i\) term caputres the additive exposure component, and \(C_i\) the interaction between own-treatment and exposure (synergy or competition). These components define the corresponding direct, spillover, and interaction causal effects studied in Section~\ref{sec:estimands}.

The representation is a full-rank reparameterization of the causal response surface. In particular, it makes no assumptions on the potential outcomes and parametrizes the $2K_i$ distinct values of row $i$ using $2K_i$ linearly independent parameters. For example, under the identity exposure mapping, the number of exposure levels is \(K_i=2^{d_i}\) and the number of  parameters in the representation $Y_i(\zvec) = \alpha_i + \beta_i z_i + B_i(z_{N_i}) + z_i C_i(z_{N_i})$ for each unit $i$ is \(2^{d_i+1}\). Similarly, under arbitrary interference, $K_i = 2^{n-1}$ and Theorem~\ref{thm:decomposition} gives the non-parametric model \(Y_i(\zvec)=\alpha_i+\beta_i z_i+B_i(\zvec_{-i})+z_i C_i(\zvec_{-i})\), which has \(2^n\) parameters for unit \(i\).

\subsection{Outcome model}
\label{sec:structural}

An outcome model imposes functional constraints on the $2K_i$ distinct values of the causal response surface via specifying semi-parametric or parametric restrictions on $\{A_i, B_i, C_i\}$ in the decomposition \eqref{eq:nonparametric}; it does not operate by further merging assignment columns of $\mathbb{T}$.  Such constraints reduce the number of free parameters and may arise from scientific knowledge (e.g., a linear response) or be used to express sharp null hypotheses.  Unlike exclusion restrictions, which operate within a single row of $\mathbb{T}$ before forming $\mathbb S$, outcome models constrain the distinct potential outcomes of $\mathbb S$ and can also link entries across rows --- homogeneity assumptions, for instance, require $\beta_i$ to be common across units.

\paragraph{Additive effects.}
A natural and commonly used restriction is that the direct and indirect effects enter additively, without interaction. In terms of the decomposition in \eqref{eq:nonparametric}, this corresponds to
\[
C_i(e) = 0 \quad\text{for all } i \text{ and all } e\in \mathcal{E}_i.
\]
The potential outcomes then simplify to $Y_i(z,e) = A_i(z) + B_i(e)$, and each unit has $K_i+2$ free parameters instead of $2K_i$. %

\paragraph{Constant (homogeneous) effects.}
In some settings, it may be reasonable to assume that treatment and exposure effects do not vary across units. This can be written as
\[
A_i(z) = A(z),\qquad B_i(e) = B(e),\qquad C_i(e) = C(e), \quad \text{for all } i.
\]
This links entries across rows of $\mathbb{T}$: potential outcomes from different units are constrained to share the same effect functions, reducing the parameter count from $O(n K_i)$ to $O(K)$.

\paragraph{Linear effects in exposure.}
When exposure levels are ordered (e.g., counts of treated neighbors), it may be plausible that the interference effect depends linearly on $e$. One simple example is
\[
B_i(e) = \gamma_i \, e, \qquad C_i(e) = \theta_i \, e,
\]
which yields $Y_i(z,e) = \alpha_i + \beta_i z_i + \gamma_i e + \theta_i z e$. If $\alpha_i, \beta_i, \gamma_i$ and $\theta_i$ are further assumed constant across units, the model reduces to a four-parameter linear dose-response formulation.

\paragraph{Sharp null hypotheses.}
Outcome models can also be used to express sharp null hypotheses for randomization testing.  The linear decomposition allows these to be stated precisely.  For instance, a sharp null of no treatment effect under interference is
$$\beta_i  = 0  \mbox{ and } B_i(e_i) = C_i(e_i) = 0  \mbox{ for all } e_i,i$$ 
Similarly, the condition 
\[
Y_i(1,e) = Y_i(0,e) \quad \text{for all } i \text{ and all } e\in\mathcal{E}_i,
\] is equivalent to the hypothesis $A_i(1) + C_i(e_i) = A_i(0)$ and implies sharp null of no treatment effect only under additivity (i.e. $C_i(e_i) = 0$). More targeted nulls, such as absence of spillover effects or no interaction effects, can be expressed by setting subsets of $\{B_i(e),C_i(e)\}$ to zero.  See \cite{rosenbaum2007interference,aronow2012general,pouget2019testing,athey2018exact,basse2019randomization,puelz2022graph,rigdon2015exact} for construction of statistical tests under various such hypotheses.

\paragraph{Relation to estimation.}
Outcome modeling assumptions are not required to define causal estimands. However, they can be used for model-assisted design and estimation. In the absence of such assumptions, the parameters \(\{A_i, B_i, C_i\}\) vary freely across units and exposure levels. Imposing them constrains these parameters, so that a response observed under one treatment-exposure combination can inform missing responses under other combinations. If correctly specified, this model-assisted approach can significantly reduce variance at either the design or analysis stage.%

\subsection{Canonical model classes}
\label{sec:examplePOmodels}
Different choices of the interference neighborhood $N_i$, the exposure mapping $f_i$, and the outcome model restrictions imposed on $(A_i,B_i,C_i)$ lead to different potential-outcome models. The decomposition in Theorem~\ref{thm:decomposition} encompasses the full range of models used in the literature. A \emph{non-parametric} model imposes no restrictions on $A_i$, $B_i$, $C_i$ beyond the interference neighborhood and exposure mapping; the functions themselves are unrestricted. A \emph{semi-parametric} model imposes qualitative restrictions on $A_i$, $B_i$, $C_i$---such as additivity ($C_i\equiv 0$), symmetry, or monotonicity. A \emph{parametric} model specifies finite-dimensional functional forms for $A_i$, $B_i$, $C_i$, for example linear functions of exposure or homogeneous effects across units. We present some commonly used examples below. %

\begin{example}[Arbitrary interference]
Let $N_i = \{1,\ldots,n\}\setminus\{i\}$, and let $f_i(\zvec_{N_i}) = \zvec_{N_i}$. Then
\[
Y_i(\zvec)= Y_i(z_i, \zvec_{-i}) = \alpha_i + \beta_i z_i + B_i(\zvec_{-i}) + z_i C_i(\zvec_{-i}).
\]
Row $i$ of $\mathbb{T}$ takes $2^n$ distinct values; Assumption~\ref{assump:interference-exclusion} places no constraint.
\end{example}

\begin{example}[Neighborhood interference]
Let $N_i \subsetneq \{1,\ldots,n\}\setminus\{i\}$ be a specified proper subset, and let $f_i(\zvec_{N_i}) = \zvec_{N_i}$.  Then
\[
Y_i(\zvec) = Y_i(z_i, \zvec_{N_i}) = \alpha_i + \beta_i z_i + B_i(\zvec_{N_i}) + z_i C_i(\zvec_{N_i}).
\]
Row $i$ of $\mathbb{T}$ takes $2^{d_i+1}$ distinct values, where $d_i = |N_i|$; the exponent reduces from $n-1$ to the local neighborhood size.
\end{example}

\begin{example}[No interference]
Let $N_i = \emptyset$. Then $Y_i(\zvec) = Y_i(z_i)$, which is the classical no-interference case. Each unit has two own-treatment potential outcomes.
\end{example}

\begin{example}[Count or fractional exposure mappings]
Let $e_i = f_i(\zvec_{N_i}) = \sum_{j \in N_i} z_j$ denote the number of treated neighbors of unit $i$, or alternatively let $e_i =\sum_{j \in N_i} z_j/d_i$ denote the proportion of treated neighbors of unit $i$. Then
\begin{align}
\label{eq:symmetricExposure}
Y_i(z_i,e_i) &= \alpha_i + \beta_i z_i + B_i(e_i) + z_iC_i(e_i),
\end{align}
where $e_i \in \{0,1,\ldots, d_i\}$ in the count case and $e_i \in \{0,1/d_i,2/d_i,\ldots, 1\}$ in the fractional case. Under this model, the number of free parameters per unit is $2(d_i+1)$. If additivity is imposed, $C_i(e_i)=0$ for all $e_i$, and the number of free parameters per unit reduces to $d_i+2$.
\end{example}

\begin{example}[Linear-in-means model]
Starting from equation \eqref{eq:symmetricExposure}, let $B_i(e_i) = \gamma_i e_i$ and $C_i(e_i) = \theta_i e_i$. Then
\begin{align}
\label{eq:LinearSymmExposure}
Y_i(z_i,e_i)
&= \alpha_i + \beta_i z_i + \gamma_i e_i  + \theta_i z_i e_i .
\end{align}
If the interaction term is zero, the model becomes
\begin{align}
\label{eq:LinearAddSymmExposure}
Y_i(z_i,e_i)
&= \alpha_i + \beta_i z_i + \gamma_i e_i .
\end{align}
This additive linear outcome model is the linear-in-means model used in the econometrics literature \citep{manski1993identification,bramoulle2009identification,graham2008identifying}.
\end{example}

\begin{example}[$2 \times 2$ exposure mapping]
Let $f_i(\zvec_{N_i}) = 0$ if all elements of $\zvec_{N_i}$ are $0$ and $f_i(\zvec_{N_i})=1$ if at least one element of $\zvec_{N_i}$ is $1$. This gives the two-by-two response-surface model, or binary exposure mapping,
\begin{align}
\label{eq:BinaryExposure}
Y_i(z_i,e_i) = \alpha_i + \beta_i z_i + \gamma_i e_i + \theta_i z_i e_i,
\end{align}
where $e_i \in \{0,1\}$ and $z_i \in \{0,1\}$. Each unit has four distinct potential outcomes. Setting $\theta_i=0$ gives the additive binary exposure mapping,
\begin{align}
\label{eq:addBinaryExposure}
Y_i(z_i,e_i) = \alpha_i + \beta_i z_i + \gamma_i e_i,
\end{align}
with three free parameters per unit.
\end{example}

\begin{example}[Weighted exposure mappings]
In non-anonymous interference, the exposure depends on the identity or covariates of the neighbors. For example, one may specify
\[
e_i=\sum_{j\in N_i}w_{ij}z_j,
\]
and then write
\[
Y_i(z_i,e_i)=\alpha_i+\beta_i z_i+B_i(e_i)+z_iC_i(e_i).
\]
The additive version of this model has been used for total treatment effects under network interference \citep{yu2022pnasTTE}. Additional non-anonymous and spatial examples are given in Appendix~\ref{app:additional-framework-examples}.
\end{example}

The three canonical non-parametric representations---arbitrary interference, neighborhood interference, and exposure mapping (illustrated by examples 4--7 above)---are compared in Section~\ref{sec:nonparametric-estimation} with respect to their estimation bottlenecks.

\section{Defining causal estimands under network interference}
\label{sec:estimands}
Under treatment interference, there are many non-equivalent ways to define causal effects, and the appropriate estimand depends on the scientific question of interest. Causal estimands in the framework of Section~\ref{sec:POmodels} are summaries of the causal response surface \(Y_i(z,e)\). We study two broad classes: \emph{average causal effects} (ACEs), which are contrasts of fixed potential outcomes and do not involve any distribution over treatment allocations; and \emph{expected causal effects} (ECEs), which are contrasts of expectations of potential outcomes under a specified allocation policy.

\subsection{Average causal effects under network interference}
\label{sec:ace}

\emph{Average causal effects} (ACEs) are defined as contrasts between two fixed
potential outcomes.  Once an interference neighborhood and exposure mapping have been specified, the
treatment--exposure space for unit \(i\) is
\[
\tespace_i=\{0,1\}\times\mathcal E_i.
\]
For an assignment \(\zvec\), let
\[
\Dcond_i(\zvec)=(z_i,e_i(\zvec))\in\tespace_i
\]
denote the treatment--exposure combination realized by unit \(i\).  A generic ACE
compares two fixed treatment--exposure combinations
\[
d_i^+=(z_i^+,e_i^+),\qquad d_i^-=(z_i^-,e_i^-),
\qquad d_i^+,d_i^-\in\tespace_i .
\]
Let \(d^+=\{d_i^+\}_{i=1}^n\) and \(d^-=\{d_i^-\}_{i=1}^n\).  The corresponding ACE is
defined as
\begin{equation}
\tau_{\mathrm{ACE}}(d^+,d^-)
\,=\,\frac{1}{n}\sum_{i=1}^n
\Big(Y_i(d_i^+) - Y_i(d_i^-)\Big).
\label{eq:ACE-general}
\end{equation}
This definition directly generalizes the average treatment effect under no-interference assumptions, and does not require assumptions on the functional form of the response surface. Different choices of \(d_i^+\) and \(d_i^-\) correspond to different causal questions about treatment and interference. %

This definition also includes arbitrary interference as a special case: take \(N_i=\{1,\ldots,n\}\setminus\{i\}\) and use the identity exposure mapping \(e_i=\zvec_{-i}\). Then \(\Dcond_i(\zvec)=(z_i,\zvec_{-i})\), and equation~\eqref{eq:ACE-general} defines ACEs directly in terms of the full treatment vector. %

We now give a few examples of ACEs. Recall that the exposure levels $e_i=0$ and $e_i=1$ are special values defined to represent the situations when a unit is not exposed or fully exposed respectively.

\paragraph{Direct treatment effect.}
The direct treatment effect compares the average outcome under treatment versus control when exposure is held at the baseline level of $0$:
\begin{equation}
\tau_{\mathrm{DTE}}
\,=\, \tau_{\mathrm{ACE}}(d^+,d^-)
\,=\,\frac{1}{n}\sum_{i=1}^n
\Big(Y_i(1,0)-Y_i(0,0)\Big).
\label{eq:ACE-direct}
\end{equation}
where \(d_i^+=(1,0)\) and \(d_i^-=(0,0)\) for all \(i\).
The direct treatment effect is the interference analogue of the average treatment effect in the no-interference case. It measures the average effect of treatment on a unit when the exposure is fixed at the baseline level.

\paragraph{Average indirect effects.}
For two exposure levels $e_a,e_b\in\mathcal{E}$ and own treatment $z\in\{0,1\}$, define the average indirect effect
\begin{equation}
\tau_{\mathrm{IE}}(z;e_a,e_b)
\,=\,\frac{1}{n}\sum_{i=1}^n
\Big(Y_i(z,e_b)-Y_i(z,e_a)\Big),
\label{eq:ACE-spill-generic}
\end{equation}
which holds the unit's own treatment fixed and varies only the exposure regime between the two levels. The pair $(e_a,e_b)$ is part of the estimand: spillover questions can concern any two exposure levels, and the baseline $e_a$ need not be the no-exposure level. The special case $e_a=0$ recovers the level-specific indirect effects $\tau_{\mathrm{IEC}}(e)=\tau_{\mathrm{IE}}(0;0,e)$ and $\tau_{\mathrm{IET}}(e)=\tau_{\mathrm{IE}}(1;0,e)$ on control and treated units, respectively. Writing $\tau_{\mathrm{IEC}}=\tau_{\mathrm{IEC}}(e_i^{\mathbf 1})$ and $\tau_{\mathrm{IET}}=\tau_{\mathrm{IET}}(e_i^{\mathbf 1})$ for the full-exposure versions at the all-treated level $e=e_i^{\mathbf 1}$,
\begin{equation}
\tau_{\mathrm{IEC}}
\,=\,\frac{1}{n}\sum_{i=1}^n
\Big(Y_i(0,1)-Y_i(0,0)\Big),
\label{eq:ACE-spill-0}
\end{equation}
\begin{equation}
\tau_{\mathrm{IET}}
\,=\,\frac{1}{n}\sum_{i=1}^n
\Big(Y_i(1,1)-Y_i(1,0)\Big).
\label{eq:ACE-spill-1}
\end{equation}
A natural two-level exposure pair arises from the \emph{any-treated} exposure map
\begin{equation}
U_i=\mathds 1\Big\{\textstyle\sum_{j\in N_i}Z_j\ge 1\Big\},
\label{eq:any-treated}
\end{equation}
which records whether at least one neighbor of unit $i$ is treated. With $e_a=0$ and $e_b=1$, the contrast $\tau_{\mathrm{IE}}(z;0,1)$ measures the average effect of moving a unit with own treatment $z$ from an environment free of treated neighbors to one with at least one treated neighbor. Held at $z=0$, this is the \emph{influencer spillover effect} estimated in influencer-seeding experiments in social-network settings: the effect on an average untreated unit of the presence of a treated influencer among its neighbors \citep{paluck2016changing,aralwalker2012identifying,bond201261}. Section~\ref{sec:is-indirect} shows that this two-level estimand is estimable at constant-order focal propensities on bounded-degree graphs by a sparse-seeding design, in contrast to the exponentially small propensities of its extreme full-exposure special cases.

By Proposition~\ref{prop:estimand-representation} below, $\tau_{\mathrm{IE}}(0;e_a,e_b)=n^{-1}\sum_i\{B_i(e_b)-B_i(e_a)\}$ captures the additive exposure component, while $\tau_{\mathrm{IE}}(1;e_a,e_b)$ adds the treatment--exposure interaction; the two coincide under additivity ($C_i\equiv 0$).

\begin{remark}
We use the term average indirect effect for the fixed-exposure ACE estimands above. This is different from the average indirect effect introduced in \cite{huLiWagerBiometrika}, which we define as the expected indirect effect as it belongs to the expected causal effects class.
\end{remark}

\paragraph{Total treatment effect.}
The total treatment effect compares the fully treated and fully exposed condition to the fully untreated and unexposed condition:
\begin{equation}
\tau_{\mathrm{TTE}}
\,=\, \tau_{\mathrm{ACE}}(d^+,d^-)
\,=\, \frac{1}{n}\sum_{i=1}^n
\Big(Y_i(1,1)-Y_i(0,0)\Big).
\label{eq:ACE-total}
\end{equation}
where \(d_i^+=(1,1)\) and \(d_i^-=(0,0)\) for all \(i\).

\subsection{Expected causal effects under network interference}
\label{sec:ece}

Average causal effects compare two fixed potential outcomes. In many cases, however, the exposure of a unit is random under a policy of scientific interest. Expected causal effects are defined as contrasts of allocation-averaged responses, where entries of the causal response surface are averaged over one or more allocation distributions specified by the estimand. The target allocation policy is part of the causal question itself. 
A generic expected causal effect contrasts random treatment--exposure combinations induced by an allocation policy.  Let $\Qpol^+$ and $\Qpol^-$ denote two allocation policies. For
each unit \(i\), let
\[
\Dcond_i^+(\Zvec),\qquad \Dcond_i^-(\Zvec)
\]
be two corresponding random elements of \(\tespace_i\).  A generic ECE is any contrast of the form
\begin{equation}
\tau_{\mathrm{ECE}}(\Dcond^+,\Dcond^-;\Qpol^+,\Qpol^-)
=
\frac{1}{n}\sum_{i=1}^n
\left[
\mathbb E_{\Qpol^+}
Y_i\{\Dcond_i^+(\Zvec)\}
-
\mathbb E_{\Qpol^-}
Y_i\{\Dcond_i^-(\Zvec)\}
\right].
\label{eq:ECE-general}
\end{equation}
The distributions \(\Qpol^+, \Qpol^-\) can be different from the experimental design \(P\), but are often induced by \(P\). For example, a study may use a randomized saturation design with several treatment probabilities to estimate the expected effect of a single target rollout policy, such as treating 70 percent of units; the rollout policy defines the estimand, while the saturation design is chosen to provide overlap and precision.
For the special case where \(\Qpol^+=\Qpol^-=\Qpol\), we write
\(\tau_{\mathrm{ECE}}(\Dcond^+,\Dcond^-;\Qpol)\).  
We now consider some examples of expected causal effects used in the literature. Let
\(P_{\Qpol}(E_i=e)\) denote the marginal distribution of the exposure level \(E_i\)
under allocation policy \(\Qpol\). The expected response of unit \(i\) under
own treatment \(z \in \{0,1\}\) is
\begin{equation}
\mu_i(z;\Qpol)
\,=\, \sum_{e \in \mathcal{E}} Y_i(z,e)\,P_{\Qpol}(E_i=e).
\label{eq:mu-marginal}
\end{equation}

\paragraph{Expected (average) direct effect.}
The expected direct effect compares the average outcome under treatment versus control, averaging over the exposure distribution induced by the allocation policy:
\begin{equation}
    \tau_{\mathrm{EDE}}(\Qpol) =
    \mathbb{E}_{\Qpol} \left[ \frac{1}{n} \sum_{i=1}^n
    \left( Y_i(1, E_i) - Y_i(0, E_i) \right) \right]
    \,=\,\frac{1}{n}\sum_{i=1}^n \Big(\mu_i(1;\Qpol)-\mu_i(0;\Qpol)\Big).
\label{eq:ECE-direct}
\end{equation}
This is the one-contrast special case of~\eqref{eq:ECE-general} with
\(\Dcond_i^+(\Zvec)=(1,E_i)\), \(\Dcond_i^-(\Zvec)=(0,E_i)\), and common policy
\(\Qpol\).
The expected average direct effect is also known as the Average Direct Effect (ADE) in the literature, see \cite{savje2017average}. We use the term expected direct effect to emphasize that it is an ECE and not an ACE.

\paragraph{Expected (average) indirect effect.}
The expected indirect effect compares the expected outcome of unit $i$ when unit $j$'s treatment is changed, keeping all other treatment assignments the same.
For $j\in N_i$, define
\[
E_i^{+j}=f_i(\Zvec_{N_i\setminus\{j\}},Z_j = 1),
\qquad
E_i^{-j}=f_i(\Zvec_{N_i\setminus\{j\}},Z_j = 0),
\]
as the exposure of unit $i$ when neighbor $j$ is treated or untreated, holding all others fixed. Then,
\begin{equation}
\tau_{\mathrm{EIE}}(\Qpol)
=\frac1n\sum_{i=1}^n
\sum_{j\in N_i}
\mathbb E_{\Qpol}\!\Big[
Y_i\!\big(Z_i,E_i^{+j})\big)
-
Y_i\!\big(Z_i,E_i^{-j}\big)
\Big].
\label{eq:ECE-neighbor-indirect}
\end{equation}
This is an aggregation of neighbor-specific ECE contrasts.  For each ordered pair
\((i,j)\) with \(j\in N_i\), define
\[
\Dcond_{ij}^+(\Zvec)=(Z_i,E_i^{+j}),
\qquad
\Dcond_{ij}^-(\Zvec)=(Z_i,E_i^{-j}).
\]
The expected indirect effect is also known as the Average Indirect Effect. The arbitrary interference version of this estimand was introduced in \cite{huLiWagerBiometrika}.

\begin{proposition}[Policy derivative representation]
\label{prop:HuLiWager}
Under a Bernoulli policy $\Qpol_\pi$ with $\Zvec\sim\prod_i\mathrm{Bern}(\pi_i)$, writing $\mu(\pi)=n^{-1}\sum_i \mathbb E_\pi[Y_i(Z_i,e_i(\Zvec_{N_i}))]$ for the expected population outcome, the total policy derivative $\tau_{\mathrm{TPE}}(\pi)=\sum_j \partial\mu(\pi)/\partial\pi_j$ decomposes as $\tau_{\mathrm{TPE}}(\pi)=\tau_{\mathrm{EDE}}(\pi)+\tau_{\mathrm{EIE}}(\pi)$, recovering the Average Direct Effect / Average Indirect Effect decomposition of \citet{huLiWagerBiometrika} under arbitrary interference as a special case.  Under symmetric exposure mappings, EIE has a simpler finite-difference form; details are in Appendix~\ref{app:policy-derivatives}.
\end{proposition}

\paragraph{Expected policy effect.}
The general definition also allows the two arms of the contrast to be averaged under
different allocation policies.  Let \(\Qpol^+\) and \(\Qpol^-\) be two policies, for
example Bernoulli policies with treatment probabilities \(p_+\) and \(p_-\).  The
expected policy effect is
\begin{equation}
\tau_{\mathrm{policy}}(\Qpol^+,\Qpol^-)
=
\frac1n\sum_{i=1}^n
\left[
\mathbb E_{\Qpol^+}Y_i\{\Dcond_i(\Zvec)\}
-
\mathbb E_{\Qpol^-}Y_i\{\Dcond_i(\Zvec)\}
\right].
\label{eq:ECE-policy}
\end{equation}
This is the special case of~\eqref{eq:ECE-general} with
\(\Dcond_i^+(\Zvec)=\Dcond_i^-(\Zvec)=\Dcond_i(\Zvec)\), but with
\(\Qpol^+\neq \Qpol^-\).  It represents questions such as comparing an allocation policy with
50 percent treatment probability to one with 20 percent treatment probability.

Mixed effects that combine policy-averaged and fixed-baseline potential outcomes (EIC, EIT, ETE) are defined in Appendix~\ref{app:mixed-effects}.

\subsection{Algebraic representation of ACEs and ECEs}
\label{sec:estimand-representation}
We now represent the two classes of estimands using the non-parametric linear decomposition of the causal response surface.
\begin{proposition}[Representation of fixed and expected effects]
\label{prop:estimand-representation}
Under the non-parametric linear decomposition of Theorem \ref{thm:decomposition},
\[
Y_i(z,e)=\alpha_i+\beta_i z+B_i(e)+zC_i(e),
\qquad B_i(0)=C_i(0)=0,
\]
the average causal effects can be written as
    \begin{align*}
    \tau_{\mathrm{DTE}} &= \frac{1}{n}\sum_i \beta_i, \\
    \tau_{\mathrm{IE}}(z;e_a,e_b) &= \frac{1}{n}\sum_i\{B_i(e_b)-B_i(e_a)+z\,(C_i(e_b)-C_i(e_a))\}, \\
    \tau_{\mathrm{IEC}}(e) &= \frac{1}{n}\sum_iB_i(e), \\
    \tau_{\mathrm{IET}}(e) &= \frac{1}{n}\sum_i\{B_i(e)+C_i(e)\}, \\
    \tau_{\mathrm{TTE}} &= \frac{1}{n}\sum_i \{\beta_i+B_i(1)+C_i(1)\}.
    \end{align*}
For an allocation policy \(\Qpol\), let
\(\bar B_{\Qpol}=n^{-1}\sum_i \mathbb E_{\Qpol}[B_i(E_i)]\) and
\(\bar C_{\Qpol}=n^{-1}\sum_i \mathbb E_{\Qpol}[C_i(E_i)]\). Then
\[
\tau_{\mathrm{EDE}}(\Qpol) = \frac{1}{n}\sum_i \beta_i+\bar C_{\Qpol}.
\]
The mixed effects EIC, EIT, and ETE satisfy analogous expressions: $\tau_{\mathrm{EIC}}(\Qpol)=\bar B_{\Qpol}$, $\tau_{\mathrm{EIT}}(\Qpol)=\bar B_{\Qpol}+\bar C_{\Qpol}$, and $\tau_{\mathrm{ETE}}(\Qpol)=n^{-1}\sum_i\beta_i+\bar B_{\Qpol}+\bar C_{\Qpol}$; see Appendix~\ref{app:mixed-effects}.
\end{proposition}

These representations connect the estimands directly to the components of the non-parametric response-surface decomposition. While fixed-exposure ACEs 
select values of $B_i$ and $C_i$ at specific exposure levels, ECEs average these same components 
over the exposure distribution induced by the allocation policy. 
Specifically, $\tau_{\mathrm{DTE}}$ isolates the direct effect component 
$\beta_i$, whereas $\tau_{\mathrm{IEC}}$ and $\tau_{\mathrm{IET}}$ capture 
the pure indirect and indirect-plus-interference components, respectively. 
Notably, the expected direct effect encapsulates both the pure direct 
component and the treatment--interference interaction component, the latter being $0$ under additivity. This yields the following corollary:
\begin{corollary}
\label{cor:additive-ede-dte}
Under additivity of the causal response surface, $\tau_{\mathrm{EDE}}(\Qpol)=\tau_{\mathrm{DTE}}$ for every allocation policy \(\Qpol\).
\end{corollary}

Without additivity, the expected direct effect contains the policy-weighted interaction component $\bar C_{\Qpol}$ and is policy-dependent: estimators that target EDE do not target the fixed-exposure DTE. The total treatment effect also has a decomposition:

\begin{proposition}[ACE total effect decomposition]
\label{prop:ace-decomposition}
The total treatment effect decomposes as
\[
\tau_{\mathrm{TTE}} = \tau_{\mathrm{DTE}} + \tau_{\mathrm{IET}}.
\]
Under additivity ($C_i\equiv 0$), $\tau_{\mathrm{IET}}=\tau_{\mathrm{IEC}}$, so $\tau_{\mathrm{TTE}}=\tau_{\mathrm{DTE}}+\tau_{\mathrm{IEC}}$.
\end{proposition}

Decompositions involving the mixed effects EIC, EIT, and ETE are given in Appendix~\ref{app:mixed-effects}.

\subsection{Relevant and Nuisance Potential Outcomes}
\label{sec:nuisance-pos}

Another distinction between average causal effects (ACEs) and expected causal effects (ECEs) ties to which entries of $\mathbb S$ define the estimand. %
In a non-parametric framework, no outcome model is imposed on the distinct potential outcomes: observing one element of $\mathbb S$ provides no information about another distinct element of $\mathbb S$. Thus, potential outcomes that do not enter the definition of an estimand are uninformative for its non-parametric estimation, regardless of how often an experimental design reveals them. We formalize this distinction by partitioning the causal response surface relative to a target estimand. 

\begin{definition}[Relevant and nuisance potential outcomes]\label{def:relevant-potential-outcomes}
Fix a causal estimand $\tau$, defined as a functional of the causal response surface $\mathbb S$.
A potential outcome $Y_i(z,e)$ is relevant for $\tau$ if the functional defining $\tau$ depends on $Y_i(z,e)$.
The relevant set is
\[
\mathcal{R}_\tau
=
\{Y_i(z,e)\in\mathbb{S}: \tau \text{ depends on } Y_i(z,e)\}.
\]
The complement $\mathcal{R}_\tau^c=\mathbb{S}\setminus\mathcal{R}_\tau$ is the set of nuisance potential outcomes for $\tau$.
\end{definition}
For ACEs, the relevant set $\mathcal{R}_\tau$ is typically a sparse subset of the causal response surface. For example, a direct treatment effect depends only on
\[
\{Y_i(1,0), Y_i(0,0)\}_{i=1}^n,
\]
whereas a total treatment effect depends only on
\[
\{Y_i(1,1), Y_i(0,0)\}_{i=1}^n.
\]
All other entries of $\mathbb S$ are nuisance for non-parametric estimation of these estimands. For example, if a design exposes a unit to an intermediate exposure level $e$, revealing $Y_i(1,e)$, that observation provides no non-parametric information about target potential outcomes such as $Y_i(1,1)$ or $Y_i(1,0)$. Without an outcome model relating distinct potential outcomes of $\mathbb S$ across treatment or exposure levels, observations from nuisance potential outcomes cannot reduce variance for the target estimand. They may instead waste probability mass, contribute bias, or require outcome modeling assumptions that go beyond non-parametric estimation.

For ECEs, the allocation policy is part of the definition of the estimand. Any treatment--exposure combination that is assigned positive probability under the allocation policy is included in the relevant set $\mathcal{R}_\tau$ of an ECE. For example, the relevant set for the expected direct effect $\tau_{\mathrm{EDE}}(\Qpol)$ is the set of $Y_i(1,e)$ and $Y_i(0,e)$ for each exposure level $e$ in the support of the exposure distribution induced by policy $\Qpol$. When the policy induces positive probability on every treatment--exposure combination in the specified representation, the relevant set is the entire causal response surface.

This difference in relevant potential outcomes leads to different non-parametric estimation problems, as discussed in Section~\ref{sec:nonparametric-estimation}. For ACEs, the design must reveal the relevant potential outcomes. Non-parametric estimation thus requires the propensity scores for those combinations to be bounded away from zero. When the probability of revealing the relevant potential outcomes is rare under the design, inverse-probability weights become unstable and variance can be large. Consequently, non-parametric estimation of ACEs often requires specialized designs that place sufficient probability mass on the relevant potential outcomes; such designs are discussed in Section~\ref{sec:implications}. For ECEs, the estimand instead averages over the treatment--exposure combinations induced by the policy. The relevant support is therefore determined by the policy itself. In this case, the main variance bottleneck is not the probability of observing a narrow set of fixed treatment--exposure entries, but the overlap between the observed assignment design and the policy. Importance-weighted HT estimators use outcomes across the support of the policy, with precision governed by policy-to-design weights, network dependence, and the extent of support overlap.

\subsection{Conditional expected effects}
\label{subsec:other_estimands}

The two classes introduced in Sections~\ref{sec:ace} and~\ref{sec:ece} also clarify conditional estimands that have appeared in the literature on partial interference. \citet{hudgens2012toward} introduced causal effects under partial interference based on conditional expectations of potential outcomes. In their setting, units are partitioned into disjoint interference groups, and causal effects are defined by conditioning on the treatment assignment within the group. Formally, one considers conditional potential outcomes of the form
\[
\mathbb{E}\!\bigl[Y_i(z,e) \,\big|\, Z_i=z\bigr],
\]
where the expectation is taken with respect to the distribution of exposure in the group, conditional on the individual treatment. Estimands are then defined as contrasts between such conditional expectations. These estimands can be viewed as \emph{conditional expected causal effects}, which differ from the marginal expected effects defined in Section~\ref{sec:ece}. Conditional estimands depend on the randomization scheme within groups, and are most natural under two-stage designs with partial interference in which treatment probabilities are assigned at the group level. In arbitrary network interference, or under designs where a unit's treatment and exposure are dependent, conditioning on $Z_i$ also changes the exposure distribution. The resulting estimand remains well defined, but its interpretation depends on the conditional allocation distribution; see \cite{savje2017average} for a discussion.

\section{Estimation of ACEs and ECEs}
\label{sec:nonparametric-estimation}
In this section, we study design-unbiased estimators of ACEs and ECEs. In 
particular, under a fixed experimental design $P$, we derive upper bounds 
for the variance of canonical Horvitz--Thompson (HT) estimators as well as  lower bounds on the design variance of any unbiased estimator. Finally, we establish the 
conditions under which these estimators achieve design consistency.
Under a growing sequence of finite populations 
indexed by $n$, an estimator $\hat\tau_n$ is design-MSE consistent for 
$\tau_n$ if 
\[ \mathbb E_{P_n}\{(\hat\tau_n-\tau_n)^2\}\to 0. \] 
The estimation machinery applies to any choice of interference graph and exposure mapping. As examples, we study estimation under the three examples 
introduced in Section~\ref{sec:POmodels}: arbitrary interference  \((Z_i,\Zvec_{-i})\), 
neighborhood interference \((Z_i,\Zvec_{N_i})\), and exposure mappings \((Z_i,e_i(\Zvec_{N_i}))\). Throughout this section, 
no outcome modeling assumptions are imposed. Once the interference neighborhood and exposure mapping have been fixed, non-parametric estimation treats the distinct potential outcomes of \(\mathbb S\) as unrestricted. Observing one distinct potential outcome \(Y_i(z,e)\) does not, without an outcome model, identify or reduce uncertainty about another distinct potential outcome \(Y_i(z',e')\).

\subsection{Dependence from interference and design}
\label{sec:dependence-graphs-estimation}

Consider a generic estimator \(\hat\tau_n\) that is a function of random variables
\(X_1,\ldots,X_n\).  In the design-based setting, the potential-outcome table is fixed,
so any observed-outcome term \(Y_i^{obs}=Y_i(\Zvec)\) is a measurable function of the
treatment assignment vector.  Thus each \(X_i\) can be written generally as \(X_i=g_i(\Zvec)\).
The variance bounds in the sequel use dependence graphs for the random variables
\(\{X_i\}\).

\begin{definition}[Dependence graph and average dependence]
Consider random variables \(\{X_i\}_{i=1}^n\).  A graph
\(G_{\mathrm{dep}}=(V,E_{\mathrm{dep}})\) is a dependence graph of $\{X_i\}_{i=1}^n$ if the following holds:
whenever two disjoint sets of vertices \(A,B\subseteq V\) have no edge between them,
the collections \(\{X_i:i\in A\}\) and \(\{X_j:j\in B\}\) are independent.  We write
\(c_{ij}=1\) when \(i\) and \(j\) are adjacent in this dependence graph and
\(c_{ij}=0\) otherwise.  The dependence degree of \(i\) is
\[
c_i=\sum_{j\ne i}c_{ij},
\]
and the average dependence degree is
\[
c_{\mathrm{avg}}
=
\frac1n\sum_{i=1}^n c_i
=
\frac1n\sum_{i=1}^n\sum_{j\ne i}c_{ij}.
\]
\end{definition}

\begin{remark}[Estimator dependence versus interference neighborhoods]
\label{rem:estimator-dependence-sets}
The dependence graph is an object associated with the estimator $\hat tau_n$ and the random variables \(X_i\), not
directly with the causal estimand.  If \(X_i\) depends only on a subvector of the
treatment assignment, we may write \(X_i=g_i(\Zvec_{A_i})\), where
\(A_i\subseteq\{1,\ldots,n\}\) is an assignment-dependence set for that estimator
variable.  The set \(A_i\) includes the treatment assignment coordinates used by the observed outcome and by
any random weights or normalizing factors.  When an interference graph with neighborhoods
\(N_i\) is specified, write \(\mathcal C_i=\{i\}\cup N_i\) for the closed interference
neighborhood; this includes arbitrary interference as the special case
\(N_i=\{1,\ldots,n\}\setminus\{i\}\).  For the local HT/IPW estimators below, the observed
outcome has the form \(Y_i(Z_i,\Zvec_{N_i})\) and the weights are also a function of $(Z_i, \Zvec_{N_i})$.  Hence \(A_i\subseteq\mathcal C_i\).  Ratio estimators such as Hajek estimators
may introduce additional assignment dependence because their weights depend on the full treatment assignment vector $\Zvec$.
\end{remark}

Typically, the variables $X_i$ and $X_j$ can be dependent for two reasons:
their treatment-assignment index sets $A_i$ and $A_j$ can overlap because of interference, and the
randomization design can directly induce dependence among treatment assignments, even when $A_i$ and $A_j$ do not overlap.
Below, we separate these two sources into interference-overlap dependence and residual
design dependence.

Under independent treatment assignment, there is no residual design dependence.  The
remaining dependence is induced by overlap of the assignment coordinates on which the
variables \(X_i\) depend: if \(X_i=g_i(\Zvec_{A_i})\), then
\(\mathds{1}\{A_i\cap A_j\neq\emptyset\}\) gives a valid dependence edge under
independent assignment.  The following lemma records the closed-neighborhood
specialization used for the local HT/IPW estimators studied below.  %

\begin{lemma}[Dependence induced by local estimator variables]
\label{lem:interference-dependence-graph}
Suppose the treatment assignments are mutually independent and \(X_i\) is a function of
\(\Zvec_{\mathcal C_i}\), where \(\mathcal C_i=\{i\}\cup N_i\) is the closed interference neighborhood of unit \(i\).  Then a valid
dependence graph of $\{X_i\}$ is obtained by setting
\[
c_{ij}=\mathds{1}\{\mathcal C_i\cap \mathcal C_j\neq\emptyset\},
\qquad i\neq j.
\]
If the interference graph is undirected, \(d_i=|N_i|\), and \(B_2(i)\) denotes the
two-hop interference neighborhood of \(i\), including \(i\), then
\[
c_i=|B_2(i)|-1
\]
and
\[
b_{2,\mathrm{avg}}
:=
\frac1n\sum_{i=1}^n\{|B_2(i)|-1\}
=
c_{\mathrm{avg}},
\qquad
\bar d
\le
b_{2,\mathrm{avg}}
\le
\min\left\{n-1,\frac1n\sum_{i=1}^n d_i^2\right\},
\qquad
\bar d=\frac1n\sum_{i=1}^n d_i .
\]
\end{lemma}

Thus, for independent assignment on an undirected interference graph, the average
dependence degree of estimators that use local weights is controlled by the second moment of the degree distribution of the interference graph rather
than by the average degree alone.  For bounded-degree and sparse
Erd\H{o}s--Renyi interference graphs with bounded expected degree, \(c_{\mathrm{avg}}=O(1)\).  On
locally tree-like \(d\)-regular graphs, \(c_{\mathrm{avg}}\asymp d^2\).  On a star
graph, \(\bar d=O(1)\) but \(c_{\mathrm{avg}}\asymp n\), because all leaves have
overlapping closed neighborhoods through the hub.  On complete graphs,
\(c_{\mathrm{avg}}=n-1\).

Under dependent randomization designs, two random variables $X_i$ and $X_j$ can be dependent even when their
closed interference neighborhoods do not overlap.  We measure this residual
design-induced dependence by the following coefficient.

\begin{definition}[Design dependence coefficient]
\label{def:main-design-dependence-coefficient}
For two assignment-index sets \(A,B\subseteq\{1,\ldots,n\}\), define
\[
\rho_P(A,B)
=
\sup_{\|f\|_\infty\le 1,\,\|g\|_\infty\le 1}
\left|
\operatorname{Cov}_P\{f(\Zvec_A),g(\Zvec_B)\}
\right|,
\]
where the supremum is over all measurable functions of the indicated subvectors.
\end{definition}
For the local HT/IPW variance bounds below, we use
\[
\rho_{ij}=\rho_P(\mathcal C_i,\mathcal C_j),
\]
and
\[
r_{\mathrm{avg}}
=
\frac1n\sum_{i=1}^n\sum_{j\ne i}(1-c_{ij})\rho_{ij}.
\]
Here \(\rho_{ij}\) is the pairwise design-dependence coefficient between the two
closed-neighborhood assignment subvectors, while \(r_{\mathrm{avg}}\) averages only the
residual design dependence among pairs whose closed neighborhoods do not overlap.
If the design is independent assignment and \(\mathcal C_i\cap\mathcal C_j=\emptyset\),
then \(\rho_{ij}=0\).  Complete randomization, cluster randomization, paired designs,
and other dependent designs can have \(\rho_{ij}>0\) even for disjoint closed
neighborhoods.
These quantities separate overlap of assignment coordinates from dependence induced
directly by the randomization design.  The variance bounds below therefore separate
three sources of variation: the diagonal scale of the HT or importance weights,
dependence generated by overlapping closed interference neighborhoods, and residual dependence
generated by the randomization design.  %

\subsection{ACE estimation}
\label{sec:ace-propensity-scores}

We first consider nonparametric estimation of an ACE
\[
\tau_{\mathrm{ACE}}(d^+,d^-)
=
\frac{1}{n}\sum_{i=1}^n
\{Y_i(d_i^+)-Y_i(d_i^-)\}.
\]
Write \(\pi_i(d)=P\{\Dcond_i(\Zvec)=d\}\) for the
design-induced propensity that unit \(i\) is observed in treatment--exposure combination
\(d\).  The generic Horvitz--Thompson estimator is
\[
\hat\tau_{\mathrm{ACE}}^{HT}(d^+,d^-)
=
\frac1n\sum_{i=1}^n
Y_i^{obs}
\left\{
\frac{\mathds{1}\{\Dcond_i(\Zvec)=d_i^+\}}{\pi_i(d_i^+)}
-
\frac{\mathds{1}\{\Dcond_i(\Zvec)=d_i^-\}}{\pi_i(d_i^-)}
\right\}.
\]
This is the exposure-mapping Horvitz--Thompson estimator of
\citet{aronow2017estimating}, written here for the generic ACE notation.
For the variance bound, let
\[
H_i=\frac{1}{\pi_i(d_i^+)}+\frac{1}{\pi_i(d_i^-)}.
\]

\begin{proposition}[ACE Horvitz--Thompson estimator]
\label{prop:generic-ace-ht}
Assume \(\pi_i(d_i^+)>0\) and \(\pi_i(d_i^-)>0\) for all \(i\). Then
\(\hat\tau_{\mathrm{ACE}}^{HT}(d^+,d^-)\) is design-unbiased for
\(\tau_{\mathrm{ACE}}(d^+,d^-)\) for every potential-outcome table.  If, in addition,
\(|Y_i(d)|\le M\) and the \(i\)th summand depends on the assignment only through
\(\Zvec_{\mathcal C_i}\),
then
\[
\operatorname{Var}\{\hat\tau_{\mathrm{ACE}}^{HT}(d^+,d^-)\}
\le
\frac{M^2}{n^2}
\left\{
\sum_{i=1}^n H_i
+
\sum_{i=1}^n\sum_{j\ne i}c_{ij}H_iH_j
+
\sum_{i=1}^n\sum_{j\ne i}(1-c_{ij})\rho_{ij}H_iH_j
\right\},
\]
and if \(H_i\le H\) for all \(i\),
then
\[
\operatorname{Var}\{\hat\tau_{\mathrm{ACE}}^{HT}(d^+,d^-)\}
\le
M^2\left\{
\frac{H}{n}
+
H^2\frac{c_{\mathrm{avg}}+r_{\mathrm{avg}}}{n}
\right\}.
\]
Consequently, along a sequence of finite populations, the estimator is design-MSE
consistent if \(H+H^2(c_{\mathrm{avg}}+r_{\mathrm{avg}})=o(n)\).
\end{proposition}

The proof is given in Appendix~\ref{app:nonparametric-estimation-proofs}.
This proposition separates three sources of variation in the HT estimator: the inverse propensity scores
of the relevant potential outcomes \(Y_i(d_i^+)\) and \(Y_i(d_i^-)\) ($H_i$), the dependence generated by overlapping closed interference neighborhoods ($c_{\mathrm{avg}}$), and residual dependence
from the randomization design ($r_{\mathrm{avg}}$).  The next proposition shows that the inverse-propensity
part is not merely an artifact of the HT estimator.

\begin{proposition}[Inverse propensity scores for ACEs]
\label{prop:ace-inverse-propensity}
Fix the design \(P_n\) and the induced treatment--exposure combinations \(\Dcond_i(\Zvec)\).
Suppose that
\(\hat\tau_n\) is design-unbiased for \(\tau_{\mathrm{ACE}}(d^+,d^-)\) for every bounded
potential-outcome table.  Then
\[
\sup_{|Y|\le M}
\operatorname{Var}_{P_n}(\hat\tau_n)
\ge
\frac{M^2}{n^2}
\max_i
\left\{
\frac{1}{\pi_i(d_i^+)}-1,\,
\frac{1}{\pi_i(d_i^-)}-1
\right\}.
\]
\end{proposition}

The proof is given in Appendix~\ref{app:nonparametric-estimation-proofs}. This coordinate-wise bound demonstrates that no unbiased estimator of a fixed-exposure ACE can avoid the smallest propensity scores among the relevant treatment--exposure combinations. Below, we show that the diagonal term in the upper bound of Proposition~\ref{prop:generic-ace-ht} scales identically, and is usually the dominant term when compared to off-diagonal network interference and design dependence terms, giving a matching lower bound for regular graph-unaware designs.

\paragraph{Estimation of DTE and TTE under classical designs.}
We now examine when the Horvitz--Thompson estimator can consistently estimate the direct and total treatment effects under standard graph-unaware designs.  Proposition~\ref{prop:ace-inverse-propensity} provides a matching lower bound: no unbiased estimator can avoid the inverse-propensity scale $H$, so the variance conditions we derive are necessary for any design-unbiased estimator, not just the HT.  The analysis covers independent designs (Bernoulli, heterogeneous Bernoulli) and exchangeable designs (complete randomization) alike.  Proposition~\ref{prop:dte-tte-graph-unaware-propensities} (Appendix~\ref{app:ace-table-calculations}) shows they share a key property: any specific treatment configuration on $a$ units has probability $\exp\{-\Theta(a)\}$, so its inverse propensity scales as $\exp\{\Theta(a)\}$.  Here $a$ is the number of units the exposure condition constrains — $d_i+1$ for neighborhood-interference targets, $n$ for arbitrary-interference targets.  This exponential-in-$a$ scaling is the central bottleneck for fixed-exposure ACE estimation and is the only design feature needed below.  The variance bound of Proposition~\ref{prop:generic-ace-ht} formalizes this bottleneck: under independent assignment,
\[
\operatorname{Var}\{\hat\tau_{\mathrm{ACE}}^{HT}\}
\le M^2\bigl\{H/n + H^2c_{\mathrm{avg}}/n\bigr\},
\]
where $H=\max_i(1/\pi_i(d_i^+)+1/\pi_i(d_i^-))$ and $c_{\mathrm{avg}}$ is the average dependence degree (Lemma~\ref{lem:interference-dependence-graph}).  We write $\mathcal V_n(H,c)=H/n+H^2c/n$ for this scale, suppressing $M$.  An additional residual dependence term appears under exchangeable designs (Appendix~\ref{app:crd-residual-design-dependence}), but it does not change the conclusions.

Tables~\ref{tab:dte-representation-rate} and~\ref{tab:tte-representation-rate} summarize $H$ and the matching lower bound for the DTE and TTE under different representations.  The column ``HT $H$'' records the inverse-propensity scale and ``LB $H$'' the matching lower bound.  The variance scale column combines $H$ with the dependence degree $b_{2,\mathrm{avg}}$ (which equals \(c_{\mathrm{avg}}\) under independent assignment).  When \(H\) is exponential, the polynomial dependence is negligible by comparison.  For the exposure-mapping rows, let \(S_i=\sum_{j\in N_i}Z_j\) denote the number of treated neighbors and \(U_i=\mathds 1\{S_i\ge 1\}\) the binary any-treated indicator.

\begin{table}[t]
\centering
\caption{Direct-effect ACEs under Bernoulli assignment.  The rows show the DTE
target under arbitrary interference, neighborhood interference, and count/fraction or
binary exposure representations.}
\label{tab:dte-representation-rate}
\scriptsize
\setlength{\tabcolsep}{3pt}
\renewcommand{\arraystretch}{1.15}
\resizebox{\textwidth}{!}{%
\begin{tabular}{@{}lllll@{}}
\hline
Representation & Target conditions & HT \(H\) & LB \(H\) & HT variance scale \\
\hline
Arbitrary interference
& \((1,\mathbf 0_{-i})\) vs. \((0,\mathbf 0_{-i})\)
& \(e^{\Theta(n)}\)
& \(e^{\Theta(n)}\)
& \(\mathcal V_n(e^{\Theta(n)},n)\) \\
Neighborhood interference
& \((1,\mathbf 0_{N_i})\) vs. \((0,\mathbf 0_{N_i})\)
& \(e^{\Theta(d_{\max})}\)
& \(e^{\Theta(d_{\max})}\)
& \(\mathcal V_n(e^{\Theta(d_{\max})},b_{2,\mathrm{avg}})\) \\
Count/fraction exposure
& \((Z_i,S_i)=(1,0)\) vs. \((0,0)\)
& \(e^{\Theta(d_{\max})}\)
& \(e^{\Theta(d_{\max})}\)
& \(\mathcal V_n(e^{\Theta(d_{\max})},b_{2,\mathrm{avg}})\) \\
Binary any-treated exposure
& \((Z_i,U_i)=(1,0)\) vs. \((0,0)\)
& \(e^{\Theta(d_{\max})}\)
& \(e^{\Theta(d_{\max})}\)
& \(\mathcal V_n(e^{\Theta(d_{\max})},b_{2,\mathrm{avg}})\) \\
\hline
\end{tabular}
}
\end{table}

\emph{Direct treatment effect.}  The relevant potential outcomes for the DTE are $Y_i(1,0_{N_i})$ and $Y_i(0,0_{N_i})$ (Definition~\ref{def:relevant-potential-outcomes}).  Under neighborhood interference, the treated arm $(1,0_{N_i})$ requires the unit to be treated and all its neighbors to be untreated.  Under Bernoulli assignment,
\[
\pi_i(1,0_{N_i}) = p(1-p)^{d_i},
\qquad
\pi_i(0,0_{N_i}) = (1-p)^{d_i+1},
\]
so $H_i = \Theta(e^{\gamma d_i})$ with $\gamma=-\log(1-p)$ and $H = e^{\Theta(d_{\max})}$.  The same exponential scale applies under count, fraction, and binary any-treated exposure mappings for the DTE, since all encode the no-exposure condition as \(S_i=0\) or \(U_i=0\).

\begin{table}[t]
\centering
\caption{Total-effect ACEs under Bernoulli assignment.  Exposure-mapping
rows show representative treated-exposure versus no-exposure targets.}
\label{tab:tte-representation-rate}
\scriptsize
\setlength{\tabcolsep}{3pt}
\renewcommand{\arraystretch}{1.15}
\resizebox{\textwidth}{!}{%
\begin{tabular}{@{}lllll@{}}
\hline
Representation & Target conditions & HT \(H\) & LB \(H\) & HT variance scale \\
\hline
Arbitrary interference
& \((1,\mathbf 1_{-i})\) vs. \((0,\mathbf 0_{-i})\)
& \(e^{\Theta(n)}\)
& \(e^{\Theta(n)}\)
& \(\mathcal V_n(e^{\Theta(n)},n)\) \\
Neighborhood interference
& \((1,\mathbf 1_{N_i})\) vs. \((0,\mathbf 0_{N_i})\)
& \(e^{\Theta(d_{\max})}\)
& \(e^{\Theta(d_{\max})}\)
& \(\mathcal V_n(e^{\Theta(d_{\max})},b_{2,\mathrm{avg}})\) \\
Count/fraction exposure
& \((Z_i,S_i)=(1,d_i)\) vs. \((0,0)\)
& \(e^{\Theta(d_{\max})}\)
& \(e^{\Theta(d_{\max})}\)
& \(\mathcal V_n(e^{\Theta(d_{\max})},b_{2,\mathrm{avg}})\) \\
Binary any-treated exposure
& \((Z_i,U_i)=(1,1)\) vs. \((0,0)\)
& \(e^{\Theta(d_{\max})}\)
& \(e^{\Theta(d_{\max})}\)
& \(\mathcal V_n(e^{\Theta(d_{\max})},b_{2,\mathrm{avg}})\) \\
\hline
\end{tabular}
}
\end{table}

\emph{Total treatment effect.}  The relevant potential outcomes for the TTE are $Y_i(1,\mathbf 1_{N_i})$ and $Y_i(0,\mathbf 0_{N_i})$ (Definition~\ref{def:relevant-potential-outcomes}).  The treated arm requires all neighbors to be treated and the control arm requires all neighbors to be untreated, so $H$ is exponential in $d_{\max}$ for every representation in Table~\ref{tab:tte-representation-rate}.  Crucially, even the binary any-treated exposure mapping does not remove this scale: the control arm $(0,0)$ requires all neighbors to be untreated, with propensity $(1-p)^{d_i+1}$ decaying exponentially with $d_i$.

\emph{Implications for practice.}  Under classical graph-unaware designs, the relevant potential outcomes for the DTE---$Y_i(1,0)$ and $Y_i(0,0)$ (Definition~\ref{def:relevant-potential-outcomes})---occur with probability $\exp\{-\Theta(d_i)\}$, leading to an inverse-propensity barrier exponential in the neighborhood degree.  For the DTE,
\[
H \asymp e^{\gamma d_{\max}}, \qquad \gamma = -\log(1-p),
\]
and the variance bound vanishes only when $d_{\max}\lesssim\frac{1}{2\gamma}\log n$, i.e., when the degree grows at most logarithmically in $n$.  This condition is necessary for any unbiased estimator, not just the HT (Proposition~\ref{prop:ace-inverse-propensity}).  For the TTE, the exponential barrier applies across \emph{all} exposure mappings (Table~\ref{tab:tte-representation-rate}) because both relevant potential outcomes---$Y_i(1,\mathbf 1)$ and $Y_i(0,\mathbf 0)$---require every neighbor to be in a specific treatment state.

To overcome this exponential barrier induced by classical designs, we need specialized graph aware designs to estimate ACEs. Section~\ref{sec:implications} develops such designs: independent-set designs isolate the relevant potential outcomes for the DTE on ego units, while cluster randomized designs increase the probability of the all-treated and all-control conditions needed for the TTE.

\subsection{ECE estimation}
\label{sec:ece-ht-estimators}

We now consider estimation of ECEs. Following the definition in Section~\ref{sec:ece}, let $\Qpol^+,\Qpol^-$ be two allocation policies and $\Dcond_i^+(\Zvec),\Dcond_i^-(\Zvec)\in\tespace_i$ the corresponding random treatment--exposure combinations.  Define the policy propensities scores as
\[
\lambda_i^+(d)=\Qpol^+\{\Dcond_i^+(\Zvec)=d\},\qquad
\lambda_i^-(d)=\Qpol^-\{\Dcond_i^-(\Zvec)=d\},
\]
so that the ECE can be written as
\[
\tau_{\mathrm{ECE}}=\frac1n\sum_{i=1}^n\sum_{d\in\tespace_i}Y_i(d)\{\lambda_i^+(d)-\lambda_i^-(d)\}.
\]
Its Horvitz--Thompson estimator has the same form as the ACE estimator of Section~\ref{sec:ace-propensity-scores}, with the fixed inverse propensity scores of the relevant potential outcomes replaced by importance weights that are ratios of the policy propensity scores to the design propensity scores,
\[
h_i^+(\zvec)=\frac{\lambda_i^+\{\Dcond_i(\zvec)\}}{\pi_i^P\{\Dcond_i(\zvec)\}},\qquad
h_i^-(\zvec)=\frac{\lambda_i^-\{\Dcond_i(\zvec)\}}{\pi_i^P\{\Dcond_i(\zvec)\}},
\]
with the convention $0/0=0$.  Define the diagonal scale
\[
V_i=\sum_{d:\pi_i^P(d)>0}\frac{\{\lambda_i^+(d)-\lambda_i^-(d)\}^2}{\pi_i^P(d)},
\]
and the weight bound $H_i=\sup_{\zvec}\{|h_i^+(\zvec)|+|h_i^-(\zvec)|\}$.

\begin{proposition}[ECE importance-weighted estimator]
\label{prop:generic-ece-score}
Suppose $\lambda_i^+(d)=\lambda_i^-(d)=0$ whenever $\pi_i^P(d)=0$. Then
\[
\hat\tau_{\mathrm{ECE}}^{HT}=\frac1n\sum_{i=1}^n\{h_i^+(\Zvec)-h_i^-(\Zvec)\}Y_i^{obs}
\]
is design-unbiased for $\tau_{\mathrm{ECE}}$ for every potential-outcome table. If $|Y_i(\zvec)|\le M$ and the $i$th summand depends on the assignment only through $\Zvec_{\mathcal C_i}$, then
\[
\operatorname{Var}(\hat\tau_{\mathrm{ECE}}^{HT})
\le
\frac{M^2}{n^2}
\left\{
\sum_{i=1}^n V_i
+
\sum_{i=1}^n\sum_{j\neq i}c_{ij}H_iH_j
+
\sum_{i=1}^n\sum_{j\neq i}(1-c_{ij})\rho_{ij}H_iH_j
\right\}.
\]
If $V_i\le V$ and $H_i\le H$ for all $i$, then
\[
\operatorname{Var}(\hat\tau_{\mathrm{ECE}}^{HT})
\le
M^2\left\{
\frac{V}{n}
+
H^2\frac{c_{\mathrm{avg}}+r_{\mathrm{avg}}}{n}
\right\},
\]
and the estimator is design-MSE consistent if $V+H^2(c_{\mathrm{avg}}+r_{\mathrm{avg}})=o(n)$.
\end{proposition}

The proof is given in Appendix~\ref{app:nonparametric-estimation-proofs}.
The bound has the same three terms as the ACE bound of Proposition~\ref{prop:generic-ace-ht}: a diagonal term, an interference-dependence term, and a residual design-dependence term. What changes is the diagonal scale. For an ACE, the policy is a point mass at the relevant potential outcomes $Y_i(d_i^{\pm})$. The importance weights reduce to inverse propensities, and $V_i$ and $H_i$ are of the order of $1/\pi_i(d_i^{\pm})$. For an ECE, $V_i$ and $H_i$ are governed by the ratios $\lambda_i^{\pm}(d)/\pi_i^P(d)$, and these are large only when the allocation policy places probability on combinations that the design rarely produces. When the allocation policy is the design itself, the ratios are bounded, the diagonal term is $O(1/n)$, and the variance is governed by the dependence terms alone. The next two results make this concrete for the expected direct and indirect effects under Bernoulli policy and design.

\emph{Expected direct effect.} Under Bernoulli assignment with treatment probabilities $p_i$, consider the EDE whose policy is the same Bernoulli allocation. The importance weight reduces to the classical Horvitz--Thompson weight
\[
h_i^{\mathrm{EDE}}(\Zvec)=\frac{Z_i}{p_i}-\frac{1-Z_i}{1-p_i},
\]
so $\hat\tau_{\mathrm{EDE}}^{HT}$ is the usual difference in weighted means and does not require the interference graph for computation. The weight scales are $V_i=H_i=1/p_i+1/(1-p_i)$, which are $O(1)$ when the treatment probabilities are bounded away from zero and one. No inverse propensity of a rare exposure combination appears in the bound.

\begin{proposition}[Variance bound for the EDE estimator]
\label{prop:ede-variance-bound}
Let $\Qpol_p$ be a Bernoulli allocation policy, with treatment probabilities bounded away from zero and one. Assume Bernoulli assignment, and $\sup_{\zvec}|Y_i(\zvec)|\le M$. Then $\hat\tau_{\mathrm{EDE}}^{HT}$ is design-unbiased for $\tau_{\mathrm{EDE}}(\Qpol_p)$, and
\[
\operatorname{Var}_{P_n}(\hat\tau_{\mathrm{EDE}}^{HT})
\le
C_p M^2\frac{1+c_{\mathrm{avg}}}{n},
\]
where $C_p$ depends only on the bounds on the treatment probabilities. In particular, $\hat\tau_{\mathrm{EDE}}^{HT}$ is design-MSE consistent if $c_{\mathrm{avg}}=o(n)$.
\end{proposition}

The proof is given in Appendix~\ref{app:nonparametric-estimation-proofs}.
This result should be read against the DTE rows of Table~\ref{tab:dte-representation-rate}. The two estimators use the same design and the same interference graph. The DTE estimator must isolate the treatment--exposure combination $(1,0_{N_i})$, which the design produces with probability $p(1-p)^{d_i}$, so its inverse-propensity scale is $H=e^{\Theta(d_{\max})}$ and its full variance scale from Proposition~\ref{prop:generic-ace-ht} is $H/n+H^2c_{\mathrm{avg}}/n$. The EDE estimator uses every observed outcome, and its variance scale is $(1+c_{\mathrm{avg}})/n$. On a $d$-regular locally tree-like interference graph the comparison is $(e^{2\gamma d}d^2)/n$ against $d^2/n$: the DTE is consistent only in the logarithmic-degree regime $e^{2\gamma d}d^2=o(n)$, i.e., $d\lesssim\frac{1}{2\gamma}\log n$, while the EDE HT estimator is consistent whenever $d=o(\sqrt n)$. The price of interference for the EDE is not the rarity of any treatment-exposure combination but the dependence among the summands, measured by $c_{\mathrm{avg}}$.

\emph{Expected indirect effect.} Graph-aware ECEs that contrast neighbor treatments have larger importance weights. The expected indirect effect sums the weight over neighbors, $h_i^{\mathrm{EIE},G}(\Zvec)=\sum_{j\in N_i}\{Z_j/p_j-(1-Z_j)/(1-p_j)\}$, so $H_i\le C_p d_i$ when the treatment probabilities are bounded away from zero and one. The weights grow with the degree, but linearly rather than exponentially.

\begin{corollary}[Variance bound for the graph-aware EIE estimator]
\label{cor:eie-variance-bound}
Assume the true potential outcomes satisfy neighborhood interference with known neighborhoods $N_i$. Assume Bernoulli assignment with treatment probabilities bounded away from zero and one, and assume $\sup_{\zvec}|Y_i(\zvec)|\le M$. Then $\hat\tau_{\mathrm{EIE}}^{G}$ is design-unbiased for $\tau_{\mathrm{EIE}}(\pi)$, and
\[
\operatorname{Var}(\hat\tau_{\mathrm{EIE}}^{G})
\le
\frac{C_pM^2}{n^2}
\left\{
\sum_i d_i^2
+
\sum_i\sum_{j\neq i}c_{ij}d_id_j
\right\},
\]
where $d_i=|N_i|$ and $c_{ij}=\mathds{1}\{\mathcal C_i\cap\mathcal C_j\neq\emptyset\}$. If $d_i\le d_{\max}$ for all $i$, then
\[
\operatorname{Var}(\hat\tau_{\mathrm{EIE}}^{G})
\le
C_pM^2 d_{\max}^2\frac{1+c_{\mathrm{avg}}}{n}.
\]
\end{corollary}

The proof is given in Appendix~\ref{app:nonparametric-estimation-proofs}.
The variance scale $d_{\max}^2(1+c_{\mathrm{avg}})/n$ is polynomial in the degrees: on a $d$-regular locally tree-like graph it is of order $d^4/n$, so the EIE is consistently estimable whenever $d=o(n^{1/4})$. As for the EDE, no exponential barrier due to the propensity score appears, and the dependence term dominates the bound.

Appendix~\ref{app:ace-ece-separation} extends the DTE/EDE comparison to arbitrary
heterogeneous degree sequences, where it is governed by two graph functionals:
exponential moments of the degree sequence for fixed-exposure ACEs, and the average
dependence degree for the EDE.

The variance bounds of Section~\ref{sec:nonparametric-estimation} show that classical graph-unaware designs are matched to expected causal effects and pay an exponential cost for average causal effects. Section~\ref{sec:designs} constructs designs that remove this cost, and Section~\ref{sec:implications} collects the resulting implications for classical estimators, model-assisted estimation, and the interpretation of positivity under interference.

\section{Design-Based Estimation of Average Causal Effects}
\label{sec:designs}
When classical estimators fail for average causal effects under interference, the failure arises because the realized assignment does not reveal the relevant potential outcomes with sufficient probability. Section~\ref{sec:nonparametric-estimation} shows that under classical graph-unaware designs the target treatment--exposure combinations have inverse propensities that are exponential in the interference degree, and that this cost is unavoidable for any unbiased estimator. One remedy is to modify the experimental design so that those potential outcomes are observed with non-negligible probability. We study designs that control the assignment structure rather than impose an outcome model. The common device is to use the interference graph to create positive correlation within neighborhoods, so that the event that all of a unit's neighbors are held at a common treatment occurs as a single correlated event rather than as a product of independent events. This replaces the exponential propensity $(1-p)^{d_i}$ of a classical design by a propensity that decays only polynomially in the degree $d_i$, and the variance is reduced accordingly. Throughout, we assume that the interference neighborhood is known and that an exposure mapping is specified, and we take the potential outcomes to be uniformly bounded, $\sup_{i,z,e}|Y_i(z,e)|\le M$.

 Suppose the target average causal effect has the form
\[
\tau
=
\frac{1}{n}\sum_{i=1}^n
\{Y_i(d_i^+)-Y_i(d_i^-)\},
\]
where $d_i^{+}$ and $d_i^{-}$ are two target treatment--exposure combinations. If the design gives positive probability to both combinations,
\[
\pi_i(d_i^+)=\Pr\{(Z_i,E_i)=d_i^+\}>0,
\qquad
\pi_i(d_i^-)=\Pr\{(Z_i,E_i)=d_i^-\}>0,
\]
then the treatment--exposure Horvitz--Thompson estimator
\[
\hat\tau_{\mathrm{HT}}^{ZE}
=
\frac{1}{n}\sum_{i=1}^n
\left\{
\frac{\mathds{1}\{(Z_i,E_i)=d_i^+\}Y_i^{\mathrm{obs}}}{\pi_i(d_i^+)}
-
\frac{\mathds{1}\{(Z_i,E_i)=d_i^-\}Y_i^{\mathrm{obs}}}{\pi_i(d_i^-)}
\right\}
\]
is unbiased for $\tau$. The role of the design is therefore to make the relevant treatment--exposure combinations occur with positive and reasonably large propensity scores.

\begin{proposition}[Treatment--exposure HT unbiasedness for ACEs]
\label{prop:ze-ht-unbiased}
Suppose the observed outcome satisfies $Y_i^{\mathrm{obs}}=Y_i(Z_i,E_i)$. Let
\[
\tau
=
\frac{1}{n}\sum_{i=1}^n
\{Y_i(d_i^+)-Y_i(d_i^-)\},
\]
and suppose $\pi_i(d_i^+)>0$ and $\pi_i(d_i^-)>0$ for all $i$. Then $\hat\tau_{\mathrm{HT}}^{ZE}$ is design-unbiased for $\tau$.
\end{proposition}

\noindent The proof is given in Appendix~\ref{app:design-proofs}.

We develop two graph-aware constructions based on independent sets. The first targets the direct treatment effect and uses a distance-one independent set; the second targets the total treatment effect and uses a distance-two independent set. In each case a randomized greedy construction gives focal inclusion probabilities in closed form, and the resulting Horvitz--Thompson estimator has mean squared error that is polynomial, rather than exponential, in the interference degree. We then discuss cluster randomized designs, whose polynomial behavior is conditional on the clustering being aligned with the graph.

\subsection{Independent set designs for the direct treatment effect}
\label{sec:is-dte}

The direct treatment effect
\[
\tau^{D}
=
\frac{1}{n}\sum_{i=1}^n
\{Y_i(1,\mathbf 0_{N_i})-Y_i(0,\mathbf 0_{N_i})\}
\]
contrasts the treated and control potential outcomes of a unit whose neighbors are all held at control. Both arms require the no-exposure condition $Z_{N_i}=\mathbf 0$; they differ only in the unit's own treatment. Consequently two units $i$ and $j$ that share a neighbor $k\in N_i\cap N_j$ impose the same requirement $Z_k=0$, so there is no conflict between their target conditions. A distance-one independent set, in which selected units are mutually non-adjacent but may share neighbors, therefore suffices.

We construct a random independent set of \emph{ego} units by a randomized greedy rule, and randomize treatment among the egos while holding their neighbors at control. Units outside the independent set act as \emph{alter} units that buffer the egos from interference.

\paragraph{Design (IS-1).}
\begin{enumerate}[label=\textup{(\arabic*)},leftmargin=2.2em]
\item Draw $U_1,\dots,U_n$ independently from the uniform distribution on $[0,1]$.
\item Let $\mathcal I=\{i: U_i<U_k \text{ for all } k\in N_i\}$ be the set of units whose draw is the strict minimum in their closed neighborhood $\{i\}\cup N_i$. The set $\mathcal I$ is an independent set almost surely.
\item For each ego $i\in\mathcal I$, draw $B_i\sim\mathrm{Bernoulli}(q)$ independently, set $Z_i=B_i$, and set $Z_k=0$ for all $k\in N_i$.
\item Assign any unit not fixed by step (3) a treatment by a prespecified rule that does not overwrite the constraints of step (3).
\end{enumerate}

The construction is graph-aware: whether $i$ is an ego depends only on the draws in its closed neighborhood. If two egos $i$ and $j$ share a neighbor $k$, step (3) sets $Z_k=0$ from both, so the assignment is consistent regardless of how many egos claim $k$. Because the event $\{i\in\mathcal I\}$ is determined within $\{i\}\cup N_i$, the inclusion probability is available in closed form.

\begin{lemma}[Inclusion and reveal probabilities under IS-1]
\label{lem:is1-propensity}
Under the design IS-1,
\[
\Pr(i\in\mathcal I)=\frac{1}{d_i+1}.
\]
Let $F_i^{(1)}=\{i\in\mathcal I,\ B_i=1\}$ and $F_i^{(0)}=\{i\in\mathcal I,\ B_i=0\}$ be the focal reveal events. On $F_i^{(1)}$ the assignment satisfies $(Z_i,Z_{N_i})=(1,\mathbf 0)$, and on $F_i^{(0)}$ it satisfies $(Z_i,Z_{N_i})=(0,\mathbf 0)$, so
\[
\rho_i^{(1)}:=\Pr(F_i^{(1)})=\frac{q}{d_i+1},
\qquad
\rho_i^{(0)}:=\Pr(F_i^{(0)})=\frac{1-q}{d_i+1}.
\]
\end{lemma}

\noindent The proof is given in Appendix~\ref{app:design-proofs}. The inclusion probability follows from the symmetry of the $d_i+1$ independent draws in the closed neighborhood, and the reveal probabilities follow from the independent treatment coin $B_i$. The focal reveal probabilities are exact and do not require simulation. They are stated for the focal events rather than for the raw exposure events $\{Z_i=s,Z_{N_i}=\mathbf 0\}$, because non-ego units may also happen to satisfy the no-exposure condition depending on the leftover rule in step (4); the estimator below uses only the focal events, whose probabilities are known exactly. Taking $q=\tfrac12$ gives the symmetric values $\rho_i^{(1)}=\rho_i^{(0)}=1/\{2(d_i+1)\}$.

The focal Horvitz--Thompson estimator for the direct treatment effect is
\[
\hat\tau_{\mathrm{IS}}^{D}
=
\frac{1}{n}\sum_{i=1}^n
\left\{
\frac{\mathds{1}\{F_i^{(1)}\}Y_i^{\mathrm{obs}}}{\rho_i^{(1)}}
-
\frac{\mathds{1}\{F_i^{(0)}\}Y_i^{\mathrm{obs}}}{\rho_i^{(0)}}
\right\},
\]
which at $q=\tfrac12$ reduces to
\[
\hat\tau_{\mathrm{IS}}^{D}
=
\frac{2}{n}\sum_{i=1}^n (d_i+1)
\left\{
\mathds{1}\{F_i^{(1)}\}Y_i^{\mathrm{obs}}
-
\mathds{1}\{F_i^{(0)}\}Y_i^{\mathrm{obs}}
\right\}.
\]

\begin{proposition}[Unbiasedness of the IS-1 estimator]
\label{prop:is1-unbiased}
Under the design IS-1 with $q\in(0,1)$, the estimator $\hat\tau_{\mathrm{IS}}^{D}$ is design-unbiased for the direct treatment effect, $\mathbb E[\hat\tau_{\mathrm{IS}}^{D}]=\tau^{D}$.
\end{proposition}

\noindent The proof is the same Horvitz--Thompson calculation as Proposition~\ref{prop:ze-ht-unbiased}, applied to the focal reveal events rather than to the raw treatment--exposure events. On $F_i^{(1)}$ and $F_i^{(0)}$ the observed outcome equals the corresponding direct-effect potential outcome, and the focal probabilities $\rho_i^{(1)},\rho_i^{(0)}$ are positive and known.

The variance of the estimator is controlled by the closed-neighborhood inclusion probabilities and by the dependence among egos that lie within distance two of one another. Write $\delta_i=Y_i(1,\mathbf 0_{N_i})-Y_i(0,\mathbf 0_{N_i})$ and $p_i=\Pr(i\in\mathcal I)=1/(d_i+1)$.

\begin{proposition}[Variance of the IS-1 estimator]
\label{prop:is1-variance}
Under the design IS-1 with $q=\tfrac12$, the estimator satisfies the exact identity
\[
n^2\operatorname{Var}(\hat\tau_{\mathrm{IS}}^{D})
=
\sum_{i=1}^n \operatorname{Var}(W_i)
+
\sum_{\substack{i\ne j:\ \mathrm{dist}_G(i,j)\le 2}}
\delta_i\delta_j
\left(\frac{\Pr(i,j\in\mathcal I)}{p_ip_j}-1\right),
\]
where $W_i=2(d_i+1)\mathds{1}\{i\in\mathcal I\}\{Y_i(1,\mathbf 0_{N_i})B_i-Y_i(0,\mathbf 0_{N_i})(1-B_i)\}$. For non-adjacent $i,j$ sharing $s_{ij}=|N_i\cap N_j|$ common neighbors, $\Pr(i,j\in\mathcal I)/(p_ip_j)=(d_i+d_j+2)/(d_i+d_j+2-s_{ij})$, while $\Pr(i,j\in\mathcal I)=0$ when $i$ and $j$ are adjacent. Consequently, uniformly over $|Y_i(\cdot)|\le M$,
\[
\operatorname{Var}(\hat\tau_{\mathrm{IS}}^{D})
\le
\frac{4M^2}{n^2}\sum_{i=1}^n(d_i+1)
+
\frac{4M^2}{n^2}
\left[
2|E|
+
\sum_{\substack{i\ne j:\ i\not\sim j,\ s_{ij}\ge 1}}
\frac{s_{ij}}{d_i+d_j+2-s_{ij}}
\right].
\]
\end{proposition}

\noindent The proof is given in Appendix~\ref{app:design-proofs}. The first term is a diagonal contribution of order $n^{-2}\sum_i(d_i+1)$; the second collects the dependence among egos whose closed neighborhoods overlap, that is, egos within graph distance two.

\begin{corollary}[Polynomial rate for the direct treatment effect]
\label{cor:is1-rate}
If $G$ is $d$-regular and locally tree-like, in the sense that every pair of non-adjacent vertices has at most a bounded number of common neighbors ($s_{ij}=O(1)$), then
\[
\operatorname{Var}(\hat\tau_{\mathrm{IS}}^{D})
=
O\!\left(\frac{M^2 d}{n}\right).
\]
For a general $d$-regular graph, the second-moment bound gives the cruder rate
\[
\operatorname{Var}(\hat\tau_{\mathrm{IS}}^{D})
=
O\!\left(\frac{M^2 d^2}{n}\right),
\]
since on any $d$-regular graph the crude bound evaluates to $4M^2(n+nd+nd^2)/n^2$. For an arbitrary graph, the crude second-moment bound gives
\[
\operatorname{Var}(\hat\tau_{\mathrm{IS}}^{D})
\le
\frac{4M^2}{n^2}\left[n+\sum_{i}d_i+\sum_i d_i^2\right]
=
O\!\left(\frac{M^2\,(1+\bar d+\overline{d^2})}{n}\right),
\qquad
\overline{d^2}=\frac1n\sum_i d_i^2.
\]
\end{corollary}

\noindent The proof is given in Appendix~\ref{app:design-proofs}. Under a Bernoulli design the same estimand requires the no-exposure condition, whose propensity $p(1-p)^{d_i}$ is exponentially small in $d_i$, and the resulting variance is exponential in the average degree by Proposition~\ref{prop:ace-inverse-propensity}. The IS-1 design replaces this exponential cost by dependence that is polynomial in the degree sequence. The bound of Corollary~\ref{cor:is1-rate} holds for arbitrary, not necessarily regular, interference graphs: it is of order $M^2(1+\bar d+\overline{d^2})/n$, so this estimator is consistent whenever $\overline{d^2}=o(n)$. This is the same second-moment condition on the degree sequence under which the expected direct effect is consistent by the generic dependence-graph bound. On $d$-regular graphs that are locally tree-like, in the sense that non-adjacent vertices share at most a bounded number of common neighbors, the sharper rate $O(M^2\bar d/n)$ holds. On a general $d$-regular graph the wedge sum $\sum_{i\ne j:\ i\not\sim j}s_{ij}/(d_i+d_j+2-s_{ij})$ can be as large as $\Theta(nd^2)$ when closed neighborhoods overlap heavily, so the crude rate $O(M^2d^2/n)$ is the best that the second-moment bound certifies in general. Figure~\ref{fig:design_repair} illustrates the effect: on a fixed family of graphs, the error of the independent set estimator remains nearly flat in the degree, whereas under a Bernoulli design it grows exponentially. For irregular graphs with high-degree hubs the crude bound can be larger, and a hub-aware modification of the focal sampling may control it, though we do not pursue this refinement here. For extreme hub graphs such as a star, this particular randomized local-minimum IS-1 design has a non-vanishing second-moment bound; this indicates a limitation of the simple construction rather than a universal impossibility theorem.

\subsection{Independent set designs for the total treatment effect}
\label{sec:is-tte}

The total treatment effect
\[
\tau^{T}
=
\frac{1}{n}\sum_{i=1}^n
\{Y_i(1,e_i^{\mathbf 1})-Y_i(0,0)\},
\qquad
e_i^{\mathbf 1}=f_i(\mathbf 1_{N_i}),
\]
contrasts the all-treated and all-control regimes on the closed neighborhood $\{i\}\cup N_i$. Unlike the direct treatment effect, the two arms now impose opposite requirements on a shared neighbor: if egos $i$ and $j$ share a neighbor $k$ and $i$ is on the treated side while $j$ is on the control side, then $i$ requires $Z_k=1$ and $j$ requires $Z_k=0$, a direct conflict. Avoiding such conflicts requires that egos have disjoint closed neighborhoods, that is, that they be at graph distance at least three. This is a distance-two independent set, obtained by running the randomized greedy rule on the radius-two balls.

For a unit $i$ let $B_i^{(2)}=\{k:\mathrm{dist}_G(i,k)\le 2\}$ denote its radius-two ball and $b_i=|B_i^{(2)}|$ its size. Since the neighborhood is contained in the ball,
\[
d_i+1\le b_i\le 1+d_i+\sum_{j\in N_i}(d_j-1)\le 1+d_i d_{\max}.
\]
On $d$-regular locally tree-like graphs this gives $b_i=1+d+d(d-1)=d^2+1$.

\paragraph{Design (IS-2).}
\begin{enumerate}[label=\textup{(\arabic*)},leftmargin=2.2em]
\item Draw $U_1,\dots,U_n$ independently from the uniform distribution on $[0,1]$.
\item Let $\mathcal I_2=\{i: U_i<U_k \text{ for all } k\in B_i^{(2)}\setminus\{i\}\}$ be the set of units whose draw is the strict minimum in their radius-two ball. The set $\mathcal I_2$ is a distance-two independent set almost surely.
\item For each ego $i\in\mathcal I_2$, draw $R_i\sim\mathrm{Bernoulli}(\tfrac12)$ independently, and set $Z_k=R_i$ for all $k\in\{i\}\cup N_i$.
\item Assign any unit not fixed by step (3) an independent $\mathrm{Bernoulli}(\tfrac12)$ treatment.
\end{enumerate}

Because egos are at distance at least three, their closed neighborhoods are disjoint and step (3) is free of conflicts. Since the outcome of an ego depends only on the assignment of its closed neighborhood, which is entirely fixed in step (3), the leftover assignments in step (4) do not affect the revealed outcomes of egos.

\begin{lemma}[Inclusion and reveal probabilities under IS-2]
\label{lem:is2-propensity}
Under the design IS-2,
\[
\Pr(i\in\mathcal I_2)=\frac{1}{b_i},
\qquad
\rho_i^{(1)}:=\Pr(i\in\mathcal I_2,R_i=1)=\frac{1}{2b_i},
\qquad
\rho_i^{(0)}:=\Pr(i\in\mathcal I_2,R_i=0)=\frac{1}{2b_i}.
\]
On these focal events the assignment reveals the corresponding all-treated or all-control potential outcome. In particular, if the radius-two balls are polynomially bounded in the interference degree scale, the inverse focal probabilities are polynomial as well.
\end{lemma}

\noindent The proof is given in Appendix~\ref{app:design-proofs}. The focal Horvitz--Thompson estimator for the total treatment effect is
\[
\hat\tau_{\mathrm{IS}}^{T}
=
\frac{2}{n}\sum_{i\in\mathcal I_2} b_i
\left\{
\mathds{1}\{R_i=1\}Y_i^{\mathrm{obs}}
-
\mathds{1}\{R_i=0\}Y_i^{\mathrm{obs}}
\right\}.
\]

\begin{proposition}[Unbiasedness and variance of the IS-2 estimator]
\label{prop:is2-variance}
Under the design IS-2, the estimator $\hat\tau_{\mathrm{IS}}^{T}$ is design-unbiased for the total treatment effect. The unbiasedness is with respect to the focal reveal events $(i\in\mathcal I_2,R_i=1)$ and $(i\in\mathcal I_2,R_i=0)$, each of which implies the corresponding treatment--exposure condition. Writing $\Delta_i=Y_i(1,e_i^{\mathbf 1})-Y_i(0,0)$ and $q_i=1/b_i$, it satisfies the exact identity
\[
n^2\operatorname{Var}(\hat\tau_{\mathrm{IS}}^{T})
=
\sum_{i=1}^n \operatorname{Var}(V_i)
+
\sum_{\substack{i\ne j:\ B_i^{(2)}\cap B_j^{(2)}\ne\emptyset}}
\Delta_i\Delta_j
\left(\frac{\Pr(i,j\in\mathcal I_2)}{q_iq_j}-1\right),
\]
where $V_i=2b_i\mathds{1}\{i\in\mathcal I_2\}\{Y_i(1,e_i^{\mathbf 1})R_i-Y_i(0,0)(1-R_i)\}$. If $\mathrm{dist}_G(i,j)\le2$, then $\Pr(i,j\in\mathcal I_2)=0$; if $\mathrm{dist}_G(i,j)\in\{3,4\}$ and $s=|B_i^{(2)}\cap B_j^{(2)}|\ge 1$, then the ratio equals $(b_i+b_j)/(b_i+b_j-s)$. Consequently, uniformly over $|Y_i(\cdot)|\le M$,
\[
\operatorname{Var}(\hat\tau_{\mathrm{IS}}^{T})
\le
\frac{4M^2}{n^2}\sum_{i=1}^n b_i
+
\frac{4M^2}{n^2}
\sum_{\substack{i\ne j:\ \mathrm{dist}_G(i,j)\le2}}1
+
\frac{4M^2}{n^2}
\sum_{\substack{i\ne j:\ \mathrm{dist}_G(i,j)\in\{3,4\},\ s_{ij}^{(2)}\ge 1}}
\frac{s_{ij}^{(2)}}{b_i+b_j-s_{ij}^{(2)}},
\qquad
s_{ij}^{(2)}=|B_i^{(2)}\cap B_j^{(2)}|.
\]
\end{proposition}

\noindent The proof is given in Appendix~\ref{app:design-proofs}.

\begin{corollary}[Polynomial rate for the total treatment effect]
\label{cor:is2-rate}
On $d$-regular graphs that are locally tree-like, the diagonal and distance-two exclusion terms are of order $M^2d^2/n$, and the positive radius-two overlap term contributes at most the same order (distance-three partners contribute $M^2d/n$ and distance-four partners $M^2d^2/n$), so
\[
\operatorname{Var}(\hat\tau_{\mathrm{IS}}^{T})
=
O\!\left(\frac{M^2 d^2}{n}\right).
\]
For an arbitrary graph, the universal bound is $\operatorname{Var}(\hat\tau_{\mathrm{IS}}^{T})=O(M^2\overline{b^2}/n)$ with $\overline{b^2}=n^{-1}\sum_i b_i^2$.
\end{corollary}

Since $d_i\le d_{\max}$ and $b_i\le 1+d_i d_{\max}$, the bounds in Corollaries~\ref{cor:is1-rate} and~\ref{cor:is2-rate} give variance bounds of order $O(M^2(1+d_{\max}^2)/n)$ for IS-1 and $O(M^2(1+d_{\max}^4)/n)$ for IS-2 on every interference graph. The proof is given in Appendix~\ref{app:design-proofs}. Under a Bernoulli design the total treatment effect requires the all-treated condition on the closed neighborhood, whose propensity $p^{d_i+1}$ is again exponentially small in $d_i$. On locally tree-like $d$-regular graphs, the IS-2 design attains the polynomial rate $O(M^2d^2/n)$, the same degree order as the expected direct effect under a Bernoulli design established in Proposition~\ref{prop:ede-variance-bound}. On arbitrary graphs the rate is governed by the radius-two ball sizes $b_i$. The exponential separation between average and expected causal effects under classical designs is therefore not intrinsic to the estimands: with a design aligned to the estimand, the total treatment effect can be estimated at polynomial rates. The direct treatment effect is easier still, since Corollary~\ref{cor:is1-rate} gives the smaller rate $O(M^2\bar d/n)$ on locally tree-like regular graphs.

\subsection{Independent set designs for indirect effects}
\label{sec:is-indirect}

Indirect (spillover) effects ask how the outcome of a unit is affected by the treatment of others, its own treatment held fixed. A recurring target in network experiments is the effect of treating a \emph{selected set of influencers} on an average unit: seeding experiments assign treatment to a sparse set of influential units and measure the response of otherwise untreated peers exposed to a treated influencer \citep{paluck2016changing,aralwalker2012identifying,aral2013engineering,bond201261}. Under the any-treated exposure map \eqref{eq:any-treated}, which records $U_i=1$ whenever at least one neighbor of $i$ is treated, this target is the two-level average indirect effect $\tau_{\mathrm{IE}}(z;0,1)=n^{-1}\sum_i\{Y_i(z,1)-Y_i(z,0)\}$ of Section~\ref{sec:ace}: a comparison of the no-exposure level $e_a=0$ and the any-treated level $e_b=1$ with the unit's own treatment $z$ fixed ($z=0$ for an untreated peer). Estimators of spillover effects under two-stage and independent-set randomization are developed in \citet{baird2018optimal,basse2018design,cai2024independent_set}, and two-level exposure contrasts of this form are the fixed-exposure analogue of the allocation-averaged estimand class studied by \citet{viviano2024design}. The uncompressed special cases contrast the fully exposed and unexposed neighborhood regimes, with the unit's own treatment held fixed:
\[
\tau_{\mathrm{IEC}}=\frac1n\sum_{i=1}^n\{Y_i(0,e_i^{\mathbf 1})-Y_i(0,0)\},
\qquad
\tau_{\mathrm{IET}}=\frac1n\sum_{i=1}^n\{Y_i(1,e_i^{\mathbf 1})-Y_i(1,0)\}.
\]

Both arms of an indirect contrast fix an extreme neighborhood regime, and the exponential barrier of Section~\ref{sec:nonparametric-estimation} applies exactly as it does to the direct and total effects. Under a Bernoulli design with probability $p$, the unexposed arm has propensity $\Pr(Z_i=z^\dagger)(1-p)^{d_i}$, the any-treated arm has propensity $\Pr(Z_i=z^\dagger)(1-(1-p)^{d_i})$, and the full-exposure arm has propensity $\Pr(Z_i=z^\dagger)\,p^{d_i}$, where $z^\dagger$ is the own-treatment level of the estimand and all but the any-treated propensity are exponentially small in the degree. Every indirect effect that fixes a rare exposure regime is therefore exponentially costly under a graph-unaware design.

The distance-two independent set construction of Section~\ref{sec:is-tte} removes this barrier for the indirect effects with a single modification. On the ego set $\mathcal I_2$, hold each ego's own treatment at the level $z^\dagger$ required by the estimand ($z^\dagger=0$ for $\tau_{\mathrm{IEC}}$ and for control units; $z^\dagger=1$ for $\tau_{\mathrm{IET}}$), and assign the ego's neighborhood the common value $R_i\in\{0,1\}$ by an independent fair coin. An ego then reveals $(z^\dagger,e_i^{\mathbf 1})$ when $R_i=1$ and $(z^\dagger,0)$ when $R_i=0$, each with focal probability $1/(2b_i)$.

\begin{proposition}[Independent set designs for indirect effects]
\label{prop:is-indirect}
Under the modified distance-two independent set design above, the focal Horvitz--Thompson estimator is design-unbiased, with focal reveal probabilities $1/(2b_i)$ for both arms, for the any-treated indirect effects $\tau_{\mathrm{IE}}(0;0,1)$ and $\tau_{\mathrm{IE}}(1;0,1)$ under the exposure map \eqref{eq:any-treated}, and for the full-exposure versions $\tau_{\mathrm{IEC}}$ ($z^\dagger=0$) and $\tau_{\mathrm{IET}}$ ($z^\dagger=1$). Its variance satisfies the bound of Proposition~\ref{prop:is2-variance}, and on locally tree-like $d$-regular graphs
\[
\operatorname{Var}(\hat\tau_{\mathrm{IS}})=O\!\left(\frac{M^2\overline{d^2}}{n}\right),
\]
matching the rate of the total treatment effect and of the expected direct effect on this graph class.
\end{proposition}

\noindent The proof is identical to that of Proposition~\ref{prop:is2-variance}: fixing the ego's own treatment changes neither the inclusion probabilities nor the covariance structure, and only the pair of revealed treatment--exposure arms differs. No modification of the design is needed for the coarsened two-level estimand. Because the all-treated arm realizes the any-treated level $U_i=1$ and the all-control arm realizes $U_i=0$ under \eqref{eq:any-treated}, the polarized design estimates $\tau_{\mathrm{IE}}(z^\dagger;0,1)$ with the same focal probabilities $1/(2b_i)$. More generally, any monotone exposure map that takes the value $1$ on the all-treated vector and $0$ on the all-control vector yields the same two realized levels, so the same estimator applies to the corresponding coarsened two-level indirect effect. The coarsening is therefore free for the polarized design: it changes only the exposure interpretation, not the assignment, and it can only shorten the distance between the two levels of the contrast.

\paragraph{Sparse seeding with the distance-one independent set.}
Influencer-seeding experiments treat a sparse, well-separated set of units and measure the responses of their nontreated alters. The distance-one independent set $\mathcal I$ of Section~\ref{sec:is-dte} is exactly such a set: egos are mutually non-adjacent, and on typical graphs each alter is adjacent to a controlled number of egos. Treating each ego with an independent coin $B_i\sim\mathrm{Bernoulli}(q)$ and setting all non-ego units to control yields the \emph{sparse seeding design}: the treated influencers are the egos with $B_i=1$, and an alter's any-treated indicator \eqref{eq:any-treated} equals $1$ precisely when at least one of its neighbors is a treated ego. Let $K_i=|\mathcal I\cap N_i|$ be the number of ego neighbors of unit $i$. Because $I$ is an independent set, an ego has no ego neighbors; the unit $i$ consequently reveals $(0,1)$ on the event
\[
F_i^{(1)}=\Big\{\textstyle\sum_{j\in N_i}B_j\mathds 1\{j\in\mathcal I\}\ge 1\Big\}
\qquad\text{i.e.}\qquad
\{Z_i=0,\ U_i=1\},
\]
and reveals $(0,0)$ on $F_i^{(0)}=\{Z_i=0\}\setminus F_i^{(1)}$, with $F_i^{(1)}\cup F_i^{(0)}=\{Z_i=0\}$. The focal probabilities are
\begin{equation}
\rho_i^{(1)}=\mathbb E\big[1-(1-q)^{K_i}\big],
\qquad
\rho_i^{(0)}=\mathbb E\big[(1-q)^{K_i}\big]-\frac{q}{d_i+1}.
\label{eq:seeding-focal}
\end{equation}
Both are positive for every unit with degree at least one.

\begin{proposition}[Sparse seeding estimates the influencer spillover effect]
\label{prop:is-seeding}
Under the sparse seeding design with $q\in(0,1)$ and no isolated units, the focal Horvitz--Thompson estimator
\[
\hat\tau_{\mathrm{seed}}
=\frac1n\sum_{i=1}^n\Big\{
\frac{\mathds 1\{F_i^{(1)}\}Y_i^{\mathrm{obs}}}{\rho_i^{(1)}}
-
\frac{\mathds 1\{F_i^{(0)}\}Y_i^{\mathrm{obs}}}{\rho_i^{(0)}}
\Big\}
\]
is design-unbiased for the influencer spillover effect $\tau_{\mathrm{IE}}(0;0,1)$ under the any-treated exposure map \eqref{eq:any-treated}. On a locally tree-like $d$-regular graph, $K_i\Rightarrow\mathrm{Poisson}(1)$ and, as $d\to\infty$,
\[
\rho_i^{(1)}\to 1-e^{-q},
\qquad
\rho_i^{(0)}\to e^{-q}:
\]
both arms have constant focal probability, independent of $n$ and of the degree, and the inverse-propensity weights $1/\rho_i^{(1)}+1/\rho_i^{(0)}$ are $O(1)$. The covariance structure is inherited from the ego draws and has the radius-two range of Proposition~\ref{prop:is1-variance}, so the variance satisfies the generic bound of Proposition~\ref{prop:generic-ace-ht}; on locally tree-like $d$-regular graphs the diagonal term is $O(M^2/n)$, removing the inverse-propensity barrier entirely for this estimand.
\end{proposition}

\noindent The proof is given in Appendix~\ref{app:design-proofs}. The constant-order focal probabilities contrast sharply with the focal probability $1/(2b_i)=O(1/d^2)$ of the full-exposure design of Proposition~\ref{prop:is-indirect}: the coarsening to the any-treated level $e_b=1$ is what converts the exponentially rare all-treated event into a condition--``at least one treated neighbor''--that has constant probability under sparse seeding. The hub caveat is that the guarantee is tied to the ego-neighborhood count $K_i$ being $O(1)$ in expectation. On a star, for example, the center faces up to $d$ simultaneously eligible ego candidates among its leaves and $\rho_c^{(0)}=\int_0^1(1-q u)^d\,du=O(1/(qd))$: the unexposed arm of the hub decays polynomially rather than staying constant, so the invertible focal weights grow linearly in the hub degree. The constant-propensity claim is therefore restricted to graph families in which ego-neighborhoods are small in expectation, such as locally tree-like $d$-regular graphs and, more generally, graphs of bounded degree. Thus the distance-one construction of Section~\ref{sec:is-dte}, run as a seeding experiment, estimates the influencer spillover effect at constant focal propensities on exactly the bounded-degree families where the intervention is well-defined. In every case the exponential separation between average and expected causal effects under classical designs is a consequence of graph-unaware assignment, not a property of the estimand.

\subsection{Cluster randomized designs}
\label{sec:cluster-designs}

A cluster randomized design partitions the units into disjoint clusters $\mathcal C=\{C_1,\dots,C_K\}$ and assigns a common treatment to all units in a cluster, drawing the cluster-level indicators $W_1,\dots,W_K$ according to a specified scheme. A unit $i$ reveals $Y_i(1,e_i^{\mathbf 1})$ when its cluster and all clusters meeting its neighborhood are treated, and reveals $Y_i(0,0)$ when they are all controlled. Provided the induced propensities $\pi_i(1,e_i^{\mathbf 1})$ and $\pi_i(0,0)$ are positive, the treatment--exposure Horvitz--Thompson estimator of Proposition~\ref{prop:ze-ht-unbiased} is unbiased for the total treatment effect; explicit expressions for these propensities under graph cluster randomization are given by \citet{ugander2013graph}.

The polynomial behavior of cluster randomization is, however, conditional on the clustering being aligned with the graph. Let $m_i$ denote the number of distinct clusters met by the closed neighborhood $\{i\}\cup N_i$. Under independent cluster-level assignment with probability $\tfrac12$, the all-treated condition requires all $m_i$ of these clusters to be treated, so
\[
\pi_i(1,e_i^{\mathbf 1})=\Big(\tfrac12\Big)^{m_i},
\]
and the inverse propensity is $2^{m_i}$, exponential in the number of clusters the neighborhood spans rather than in the degree. When the clustering places each closed neighborhood inside a bounded number of clusters, $m_i=O(1)$ and the propensity is bounded below by a constant. Low-boundary or low-conductance partitions can have this property in clustered graphs, but the relevant design quantity is the neighborhood cluster count $m_i$ itself. On graphs for which every practical balanced clustering splits many neighborhoods across a growing number of clusters, $m_i$ grows with the degree, and the exponential cost reappears. Cluster randomization therefore trades an exponential dependence on degree for an exponential dependence on the neighborhood boundary, and removes the estimation barrier only when the chosen clustering is aligned with the interference graph. The independent set design IS-2 gives a graph-agnostic polynomial upper bound for the total treatment effect, while cluster designs can be sharper on graphs with strong community structure.

\section{Implications for Designs, Estimators, and Models}
\label{sec:implications}

The variance bounds of Section~\ref{sec:nonparametric-estimation} have several implications for the design and analysis of experiments under interference. We collect them here. 
\subsection{Classical designs and estimators target expected causal effects}
\label{sec:implications-classical}

Bernoulli and completely randomized designs, analyzed with the difference-in-means estimator, are the tools most commonly used in practice. That these classical tools estimate an expected causal effect rather than an average causal effect under interference is well known \citep{hudgens2012toward,savje2017average}: the difference-in-means estimator targets the expected direct effect, obtained by averaging the unit-level direct effect over the exposure induced by the design. Under a Bernoulli design, the difference-in-means estimator is design-unbiased for the expected direct effect; by Proposition~\ref{prop:ede-variance-bound}, its variance is of order $(1+c_{\mathrm{avg}})/n$, and its computation does not require the knowledge of interference graph. However, the expected direct effect is policy dependent, and it mixes direct treatment effect and interference effects, as shown in Proposition \ref{prop:estimand-representation} which uses the non-parametric linear decomposition.  %

The same design and estimator are ill-suited to non-parametrically estimate average causal effects. The direct treatment effect is a contrast of the potential outcomes $Y_i(1,\mathbf 0_{N_i})$ and $Y_i(0,\mathbf 0_{N_i})$, and under a Bernoulli design the cell $(1,\mathbf 0_{N_i})$ is observed with probability $p(1-p)^{d_i}$. By Proposition~\ref{prop:ace-inverse-propensity}, the variance of any graph unaware design-unbiased estimator of the direct treatment effect is then exponential in the interference degree, a similar result holds for the total treatment effect. Moreover, the difference-in-means estimator is not consistent for the direct treatment effect: its probability limit is the expected direct effect, so it carries a bias equal to the expected treatment--exposure interaction $\overline{C}$, which is zero only under additivity. Classical estimators are therefore appropriate for expected causal effects such as the expected direct effect, but not for average causal effects such as the direct and total treatment effects.

\subsection{Positivity and overlap for average causal effects}
\label{sec:implications-designs}

Changing the exposure mapping does not provide an escape. As shown in Section~\ref{sec:nonparametric-estimation}, the direct treatment effect requires the no-exposure potential outcome under any mapping, so its propensity remains exponentially small in the degree. The difficulty is a failure of positivity for the relevant treatment--exposure combination, the analogue under interference of the overlap condition in observational studies. On interference graphs with high-degree units this failure is severe, and a fixed-exposure average causal effect can be identified only on the subpopulation of units whose neighborhoods are small enough to provide adequate overlap.

\subsection{Model-assisted estimation of average causal effects}
\label{sec:implications-models}

A second strategy is available when one is willing to impose outcome model restrictions on the potential outcomes. The non-parametric decomposition $Y_i(z,e)=A_i(z)+B_i(e)+zC_i(e)$ suggests a natural class of working models: a restriction on the spillover and interaction components $B_i$ and $C_i$ relates the nuisance potential outcomes to the relevant ones, so that units that reveal nuisance potential outcomes also become informative for the estimand. The simplest such restriction is additivity, $C_i\equiv 0$, under which the difference-in-means estimator is consistent for the direct treatment effect, since the bias $\overline{C}$ then vanishes. More generally, a working model can be used within a model-assisted estimator, such as a generalized regression estimator, that retains a design-based target while borrowing information across treatment--exposure cells. Such estimators trade the exponential cost of the inverse propensity for sensitivity to misspecification of the working model. Model-assisted estimators of this kind under interference are developed in \citet{karwa2018systematic}.

\section{Numerical illustration}
\label{sec:numerical}
 We illustrate the estimation gap between ACEs and ECEs on simulated experiments. The interference graph, the exposure mapping, and the potential outcomes are fixed in advance, and the only source of randomness is the treatment assignment. %

We draw the interference graph from two random graph model families. %
The first is the Erd\H{o}s--R\'enyi graph $G(n,c/n)$, which is sparse and has degrees concentrated about their mean; it represents limited and roughly homogeneous interference. The second is the Barab\'asi--Albert graph, whose degree distribution is heavy-tailed and contains a few high-degree hubs; it represents the heterogeneous interference common in social and economic networks. We use $n=2000$ units in the Erd\H{o}s--R\'enyi experiments and $n=3000$ in the Barab\'asi--Albert experiment.

The exposure mapping is the fraction of treated neighbors, \(e_i=d_i^{-1}\sum_{j\in N_i}z_j\), with \(e_i=0\) when \(d_i=0\). Potential outcomes follow the decomposition \(Y_i(z,e)=\mu_i+A_i(z)+B_i(e)+zC_i(e)\), with \(A_i(z)=\alpha_i z\), \(B_i(e)=\beta_i e\), and \(C_i(e)=\gamma_i e\). The unit-level coefficients are drawn independently once, with \(\mu_i\sim \operatorname{Unif}(-0.2,0.2)\), \(\alpha_i\sim \operatorname{Unif}(0.3,0.7)\), \(\beta_i\sim \operatorname{Unif}(0.2,0.6)\), and \(\gamma_i\sim \operatorname{Unif}(0,0.4)\), and then held fixed. Thus the response surface is bounded and heterogeneous across units. The interaction coefficients have positive mean, so the direct and expected direct effects differ. The figures focus on three estimands: the direct treatment effect, which contrasts \(Y_i(1,0)\) and \(Y_i(0,0)\); the expected direct effect under the Bernoulli allocation policy; and the expected indirect effect under the same policy. Because the treated-neighbor count is binomial conditional on the graph under a Bernoulli design, each target is available in closed form and estimation error is measured against the exact finite-population value.

\begin{figure}[t]
\centering
\includegraphics[width=0.62\textwidth]{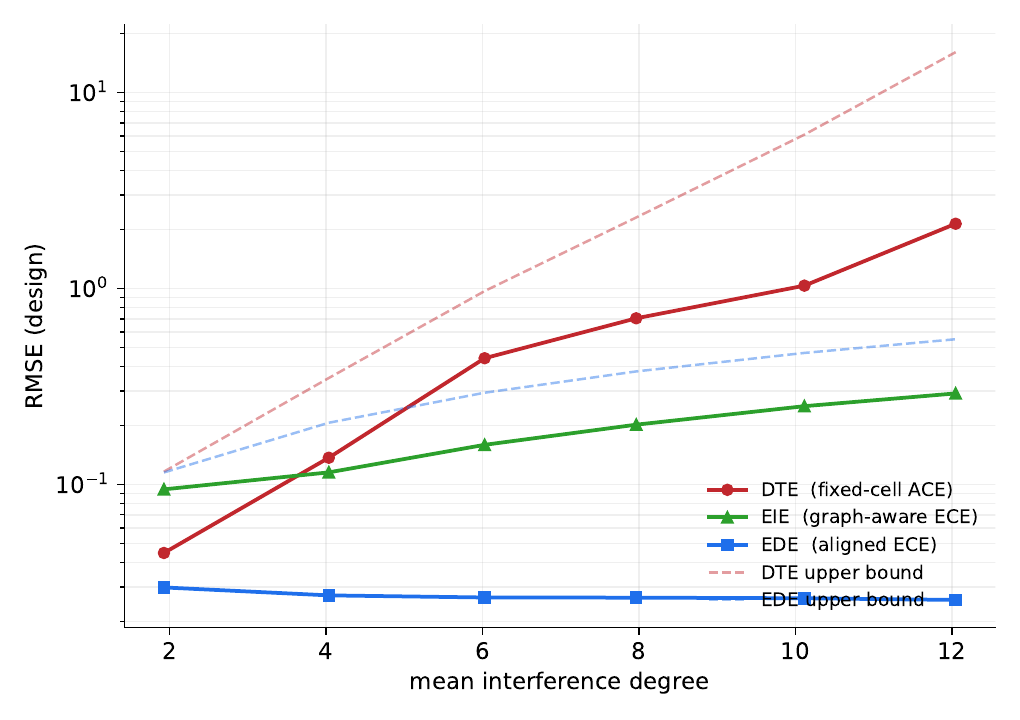}
\caption{Design root-mean-square error of the Horvitz--Thompson estimator under a
Bernoulli design ($p=1/2$) against the mean interference degree, on
Erd\H{o}s--R\'enyi graphs ($n=2000$); the vertical axis is logarithmic. The
direct treatment effect (fixed-exposure ACE, red) follows the exponential
inverse-propensity scale (red dashed; Proposition~\ref{prop:generic-ace-ht}); the
expected direct effect (aligned ECE, blue) is horizontal; the expected indirect effect
(graph-aware ECE, green) grows polynomially.}
\label{fig:separation}
\end{figure}

\begin{figure}[t]
\centering
\includegraphics[width=0.62\textwidth]{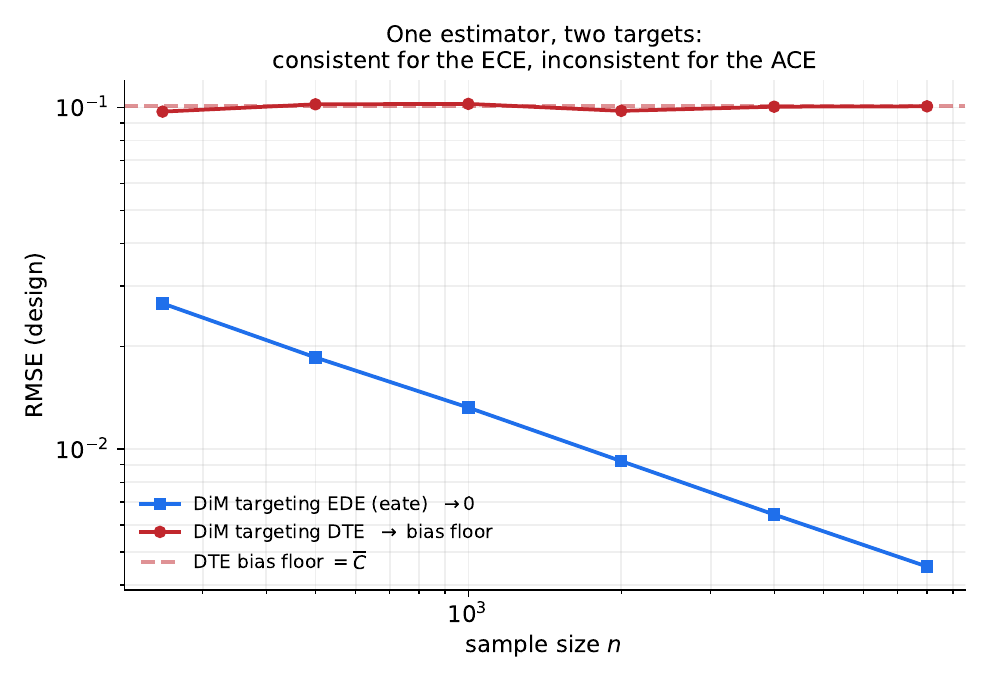}
\caption{The difference-in-means estimator against two targets as the sample
grows (Erd\H{o}s--R\'enyi, mean degree $8$, $p=1/2$). Its error for the expected
direct effect decays at the parametric rate; its error for the direct treatment
effect settles at the policy-averaged interaction floor $\bar C_{\pi}$ set by the treatment--exposure
interaction that own-treatment contrasts cannot remove.}
\label{fig:reconciliation}
\end{figure}

\begin{figure}[t]
\centering
\includegraphics[width=0.62\textwidth]{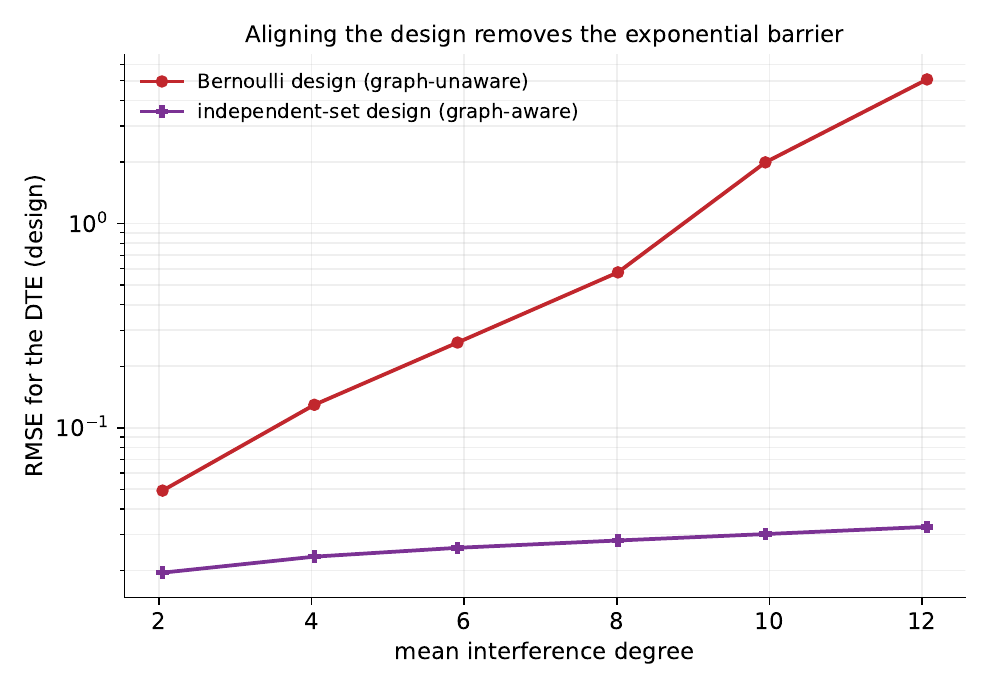}
\caption{Estimand-aligned design for the direct treatment effect (Erd\H{o}s--R\'enyi,
$n=2000$, $p=1/2$). Under a Bernoulli design the Horvitz--Thompson error grows
exponentially in the mean interference degree; under an independent-set design on
the same graphs, which reveals \(Y_i(1,0)\) and \(Y_i(0,0)\) for ego units, it stays
nearly flat.}
\label{fig:design_repair}
\end{figure}

\begin{figure}[t]
\centering
\includegraphics[width=\textwidth]{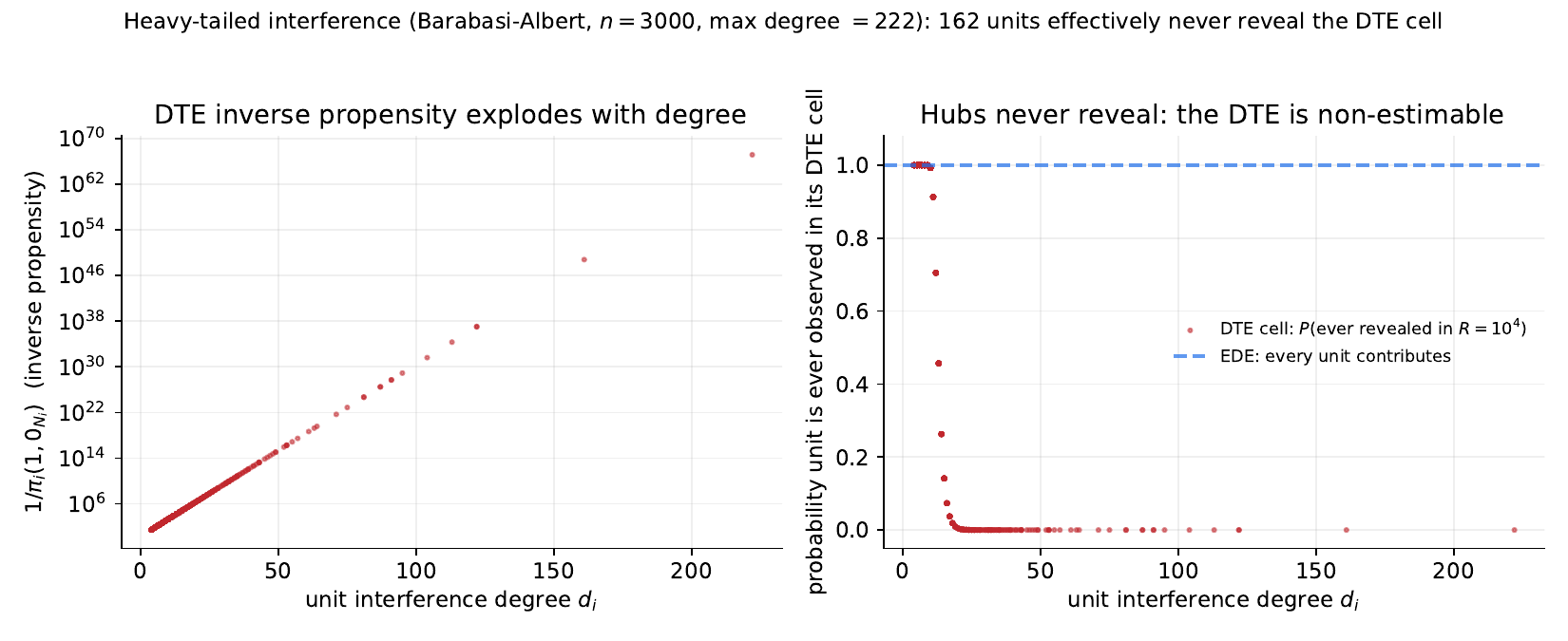}
\caption{Heavy-tailed interference (Barab\'asi--Albert, $n=3000$). Left: the
inverse propensity of the direct-effect exposure condition grows exponentially in a
unit's degree, exceeding $10^{70}$ for the highest-degree unit. Right: the
probability that a unit ever reveals the relevant direct-effect potential outcome, in
$10^{4}$ assignment draws, collapses to zero for high-degree units, whereas the
expected direct effect uses every unit.}
\label{fig:heavytail}
\end{figure}

Each effect is estimated by the corresponding Horvitz--Thompson estimator of Section~\ref{sec:nonparametric-estimation}: treatment--exposure weights for the direct treatment effect, fixed own-treatment weights for the expected direct effect, and graph-aware weights for the expected indirect effect. The usual difference-in-means estimator appears separately in Figure~\ref{fig:reconciliation}; it is the ratio form of the own-treatment weighted estimator. We consider two designs: a graph-unaware Bernoulli design with treatment probability one half, used in all but one experiment, and an independent-set design aligned with the direct treatment effect. Each reported error is the root-mean-square error over independent draws of the assignment; Figures~\ref{fig:separation}--\ref{fig:design_repair} use Monte Carlo repetitions, while Figure~\ref{fig:heavytail} reports the exact probability that the relevant potential outcome is observed at least once in \(10^4\) Bernoulli assignments. The Horvitz--Thompson estimators are design-unbiased by construction, and we confirmed that the empirical bias of each is zero up to Monte Carlo error.

Figure~\ref{fig:separation} reports the root-mean-square error against the mean interference degree on the Erd\H{o}s--R\'enyi graphs; the vertical axis is logarithmic, so a straight line is exponential growth. The direct treatment effect (red circles) follows the inverse-propensity scale for observing \(Y_i(1,0)\) or \(Y_i(0,0)\) (the red dashed line, Proposition~\ref{prop:generic-ace-ht}). The expected direct effect (blue squares) is horizontal, as Proposition~\ref{prop:ede-variance-bound} predicts, and stays near its dependence scale (the blue dashed line). The expected indirect effect (green triangles) rises only polynomially, well below the direct effect. The three estimands use the same graph, the same design, and the same outcomes. %
The separation is not only asymptotic. At bounded degree the RMSE of both classes decreases at the parametric rate \(n^{-1/2}\), but the constant differs sharply: for the direct treatment effect the effective sample size is of order \(n p(1-p)^d\), since only assignments revealing the relevant potential outcomes contribute, whereas the expected direct effect uses every unit and its constant grows only polynomially in the degree. Computed at a single sample size, Figure~\ref{fig:separation} shows that this gap in constants is already large at the sizes typical of experiments.

Figure~\ref{fig:reconciliation} follows a single estimator, the difference in means, against two targets as the sample grows. Its error for the expected direct effect (blue) decays as a straight line of slope $-1/2$ on the logarithmic axes, the parametric rate; its error for the direct treatment effect (red) flattens at a positive floor, the horizontal dashed line at $\bar C_{\pi}$. The estimand class, not the estimator, governs consistency: the same estimator is consistent for the effect the design averages over and inconsistent for the fixed-exposure effect on the same data. The decomposition $Y_i(z,e)=A_i(z)+B_i(e)+zC_i(e)$ identifies the floor exactly. It is the policy-averaged treatment--exposure interaction $\bar C_{\pi}$, equal to $p\,n^{-1}\sum_i\gamma_i$ in this simulation, and is the part of the response surface that own-treatment contrasts cannot remove; under additivity, $C_i\equiv 0$, the floor vanishes and the difference in means is consistent for the direct effect as well.

Figure~\ref{fig:design_repair} illustrates the first route around the exponential barrier for estimating ACEs. The independent-set design treats mutually non-adjacent ego units and holds their neighborhoods at control, so each ego reveals one of the two potential outcomes that define the direct treatment effect, \(Y_i(1,0)\) or \(Y_i(0,0)\). On the same Erd\H{o}s--R\'enyi graphs as Figure~\ref{fig:separation}, the error under the Bernoulli design (red) grows exponentially in the degree, while the error under the independent-set design (purple) remains nearly flat.

Figure~\ref{fig:heavytail} turns to a heavy-tailed Barab\'asi--Albert graph. The direct-effect potential outcome for a high-degree unit is observed with probability that decays exponentially in the unit's degree, so the inverse propensity of the highest-degree unit exceeds \(10^{70}\) (left panel) and the probability of seeing the relevant potential outcome at least once in \(10^4\) assignment draws is essentially zero (right panel). The fixed-exposure ACE is not merely expensive to estimate on such graphs; at feasible sample sizes it is effectively non-estimable for the hubs, while the expected direct effect uses every unit. Truncating or stabilizing the extreme inverse-propensity weights can reduce variance, but only by changing the estimator and introducing bias for the fixed finite-population target unless additional modeling assumptions are imposed. Heterogeneity in the degree distribution sharpens the separation rather than softening it: the cost of a fixed-exposure ACE is governed by the exponential moment of the degree distribution, which the hubs dominate.

\section{Discussion}
\label{sec:discussion}

For principled causal inference, the potential-outcomes framework keeps three ingredients separate: restrictions on the potential outcomes, the definition of a causal estimand, and the design and estimator used to learn it. Without interference, the first ingredient is a single clean exclusion restriction: each unit has two potential outcomes. Interference is not the negation of that restriction. It is a family of possible restrictions. One has to specify which units' treatment assignments can affect a given unit's potential outcome, how those assignments enter the potential outcome, and whether the response surface has additional structure such as additivity, homogeneity, or linearity. The non-parametric decomposition of potential outcomes given in this paper provides a unified language to specify any finite-population model of potential outcomes under interference.

Moreover, without interference there is a single canonical finite-population average treatment effect. With interference there are multiple causal effects of interest, such as direct effects, spillover effects, total effects, and policy effects. We classify estimands under interference into two broad categories: average causal effects, which contrast fixed potential outcomes, and expected causal effects, which average potential outcomes under an allocation policy. This distinction matters because the same name can refer to different functionals of the potential-outcome table. For example, a direct effect may be a fixed contrast holding neighbors' treatment fixed, as in the direct treatment effect. It may also average over the neighbor treatments generated by an allocation policy, as in the expected direct effect. These estimands coincide under additivity, but not in general.

This perspective also clarifies why several results in the literature that appear to point in different directions are compatible. Partial-interference analyses specify groups and exposure within groups, and the resulting effects are tied to the group-level allocation policy \citep{hudgens2012toward}. Exposure-mapping approaches target average causal effects by contrasting fixed treatment--exposure combinations \citep{aronow2017estimating}. Some estimators remain valid under unknown interference because their targets average over the exposure distribution induced by a policy or design; in the terminology of this paper, the average direct effect studied by \citet{savje2017average} is an expected direct effect. Other approaches recover average causal effects under unknown network structure by imposing outcome-model restrictions; for example, \citet{yu2022pnasTTE} target a total treatment effect using additivity and linearity. Estimation of these effects is not free. The graph assumption is replaced either by a policy-averaged target or by an outcome-model assumption on the potential outcomes.

The same distinction governs design. Classical graph-unaware designs are often well matched to estimate policy-averaged expected effects, because they generate the exposure distribution being averaged over. They are poorly matched to estimate fixed-exposure average effects when the defining treatment--exposure combinations are rare. In observational studies without interference, positivity requires treatment variation within the covariate strata that define the target comparison. Under interference, the analogous strata are treatment--exposure cells. A Bernoulli design gives abundant information about typical exposure states, but a direct treatment effect ACE may require the atypical cell in which a treated unit has no treated neighbors. Similarly, a total treatment effect ACE may require all-treated and all-control neighborhoods. These cells can have exponentially small probability in the interference degree under classical designs. Fixed-exposure ACEs therefore require either estimand-aligned designs or outcome-model restrictions. Coarsening the spillover boundary can collapse this cost: under the any-treated exposure map the influencer spillover effect is estimated with constant focal propensities by sparse seeding on locally tree-like regular graphs (Section~\ref{sec:is-indirect}), while the uncompressed full-exposure versions retain the exponential barrier and need the distance-two design.

The exponential-versus-polynomial gap implies that the estimand and the design cannot be chosen separately. An investigator who targets a fixed-exposure average causal effect but implements a classical design will incur large estimation errors. The estimand and a design that reveals it must be chosen together, at the planning stage. When the design cannot be controlled, as in observational studies, the situation is more severe: average causal effects are then essentially out of reach without strong outcome models or bounded or logarithmic interference degrees, whereas expected causal effects aligned with the assignment mechanism remain estimable under the conditions that make them estimable by design.

A central problem is that the graph and exposure mapping are rarely known at the design stage. If the specified graph and exposure mapping are correct, a fixed-exposure average causal effect is a contrast on the reduced potential outcomes. If the exposure mapping is misspecified, the same contrast may average over assignment features that the mapping omits; this distinction between a target defined under a working exposure mapping and the corresponding target in the full Table of Science is emphasized by \citet{savje2024causal}. The same issue arises for the interference graph. If the graph omits relevant treatment coordinates, a fixed-exposure contrast in the reduced representation averages over those omitted coordinates in the full Table of Science. It is then no longer the original fixed-exposure ACE, but a coarsened estimand defined by the misspecified representation. This is not only a matter of bias in an estimator. The target itself has changed.

Several problems follow. Experiments should be designed so that conclusions are stable to plausible changes in the interference graph and exposure mapping. When network data, pretreatment covariates, or repeated experiments are used to learn these objects, the uncertainty in that learning step should enter the final inference. Variance estimation must account for randomization dependence and interference dependence at the same time. Model-assisted estimators add another layer: the fitted response model can reduce variance or recover information from nuisance potential outcomes, but its use must be separated from the design-based correction it supports. These are the questions that arise after the estimand, the interference model, and the design have been put in the same language.
\appendix

\section{Additional examples and extensions for Section~\ref{sec:POmodels}}
\label{app:additional-framework-examples}

\subsection{Asymmetric and bipartite interference}

While many models assume symmetric interference, interference may be directional. For instance, in social hierarchies or upstream--downstream systems, the treatment of one unit can affect another without a reciprocal effect. In these cases, the in-degree represents a unit's exposure to external interference, while the out-degree measures that unit's reach within the network.

The framework also extends to bipartite interference. In such cases, the graph $G = (U \cup V, E)$ consists of a set of treatment units $U$ and outcome units $V$, where edges only exist from $U$ to $V$. This structure is useful when the units being treated differ from the units whose outcomes are measured, as in environmental and service-provider settings \citep{zigler2021bipartite}.

\subsection{Vector-valued and continuous exposures}

The exposure function may be multidimensional. For example, in a social network one may use
\[
e_i = \big(\text{number of treated friends}, \ \text{number of treated friends-of-friends}\big).
\]
In settings where interference is governed by physical proximity, the exposure mapping can be defined as a continuous function of geographic distance. For example, $N_i$ can represent all units within a relevant radius, with
\[
e_i = \sum_{j \in N_i} K(\rho_{ij}; \phi) z_j,
\]
where $\rho_{ij}$ is the distance between units $i$ and $j$, and $K(\cdot)$ is a kernel function parameterized by $\phi$ \citep{giffin2023generalized}.

\subsection{Nonlinear and threshold outcome restrictions}

The decomposition in Theorem~\ref{thm:decomposition} also accommodates nonlinear restrictions on the spillover function $B_i$ and the interaction function $C_i$. For example, under additivity ($C_i=0$), one may posit a concave interference function such as $B_i(e_i)=\gamma\log(1+e_i)$, a convex function such as $B_i(e_i)=\gamma e_i^2$, or a sigmoidal function
\[
Y_i(z_i,e_i)
=
\alpha_i+\beta_i z_i+
\frac{L}{1+\exp\{-k(e_i-e_0)\}}.
\]
Such restrictions arise in settings where the first few treated neighbors have large effects, or where spillovers occur only after a threshold has been reached \citep{aral2011,leskovec2007dynamics,degiorgi2010identification}.

Threshold and contagion models can be represented by binary exposure mappings such as
\[
e_i=\mathbb{I}\left(\sum_{j\in N_i}z_j\geq K\right),
\]
which are often used in models of collective behavior and complex contagion \citep{granovetter1978,centola2010}. Additional structural restrictions can then be imposed on $B_i(e_i)$ and $C_i(e_i)$ depending on whether individual treatment and exposure are substitutes, complements, or interact in some other way.

\subsection{Non-anonymous exposure}

In non-anonymous interference, the effect of neighbors' treatments depends on their identities or covariates. One example is
\[
e_i = \sum_{j\in N_i} w(X_j) z_j,
\]
where $X_j$ are neighbor attributes. The potential outcome can then be written as
\[
Y_i(z_i,e_i)
=
\alpha_i+\beta_i z_i+
B_i\left(\sum_{j\in N_i}w(X_j)z_j\right)
+z_iC_i\left(\sum_{j\in N_i}w(X_j)z_j\right).
\]
This includes weighted and spatial exposure mappings, including environmental settings where exposure depends on distance or direction \citep{zigler2012,patacchini2017heterogeneous,adhikari2025}.

\section{Proofs for Section~\ref{sec:POmodels}}
\subsection{Proof of Theorem~\ref{thm:decomposition}}
\paragraph{Restatement of Theorem~\ref{thm:decomposition}.}
For any interference neighborhood \(N_i\) and exposure mapping
\(e_i=f(\mathbf Z_{N_i})\), there exist functions \(A_i(z)\), \(B_i(e)\), and
\(C_i(e)\) such that, for each unit \(i\) and each
\((z,e)\in\{0,1\}\times\mathcal E_i\),
\[
Y_i(z,e)=A_i(z)+B_i(e)+zC_i(e),
\]
with the normalization \(B_i(0)=C_i(0)=0\).

\begin{proof}
Fix a unit \(i\).  Once the interference neighborhood and exposure mapping are
specified, the potential outcomes for unit \(i\) are indexed by
\((z,e)\in\{0,1\}\times\mathcal E_i\).  Define
\[
A_i(z)=Y_i(z,0),
\qquad
B_i(e)=Y_i(0,e)-Y_i(0,0),
\]
and
\[
C_i(e)=Y_i(1,e)-Y_i(1,0)-Y_i(0,e)+Y_i(0,0).
\]
Then \(B_i(0)=0\) and \(C_i(0)=0\).  If \(z=0\), then
\[
A_i(0)+B_i(e)
=
Y_i(0,0)+Y_i(0,e)-Y_i(0,0)
=
Y_i(0,e).
\]
If \(z=1\), then
\[
\begin{aligned}
A_i(1)+B_i(e)+C_i(e)
&=
Y_i(1,0)+Y_i(0,e)-Y_i(0,0)\\
&\quad +Y_i(1,e)-Y_i(1,0)-Y_i(0,e)+Y_i(0,0)\\
&=
Y_i(1,e).
\end{aligned}
\]
Thus \(Y_i(z,e)=A_i(z)+B_i(e)+zC_i(e)\) for every \(z\) and \(e\).

The representation is also a full-rank reparametrization.  For each unit \(i\), there
are \(2K_i\) distinct potential outcomes.  The parametrization contains
\[
2+(K_i-1)+(K_i-1)=2K_i
\]
free parameters:
\[
\{A_i(0),A_i(1)\},\qquad
\{B_i(e):e\neq0\},\qquad
\{C_i(e):e\neq0\}.
\]
The formulas above give the inverse map from the potential-outcome table to these
parameters, so the map is one-to-one and full rank.
\end{proof}

\section{Proofs and details for Section~\ref{sec:estimands}}
\label{app:estimands}

\subsection{Mixed causal effects: EIC, EIT, and ETE}
\label{app:mixed-effects}

Mixed causal effects contrast expected potential outcomes under a policy against fixed baseline responses.  They can be viewed as expected analogues of the ACE indirect effects.

\paragraph{Expected exposure effects.}
Expected exposure effects capture the policy-averaged indirect effect, holding the own treatment fixed.  The expected exposure effect on control units is
\begin{equation}
\tau_{\mathrm{EIC}}(\Qpol)
\,=\,\frac{1}{n}\sum_{i=1}^n
\sum_{e \in \mathcal{E}} \Big(Y_i(0,e)-Y_i(0,0)\Big)\,P_{\Qpol}(E_i=e)
= \frac{1}{n}\sum_i \mu_i(0;\Qpol) - \frac{1}{n}\sum_iY_i(0,0),
\label{eq:ECE-spill-0}
\end{equation}
and on treated units is
\begin{equation}
\tau_{\mathrm{EIT}}(\Qpol)
\,=\,\frac{1}{n}\sum_{i=1}^n
\sum_{e \in \mathcal{E}} \Big(Y_i(1,e)-Y_i(1,0)\Big)\,P_{\Qpol}(E_i=e)
=  \frac{1}{n}\sum_i \mu_i(1;\Qpol) - \frac{1}{n}\sum_iY_i(1,0).
\label{eq:ECE-spill-1}
\end{equation}
These satisfy $\tau_{\mathrm{EIC}}(\Qpol)=\mathbb E_{\Qpol}[\tau_{\mathrm{IEC}}(E)]$ and $\tau_{\mathrm{EIT}}(\Qpol)=\mathbb E_{\Qpol}[\tau_{\mathrm{IET}}(E)]$.

\paragraph{Expected total effect.}
The expected total effect compares the expectation under full treatment to the baseline condition:
\begin{equation}
\tau_{\mathrm{ETE}}(\Qpol)
\,=\,\frac{1}{n}\sum_{i=1}^n
\sum_{e \in \mathcal{E}} \Big(Y_i(1,e)-Y_i(0,0)\Big)\,P_{\Qpol}(E_i=e)
=  \frac{1}{n}\sum_i \mu_i(1;\Qpol) - \frac{1}{n}\sum_iY_i(0,0).
\label{eq:ECE-total}
\end{equation}
By Proposition~\ref{prop:estimand-representation}, these satisfy $\tau_{\mathrm{EIC}}(\Qpol)=\bar B_{\Qpol}$, $\tau_{\mathrm{EIT}}(\Qpol)=\bar B_{\Qpol}+\bar C_{\Qpol}$, and $\tau_{\mathrm{ETE}}(\Qpol)=n^{-1}\sum_i\beta_i+\bar B_{\Qpol}+\bar C_{\Qpol}$, and they decompose as $\tau_{\mathrm{ETE}}(\Qpol)=\tau_{\mathrm{EDE}}(\Qpol)+\tau_{\mathrm{EIC}}(\Qpol)=\tau_{\mathrm{DTE}}+\tau_{\mathrm{EIT}}(\Qpol)$.

\begin{proposition}[ECE mixed-effect decomposition]
\label{prop:ece-decomposition}
For any allocation policy \(\Qpol\),
\[
\tau_{\mathrm{ETE}}(\Qpol)
=
\tau_{\mathrm{EDE}}(\Qpol)+\tau_{\mathrm{EIC}}(\Qpol)
=
\tau_{\mathrm{DTE}}+\tau_{\mathrm{EIT}}(\Qpol).
\]
\end{proposition}

\subsection{Proof of Proposition~\ref{prop:estimand-representation}}
\paragraph{Restatement.}
Under the non-parametric linear representation of Theorem~\ref{thm:decomposition},
\[
Y_i(z,e)=\alpha_i+\beta_i z+B_i(e)+zC_i(e),
\qquad B_i(0)=C_i(0)=0,
\]
the ACEs and ECEs have the representations stated in
Proposition~\ref{prop:estimand-representation}.

\begin{proof}
For the fixed-exposure effects, substitute the representation into each contrast.  The
direct treatment effect is
\[
Y_i(1,0)-Y_i(0,0)
=
\{\alpha_i+\beta_i+B_i(0)+C_i(0)\}-\{\alpha_i+B_i(0)\}
=\beta_i.
\]
Similarly,
\[
Y_i(0,e)-Y_i(0,0)=B_i(e),
\qquad
Y_i(1,e)-Y_i(1,0)=B_i(e)+C_i(e),
\]
and
\[
Y_i(1,1)-Y_i(0,0)=\beta_i+B_i(1)+C_i(1).
\]
Averaging these identities over \(i\) gives the ACE formulas.

For the expected effects, take expectations with respect to the exposure distribution
induced by the allocation policy.  Since
\[
Y_i(1,E_i)-Y_i(0,E_i)=\beta_i+C_i(E_i),
\]
we have
\[
\tau_{\mathrm{EDE}}(\Qpol)
=
\frac1n\sum_i\beta_i
+\frac1n\sum_i\mathbb E_{\Qpol}[C_i(E_i)].
\]
The remaining three formulas follow from
\[
Y_i(0,E_i)-Y_i(0,0)=B_i(E_i),
\]
\[
Y_i(1,E_i)-Y_i(1,0)=B_i(E_i)+C_i(E_i),
\]
and
\[
Y_i(1,E_i)-Y_i(0,0)=\beta_i+B_i(E_i)+C_i(E_i).
\]
\end{proof}

\subsection{Proof of Proposition~\ref{prop:ace-decomposition}}
\paragraph{Restatement.}
The total treatment effect satisfies
\[
\tau_{\mathrm{TTE}}=\tau_{\mathrm{DTE}}+\tau_{\mathrm{IET}}.
\]
Under additivity, \(C_i(e)=0\) for all \(i,e\), so
\[
\tau_{\mathrm{TTE}}
=
\tau_{\mathrm{DTE}}+\tau_{\mathrm{IET}}
=
\tau_{\mathrm{DTE}}+\tau_{\mathrm{IEC}}.
\]

\begin{proof}
By Proposition~\ref{prop:estimand-representation},
\[
\tau_{\mathrm{TTE}}
=
\frac1n\sum_i\{\beta_i+B_i(1)+C_i(1)\},
\]
while
\[
\tau_{\mathrm{DTE}}+\tau_{\mathrm{IET}}
=
\frac1n\sum_i\beta_i
+
\frac1n\sum_i\{B_i(1)+C_i(1)\}.
\]
The two expressions are equal.  If \(C_i(e)=0\) for all \(i,e\), then
\(\tau_{\mathrm{IET}}=\tau_{\mathrm{IEC}}\), which gives the additive form.
\end{proof}

\subsection{Proof of Proposition~\ref{prop:ece-decomposition}}
\paragraph{Restatement.}
For any allocation policy \(\Qpol\),
\[
\tau_{\mathrm{ETE}}(\Qpol)
=
\tau_{\mathrm{EDE}}(\Qpol)+\tau_{\mathrm{EIC}}(\Qpol)
=
\tau_{\mathrm{DTE}}+\tau_{\mathrm{EIT}}(\Qpol).
\]

\begin{proof}
Using the notation of Proposition~\ref{prop:estimand-representation},
\[
\tau_{\mathrm{ETE}}(\Qpol)
=
\frac1n\sum_i\beta_i+\bar B_{\Qpol}+\bar C_{\Qpol}.
\]
Also,
\[
\tau_{\mathrm{EDE}}(\Qpol)+\tau_{\mathrm{EIC}}(\Qpol)
=
\left\{\frac1n\sum_i\beta_i+\bar C_{\Qpol}\right\}
+\bar B_{\Qpol},
\]
and
\[
\tau_{\mathrm{DTE}}+\tau_{\mathrm{EIT}}(\Qpol)
=
\frac1n\sum_i\beta_i+\bar B_{\Qpol}+\bar C_{\Qpol}.
\]
Both expressions equal \(\tau_{\mathrm{ETE}}(\Qpol)\).
\end{proof}

\subsection{Proof of Corollary~\ref{cor:additive-ede-dte}}
\paragraph{Restatement.}
Under additivity of the causal response surface, \(\tau_{\mathrm{EDE}}(\Qpol)=\tau_{\mathrm{DTE}}\)
for every allocation policy \(\Qpol\).

\begin{proof}
Under additivity, \(C_i(e)=0\) for all units and exposure levels.  Therefore
\(\bar C_{\Qpol}=n^{-1}\sum_i\mathbb E_{\Qpol}[C_i(E_i)]=0\), and
Proposition~\ref{prop:estimand-representation} gives
\[
\tau_{\mathrm{EDE}}(\Qpol)
=
\frac1n\sum_i\beta_i
=
\tau_{\mathrm{DTE}}.
\]
\end{proof}

\subsection{Proof of Proposition~\ref{prop:HuLiWager}}
\label{app:policy-derivatives}
\paragraph{Restatement.}
Under a Bernoulli assignment, the exposure mapping versions of EDE and EIE decompose the
derivative of the expected population outcome:
\[
\tau_{\text{TPE}}(\pi)
=
\tau_{\mathrm{EDE}}(\pi)
+
\tau_{\mathrm{EIE}}(\pi).
\]

\begin{proof}
Let the treatment vector be $\mathbf{Z} = (Z_1, \dots, Z_n)$, where $Z_i \sim \text{Bernoulli}(\pi_i)$ are independent. Let $e_i = f(\mathbf{Z}_{N_i})$ be the exposure mapping for unit $i$. The expected population outcome under the policy $\pi = (\pi_1,\ldots,\pi_n)$ is
\[
\mu(\pi)=\frac{1}{n}\sum_{i=1}^n
\mathbb{E}_{\pi}[Y_i(Z_i,e_i)].
\]
By the multivariate chain rule,
\[
\tau_{\mathrm{TPE}}(\pi)
=
\sum_{j=1}^n \frac{\partial}{\partial \pi_j}\mu(\pi)
=
\frac{1}{n}\sum_{i=1}^n
\left\{
\frac{\partial}{\partial \pi_i}\mathbb{E}_{\pi}[Y_i(Z_i,e_i)]
+
\sum_{j\in N_i}
\frac{\partial}{\partial \pi_j}\mathbb{E}_{\pi}[Y_i(Z_i,e_i)]
\right\},
\]
where the partial derivative is zero for $j\notin N_i\cup\{i\}$.

For the own-treatment derivative, the independence of $Z_i$ from $e_i$ under Bernoulli assignment gives
\[
\frac{\partial}{\partial \pi_i}
\mathbb{E}_{\pi}[Y_i(Z_i,e_i)]
=
\mathbb{E}_{\pi}\!\left[Y_i(1,e_i)-Y_i(0,e_i)\right].
\]
Averaging over $i$ gives $\tau_{\mathrm{EDE}}(\pi)$.

For $j\in N_i$, define
\[
e_i^{+j}=e_i(Z_{N_i\setminus j},z_j=1),
\qquad
e_i^{-j}=e_i(Z_{N_i\setminus j},z_j=0).
\]
Again using the product structure of the Bernoulli design,
\[
\frac{\partial}{\partial \pi_j}
\mathbb{E}_{\pi}[Y_i(Z_i,e_i)]
=
\mathbb{E}_{\pi}\!\left[
Y_i(Z_i,e_i^{+j})-Y_i(Z_i,e_i^{-j})
\right].
\]
Summing over $j\in N_i$ and averaging over $i$ gives $\tau_{\mathrm{EIE}}(\pi)$. Combining the own-treatment and neighbor-treatment terms proves
\[
\tau_{\mathrm{TPE}}(\pi)
=
\tau_{\mathrm{EDE}}(\pi)
+
\tau_{\mathrm{EIE}}(\pi).
\]
\end{proof}

\subsection{Symmetric exposure mappings}
\begin{proposition}[EIE under symmetric exposure]
\label{prop:aie_symmetric}
Suppose the assignment is Bernoulli with common probability $\pi$, and suppose the exposure mapping is symmetric. If
\[
e_i(\mathbf{Z}_{N_i})=\sum_{j\in N_i}Z_j,
\qquad d_i=|N_i|,
\]
then the expected indirect effect can be written as
\[
\tau_{\mathrm{EIE}}(\pi)
=
\frac1n\sum_{i=1}^n d_i\,
\mathbb E_\pi\!\left[
Y_i(Z_i,\widetilde E_i+1)-Y_i(Z_i,\widetilde E_i)
\right],
\]
where $\widetilde E_i$ is the exposure generated by the remaining $d_i-1$ neighbors. If instead
\[
e_i(\mathbf{Z}_{N_i})=\frac{1}{d_i}\sum_{j\in N_i}Z_j,
\]
then
\[
\tau_{\mathrm{EIE}}(\pi)
=
\frac1n\sum_{i=1}^n d_i\,
\mathbb E_\pi\!\left[
Y_i\left(Z_i,\widetilde E_i+\frac{1}{d_i}\right)
-Y_i\left(Z_i,\widetilde E_i\right)
\right],
\]
where $\widetilde E_i$ is the fractional exposure generated by the remaining $d_i-1$ neighbors.
\end{proposition}

\begin{proof}
For the count exposure, fix \(i\) and \(j\in N_i\).  Since the Bernoulli design has a
common treatment probability and the exposure mapping is symmetric, the distribution of
the exposure generated by \(N_i\setminus\{j\}\) is the same for every neighbor \(j\).
Moreover,
\[
e_i^{+j}=\widetilde E_i+1,
\qquad
e_i^{-j}=\widetilde E_i .
\]
Substituting these two expressions into the definition of
\(\tau_{\mathrm{EIE}}(\pi)\) and summing the identical neighbor contributions over
\(j\in N_i\) gives
\[
\tau_{\mathrm{EIE}}(\pi)
=
\frac1n\sum_{i=1}^n d_i\,
\mathbb E_\pi\!\left[
Y_i(Z_i,\widetilde E_i+1)-Y_i(Z_i,\widetilde E_i)
\right].
\]
For fractional exposure, changing \(Z_j\) from zero to one increases exposure by
\(1/d_i\).  Thus
\[
e_i^{+j}=\widetilde E_i+\frac1{d_i},
\qquad
e_i^{-j}=\widetilde E_i,
\]
and the same averaging argument gives the second formula.
\end{proof}

\section{Proofs for Section~\ref{sec:nonparametric-estimation}}
\label{app:nonparametric-estimation-proofs}

\subsection{Proof of Lemma~\ref{lem:interference-dependence-graph}}
\begin{proof}
If there is no dependence-graph edge between two disjoint vertex sets \(A\) and \(B\),
then \(\mathcal C_i\cap\mathcal C_j=\emptyset\) for every \(i\in A\) and \(j\in B\).
Hence the assignment coordinates in \(\cup_{i\in A}\mathcal C_i\) are disjoint from the
assignment coordinates in \(\cup_{j\in B}\mathcal C_j\).  Since treatment assignments are
mutually independent, the two assignment subvectors are independent.  Because each
\(X_i\) is a function of \(\Zvec_{\mathcal C_i}\), the collections
\(\{X_i:i\in A\}\) and \(\{X_j:j\in B\}\) are independent.  Thus the graph defined by
\(c_{ij}=\mathds{1}\{\mathcal C_i\cap\mathcal C_j\neq\emptyset\}\) is a valid
dependence graph.

Now suppose the interference graph is undirected.  Then \(\mathcal C_i\cap\mathcal C_j
\neq\emptyset\) if and only if \(i\) and \(j\) are at graph distance at most two.  Hence
\(c_i=|B_2(i)|-1\).  Since every neighbor of \(i\) lies in \(B_2(i)\), \(c_i\ge d_i\),
and averaging gives \(c_{\mathrm{avg}}\ge \bar d\).  For the upper bound, the vertices
within graph distance at most two from \(i\), excluding \(i\), consist of the \(d_i\)
neighbors of \(i\) and the additional vertices reached through those neighbors.  Thus
\[
c_i
\le
d_i+\sum_{\ell\in N_i}(d_\ell-1)
=
\sum_{\ell\in N_i}d_\ell .
\]
Averaging over \(i\) and using the symmetry of the undirected graph gives
\[
c_{\mathrm{avg}}
\le
\frac1n\sum_{i=1}^n\sum_{\ell\in N_i}d_\ell
=
\frac1n\sum_{\ell=1}^n d_\ell^2 .
\]
The bound \(c_{\mathrm{avg}}\le n-1\) is immediate because each unit has at most \(n-1\)
other units in its dependence neighborhood.
\end{proof}

\subsection{A generic design-dependence bound}
\label{app:generic-design-dependence}

The following auxiliary result separates dependence generated by overlapping
assignment coordinates from dependence induced by the randomization design, using the
design-dependence coefficient in Definition~\ref{def:main-design-dependence-coefficient}.

\begin{proposition}[Interference and design dependence bound]
\label{prop:local-design-dependence}
Consider a generic design-based estimator of the form
\[
\hat\Psi=\frac1n\sum_{i=1}^n\psi_i(\Zvec),
\]
where \(\psi_i(\Zvec)\) is the weighted observed-outcome contribution of unit \(i\).
Suppose that \(\psi_i\) depends on the assignment only through
\(\Zvec_{\mathcal C_i}\), and suppose \(|\psi_i(\Zvec)|\le B_i\).  Define
\[
\ell_{ij}=\mathds{1}\{\mathcal C_i\cap\mathcal C_j\neq\emptyset\},\qquad i\neq j.
\]
Define
\[
B_{2,n}^2
=
\frac1n\sum_{i=1}^n B_i^2,
\]
\[
L_B
=
\frac1n\sum_{i=1}^n\sum_{j\neq i}\ell_{ij}B_iB_j,
\]
and
\[
R_B
=
\frac1n\sum_{i=1}^n\sum_{j\neq i}(1-\ell_{ij})\rho_{ij}^P B_iB_j .
\]
Then, under an arbitrary design \(P\),
\[
\operatorname{Var}_P\left(\frac1n\sum_{i=1}^n\psi_i(\Zvec)\right)
\le
\frac1n\{B_{2,n}^2+L_B+R_B\}.
\]
\end{proposition}

\begin{proof}
Write \(\hat\Psi=n^{-1}\sum_i\psi_i(\Zvec)\).  The variance expansion gives
\[
\operatorname{Var}_P(\hat\Psi)
=
\frac1{n^2}
\left[
\sum_i \operatorname{Var}_P\{\psi_i(\Zvec)\}
+
\sum_i\sum_{j\neq i}\operatorname{Cov}_P\{\psi_i(\Zvec),\psi_j(\Zvec)\}
\right].
\]
Since \(|\psi_i|\le B_i\), \(\operatorname{Var}_P(\psi_i)\le B_i^2\).  If
\(\ell_{ij}=1\), then
\[
\left|\operatorname{Cov}_P\{\psi_i(\Zvec),\psi_j(\Zvec)\}\right|
\le
\mathbb E_P|\psi_i(\Zvec)\psi_j(\Zvec)|
\le B_iB_j.
\]
If \(\ell_{ij}=0\), then \(\mathcal C_i\cap\mathcal C_j=\emptyset\).  When
\(B_iB_j>0\), set \(f(\Zvec_{\mathcal C_i})=\psi_i(\Zvec)/B_i\) and
\(g(\Zvec_{\mathcal C_j})=\psi_j(\Zvec)/B_j\).  These functions have sup-norm at most
one, and therefore
\[
\left|\operatorname{Cov}_P\{\psi_i(\Zvec),\psi_j(\Zvec)\}\right|
\le
\rho_{ij}^P B_iB_j.
\]
If \(B_iB_j=0\), the same bound is immediate.  Substituting these three bounds into
the variance expansion proves the result.
\end{proof}

\subsection{Residual design dependence under complete randomization}
\label{app:crd-residual-design-dependence}

The residual design-dependence term can be bounded explicitly for complete
randomization.  Suppose the design assigns exactly \(m\) of the \(n\) units to
treatment.  If \(A,B\subseteq\{1,\ldots,n\}\) are disjoint assignment-coordinate sets
with \(|A|=a\) and \(|B|=b\), then
\[
\rho_P(A,B)
\le
C\min\left\{1,\frac{ab}{n-a-b}\right\},
\]
for a universal constant \(C\), whenever \(a+b<n\).  Consequently, for local
HT/IPW estimators whose \(i\)th summand depends on
\(\mathcal C_i=\{i\}\cup N_i\),
\[
r_{\mathrm{avg}}
\le
\frac{C}{n}\sum_{i=1}^n\sum_{j\ne i}
(1-c_{ij})
\min\left\{1,\frac{|\mathcal C_i|\,|\mathcal C_j|}
{n-|\mathcal C_i|-|\mathcal C_j|}\right\}.
\]
In particular, if \(\max_i|\mathcal C_i|\le n/4\), then
\[
r_{\mathrm{avg}}
\le
\frac{C}{n^2}
\sum_{i=1}^n\sum_{j\ne i}
(1-c_{ij})|\mathcal C_i|\,|\mathcal C_j|
\le
C\left(\frac1n\sum_{i=1}^n|\mathcal C_i|\right)^2.
\]
Thus on a \(d\)-regular local interference graph,
\(r_{\mathrm{avg}}=O(d^2)\), matching the order of
\(c_{\mathrm{avg}}\asymp d^2\).  For bounded-degree graphs,
\(r_{\mathrm{avg}}=O(1)\), so the residual design-dependence contribution
\(H\sqrt{r_{\mathrm{avg}}/n}\) has the same \(n^{-1/2}\) scaling as the Bernoulli
dependence term, up to the relevant propensity scale \(H\).

\begin{proof}
It suffices to bound the dependence between two disjoint subvectors.  For a fixed
configuration \(x_A\) with \(s=\sum_{\ell\in A}x_\ell\), the conditional distribution
of \(\Zvec_B\) under complete randomization is sampling without replacement from a
population of size \(n-a\) containing \(m-s\) treated labels.  The marginal
distribution of \(\Zvec_B\) is sampling without replacement from a population of size
\(n\) containing \(m\) treated labels.

Couple the two draws sequentially over the \(b\) coordinates in \(B\).  After \(k<b\)
coordinates in \(B\) have been exposed and \(r\) treated labels have appeared among
them, the two conditional probabilities that the next coordinate is treated are
\[
\frac{m-s-r}{n-a-k}
\qquad\text{and}\qquad
\frac{m-r}{n-k}.
\]
Their absolute difference is at most \(C a/(n-a-b)\), uniformly over admissible
\(s,r,k\).  The union bound over the \(b\) sequential draws gives a total-variation
bound of order \(ab/(n-a-b)\), truncated at one.  For any \(\|g\|_\infty\le1\),
the difference between the conditional and marginal expectations of \(g(\Zvec_B)\) is
at most twice this total variation distance.  Taking the supremum also over
\(\|f\|_\infty\le1\) gives the displayed bound for \(\rho_P(A,B)\).

Substituting \(A=\mathcal C_i\), \(B=\mathcal C_j\) for pairs with
\(\mathcal C_i\cap\mathcal C_j=\emptyset\) gives the first bound on
\(r_{\mathrm{avg}}\).  The simplified bound follows from
\(|\mathcal C_i|+|\mathcal C_j|\le n/2\), and the final regular and bounded-degree
statements follow by setting \(|\mathcal C_i|=d+1\) or
\(\max_i|\mathcal C_i|=O(1)\).
\end{proof}

\subsection{Proof of Proposition~\ref{prop:generic-ace-ht}}
\paragraph{Restatement.}
Assume \(\pi_i(d_i^+)>0\) and \(\pi_i(d_i^-)>0\) for all \(i\), and define
\[
\hat\tau_{\mathrm{ACE}}^{HT}(d^+,d^-)
=
\frac1n\sum_{i=1}^n
Y_i^{obs}
\left\{
\frac{\mathds{1}\{\Dcond_i(\Zvec)=d_i^+\}}{\pi_i(d_i^+)}
-
\frac{\mathds{1}\{\Dcond_i(\Zvec)=d_i^-\}}{\pi_i(d_i^-)}
\right\}.
\]
Then \(\hat\tau_{\mathrm{ACE}}^{HT}(d^+,d^-)\) is design-unbiased for
\(\tau_{\mathrm{ACE}}(d^+,d^-)\).  If \(|Y_i(d)|\le M\) and the \(i\)th summand
depends on the assignment only through \(\Zvec_{\mathcal C_i}\), then the variance is
bounded as stated.

\begin{proof}
Because \(Y_i^{obs}=Y_i\{\Dcond_i(\Zvec)\}\), the expectation of the first weighted term is
\[
\mathbb E\left[
Y_i^{obs}
\frac{\mathds{1}\{\Dcond_i(\Zvec)=d_i^+\}}{\pi_i(d_i^+)}
\right]
=
\frac{1}{\pi_i(d_i^+)}
\mathbb E\left[
Y_i\{\Dcond_i(\Zvec)\}\mathds{1}\{\Dcond_i(\Zvec)=d_i^+\}
\right]
=
Y_i(d_i^+).
\]
The same calculation gives
\[
\mathbb E\left[
Y_i^{obs}
\frac{\mathds{1}\{\Dcond_i(\Zvec)=d_i^-\}}{\pi_i(d_i^-)}
\right]
=
Y_i(d_i^-).
\]
Therefore
\[
\mathbb E\{\hat\tau_{\mathrm{ACE}}^{HT}(d^+,d^-)\}
=
\frac1n\sum_{i=1}^n\{Y_i(d_i^+)-Y_i(d_i^-)\}
=
\tau_{\mathrm{ACE}}(d^+,d^-).
\]

For the variance bound, write
\[
X_i^{ACE}
=
Y_i\{\Dcond_i(\Zvec)\}
\left\{
\frac{\mathds{1}\{\Dcond_i(\Zvec)=d_i^+\}}{\pi_i(d_i^+)}
-
\frac{\mathds{1}\{\Dcond_i(\Zvec)=d_i^-\}}{\pi_i(d_i^-)}
\right\}.
\]
Then \(\hat\tau_{\mathrm{ACE}}^{HT}(d^+,d^-)=n^{-1}\sum_iX_i^{ACE}\).  Since
\(|Y_i(d)|\le M\),
\[
|X_i^{ACE}|
\le
M\left\{
\frac{1}{\pi_i(d_i^+)}+\frac{1}{\pi_i(d_i^-)}
\right\}
=
MH_i.
\]
Because the two target events in the ACE contrast are disjoint,
\[
\mathbb E\{(X_i^{ACE})^2\}
\le
M^2\left\{
\frac{1}{\pi_i(d_i^+)}
+
\frac{1}{\pi_i(d_i^-)}
\right\}
=
M^2H_i.
\]
Thus \(\operatorname{Var}(X_i^{ACE})\le M^2H_i\).  If \(c_{ij}=1\), then
\[
|\operatorname{Cov}(X_i^{ACE},X_j^{ACE})|
\le
\mathbb E|X_i^{ACE}X_j^{ACE}|
\le
M^2H_iH_j.
\]
If \(c_{ij}=0\), then the closed interference neighborhoods are disjoint.  When \(H_iH_j>0\),
the functions \(X_i^{ACE}/(MH_i)\) and \(X_j^{ACE}/(MH_j)\) have sup-norm at most one,
so Definition~\ref{def:main-design-dependence-coefficient} gives
\[
|\operatorname{Cov}(X_i^{ACE},X_j^{ACE})|
\le
M^2\rho_{ij}H_iH_j.
\]
The same bound is immediate when \(H_iH_j=0\).
Combining the diagonal and off-diagonal terms gives
\begin{align*}
\operatorname{Var}\{\hat\tau_{\mathrm{ACE}}^{HT}(d^+,d^-)\}
&=
\frac1{n^2}
\left\{
\sum_{i=1}^n\operatorname{Var}(X_i^{ACE})
+
\sum_{i=1}^n\sum_{j\ne i}\operatorname{Cov}(X_i^{ACE},X_j^{ACE})
\right\} \\
&\le
\frac{M^2}{n^2}
\left\{
\sum_{i=1}^nH_i+
\sum_{i=1}^n\sum_{j\ne i}c_{ij}H_iH_j
+
\sum_{i=1}^n\sum_{j\ne i}(1-c_{ij})\rho_{ij}H_iH_j
\right\}.
\end{align*}
If \(H_i\le H\) for all \(i\), this becomes
\begin{align*}
\operatorname{Var}\{\hat\tau_{\mathrm{ACE}}^{HT}(d^+,d^-)\}
&\le
\frac{M^2}{n^2}
\left\{
nH
+
H^2\sum_{i=1}^n\sum_{j\ne i}c_{ij}
+
H^2\sum_{i=1}^n\sum_{j\ne i}(1-c_{ij})\rho_{ij}
\right\}
\\
&=
M^2\left\{
\frac{H}{n}
+
H^2\frac{c_{\mathrm{avg}}+r_{\mathrm{avg}}}{n}
\right\}.
\end{align*}
\end{proof}

\subsection{Proof of Proposition~\ref{prop:ace-inverse-propensity}}
\paragraph{Restatement.}
Fix the design \(P_n\) and the induced treatment--exposure combinations \(\Dcond_i(\Zvec)\).
Suppose that
\(\hat\tau_n\) is design-unbiased for \(\tau_{\mathrm{ACE}}(d^+,d^-)\) for every bounded
potential-outcome table.  Then
\[
\sup_{|Y|\le M}
\operatorname{Var}_{P_n}(\hat\tau_n)
\ge
\frac{M^2}{n^2}
\max_i
\left\{
\frac{1}{\pi_i(d_i^+)}-1,\,
\frac{1}{\pi_i(d_i^-)}-1
\right\}.
\]

\begin{proof}
Fix a unit \(i\) and a target state
\(d_i^+\).  Let
\[
A_i=\{\Dcond_i(\Zvec)=d_i^+\}, \qquad \pi_i(d_i^+)=P_n(A_i).
\]
Construct two potential-outcome tables \(Y^+\) and \(Y^-\) that agree everywhere
except at the coordinate \(Y_i(d_i^+)\), where
\[
Y_i^+(d_i^+)=M,\qquad Y_i^-(d_i^+)=-M.
\]
All other potential outcomes are the same in the two tables.  If this coordinate enters
the estimand with coefficient \(c_i\), then
\[
\Delta=\tau(Y^+)-\tau(Y^-)=2M c_i .
\]
Let
\[
W(Z)=\hat\tau_n(Z,Y^{+,\obs})-\hat\tau_n(Z,Y^{-,\obs}).
\]
By design unbiasedness, \(\mathbb E[W]=\Delta\).  On the event \(A_i^c\), the two
tables produce identical observed data, and hence \(W=0\).  Therefore
\[
\Delta=\mathbb E\{W\mathds{1}(A_i)\}.
\]
By Cauchy--Schwarz,
\[
\Delta^2
\le
\mathbb E(W^2)P_n(A_i)
=
\mathbb E(W^2)\pi_i(d_i^+),
\]
so \(\mathbb E(W^2)\ge \Delta^2/\pi_i(d_i^+)\).  Since \(\mathbb E(W)=\Delta\),
\[
\operatorname{Var}(W)
=
\mathbb E(W^2)-\Delta^2
\ge
\Delta^2\left\{\frac{1}{\pi_i(d_i^+)}-1\right\}.
\]
On the other hand,
\[
W-\Delta
=
\{\hat\tau_n(Y^+)-\tau(Y^+)\}
-
\{\hat\tau_n(Y^-)-\tau(Y^-)\}.
\]
Thus
\[
\operatorname{Var}(W)
\le
4\max\left\{
\operatorname{Var}_{P_n}(\hat\tau_n;Y^+),
\operatorname{Var}_{P_n}(\hat\tau_n;Y^-)
\right\}.
\]
Combining the last two displays gives
\[
\sup_{|Y|\le M}\operatorname{Var}_{P_n}(\hat\tau_n;Y)
\ge
\frac{\Delta^2}{4}
\left\{\frac{1}{\pi_i(d_i^+)}-1\right\}
=
c_i^2M^2
\left\{\frac{1}{\pi_i(d_i^+)}-1\right\}.
\]
For an average ACE, the coefficient of each arm is \(c_i=\pm 1/n\).  Applying the same
argument to either \(d_i^+\) or \(d_i^-\), and then taking the maximum over units and arms,
gives the claimed bound.
\end{proof}

\subsection{Proof of Proposition~\ref{prop:generic-ece-score}}
\paragraph{Restatement.}
Suppose \(\lambda_i^+(d)=\lambda_i^-(d)=0\) whenever \(\pi_i^P(d)=0\), and define
\(h_i^+(\zvec)\) and \(h_i^-(\zvec)\) as in the proposition.
Let
\[
V_i
=
\sum_{d\in\tespace_i:\,\pi_i^P(d)>0}
\frac{\{\lambda_i^+(d)-\lambda_i^-(d)\}^2}{\pi_i^P(d)}.
\]
Let
\[
\hat\tau_{\mathrm{ECE}}^{HT}
=
\frac1n\sum_{i=1}^n
\{h_i^+(\Zvec)-h_i^-(\Zvec)\}Y_i^{obs}.
\]
Then \(\hat\tau_{\mathrm{ECE}}^{HT}\) is design-unbiased for
\[
\tau_{\mathrm{ECE}}
=
\frac1n\sum_{i=1}^n
\sum_{d\in\tespace_i}Y_i(d)\{\lambda_i^+(d)-\lambda_i^-(d)\}.
\]
If \(|Y_i(\zvec)|\le M\) and
\[
X_i=\{h_i^+(\Zvec)-h_i^-(\Zvec)\}Y_i\{\Dcond_i(\Zvec)\}
\]
depends on the assignment only through \(\Zvec_{\mathcal C_i}\), then
\[
\operatorname{Var}(\hat\tau_{\mathrm{ECE}}^{HT})
\le
\frac{M^2}{n^2}
\left\{
\sum_{i=1}^n V_i
+
\sum_{i=1}^n\sum_{j\ne i}c_{ij}H_iH_j
+
\sum_{i=1}^n\sum_{j\ne i}(1-c_{ij})\rho_{ij}H_iH_j
\right\}.
\]

\begin{proof}
Because \(Y_i^{obs}=Y_i(\Zvec)\), the expectation of the \(i\)th summand is
\[
\mathbb E[\{h_i^+(\Zvec)-h_i^-(\Zvec)\}Y_i^{obs}]
=
\sum_{\zvec\in\allocspace}p(\zvec)
\{h_i^+(\zvec)-h_i^-(\zvec)\}Y_i\{\Dcond_i(\zvec)\}.
\]
Grouping allocations by the treatment--exposure combination they induce gives the
plus-arm contribution
\[
\begin{aligned}
&\sum_{\zvec\in\allocspace}p(\zvec)h_i^+(\zvec)Y_i\{\Dcond_i(\zvec)\} \\
&\quad =
\sum_{d\in\tespace_i:\,\pi_i^P(d)>0}
Y_i(d)\lambda_i^+(d)
\frac{\sum_{\zvec\in\allocspace:\,\Dcond_i(\zvec)=d}p(\zvec)}
{\pi_i^P(d)} \\
&\quad =
\sum_{d\in\tespace_i:\,\pi_i^P(d)>0}Y_i(d)\lambda_i^+(d)
=
\sum_{d\in\tespace_i}Y_i(d)\lambda_i^+(d),
\end{aligned}
\]
where the last equality uses the positivity condition.  The same calculation for the
minus arm gives
\[
\mathbb E\{h_i^-(\Zvec)Y_i^{obs}\}
=
\sum_{d\in\tespace_i}Y_i(d)\lambda_i^-(d).
\]
Summing over \(i\) and dividing by \(n\) proves design unbiasedness:
\[
\mathbb E(\hat\tau_{\mathrm{ECE}}^{HT})
=
\frac1n\sum_{i=1}^n
\sum_{d\in\tespace_i}Y_i(d)\{\lambda_i^+(d)-\lambda_i^-(d)\}
=
\tau_{\mathrm{ECE}}.
\]

For the variance bound, write
\[
X_i=\{h_i^+(\Zvec)-h_i^-(\Zvec)\}Y_i\{\Dcond_i(\Zvec)\}.
\]
Then
\[
\hat\tau_{\mathrm{ECE}}^{HT}=\frac1n\sum_{i=1}^n X_i.
\]
By boundedness, \(|X_i|\le MH_i\), so
\[
\operatorname{Var}(X_i)\le \mathbb E(X_i^2)
\le
M^2
\sum_{d\in\tespace_i:\,\pi_i^P(d)>0}
\frac{\{\lambda_i^+(d)-\lambda_i^-(d)\}^2}{\pi_i^P(d)}
=
M^2V_i.
\]
If \(c_{ij}=1\), Cauchy--Schwarz gives
\[
|\operatorname{Cov}(X_i,X_j)|
\le
\mathbb E|X_iX_j|
\le
M^2H_iH_j.
\]
If \(c_{ij}=0\), then the closed interference neighborhoods are disjoint.  When \(H_iH_j>0\),
the functions \(X_i/(MH_i)\) and \(X_j/(MH_j)\) have sup-norm at most one, so
Definition~\ref{def:main-design-dependence-coefficient} gives
\[
|\operatorname{Cov}(X_i,X_j)|
\le
M^2\rho_{ij}H_iH_j.
\]
The same bound is immediate when \(H_iH_j=0\).
Therefore
\[
\begin{aligned}
\operatorname{Var}(\hat\tau_{\mathrm{ECE}}^{HT})
&=
\frac1{n^2}
\left\{
\sum_{i=1}^n\operatorname{Var}(X_i)
+
\sum_{i=1}^n\sum_{j\ne i}\operatorname{Cov}(X_i,X_j)
\right\}\\
&\le
\frac{M^2}{n^2}
\left\{
\sum_{i=1}^n V_i
+
\sum_{i=1}^n\sum_{j\ne i}c_{ij}H_iH_j
+
\sum_{i=1}^n\sum_{j\ne i}(1-c_{ij})\rho_{ij}H_iH_j
\right\}.
\end{aligned}
\]
If \(V_i\le V\) and \(H_i\le H\) for all \(i\), this becomes
\begin{align*}
\operatorname{Var}(\hat\tau_{\mathrm{ECE}}^{HT})
&\le
\frac{M^2}{n^2}
\left\{
nV
+
H^2\sum_{i=1}^n\sum_{j\ne i}c_{ij}
+
H^2\sum_{i=1}^n\sum_{j\ne i}(1-c_{ij})\rho_{ij}
\right\}
\\
&=
M^2\left\{
\frac{V}{n}
+
H^2\frac{c_{\mathrm{avg}}+r_{\mathrm{avg}}}{n}
\right\}.
\end{align*}
\end{proof}

\subsection{Weighted dependence bound for graph-aware ECEs}
\label{app:weighted-ece-dependence}

\begin{definition}[Weighted summand and dependence complexity]
\label{def:weighted-ece-dependence}
Write \(h_i(\Zvec)=h_i^+(\Zvec)-h_i^-(\Zvec)\) and
\(\psi_i(\Zvec)=h_i(\Zvec)Y_i\{\Dcond_i(\Zvec)\}\).  For the canonical ECE estimator
\(\hat\tau_h=n^{-1}\sum_i\psi_i(\Zvec)\), define
\[
\omega_{ij}(Y)
=
\operatorname{Cov}\{\psi_i(\Zvec),\psi_j(\Zvec)\},
\qquad i\neq j,
\]
and
\[
\Gamma_h(M)
=
\sup_{|Y|\le M}
\left[
\sum_{i=1}^n\operatorname{Var}\{\psi_i(\Zvec)\}
+
\sum_{i=1}^n\sum_{j\neq i}\omega_{ij}(Y)
\right].
\]
The supremum is over all bounded potential-outcome tables satisfying
\(|Y_i(\zvec)|\le M\).
\end{definition}

\begin{proposition}[Weighted dependence bound for graph-aware ECEs]
\label{prop:weighted-ece-dependence}
Assume independent Bernoulli assignment.  Suppose the potential outcomes satisfy
neighborhood interference with known neighborhoods \(N_i\), and suppose
\(h_i(\Zvec)\) depends only on \(\Zvec_{\mathcal C_i}\).  Then a valid dependence graph
for \(\psi_i(\Zvec)=h_i(\Zvec)Y_i(\Zvec)\) is given by
\[
c_{ij}=\mathds{1}\{\mathcal C_i\cap\mathcal C_j\neq\emptyset\},
\qquad i\neq j.
\]
Moreover,
\[
\sup_{|Y|\le M}\operatorname{Var}(\hat\tau_h)
=
\frac{\Gamma_h(M)}{n^2}.
\]
In particular, if \(|h_i(\Zvec)|\le H_i\), then
\[
\operatorname{Var}(\hat\tau_h)
\le
\frac{M^2}{n^2}
\left\{
\sum_{i=1}^n H_i^2
+
\sum_{i=1}^n\sum_{j\neq i}c_{ij}H_iH_j
\right\}.
\]
\end{proposition}

This result separates two roles of the dependence graph.  The unweighted graph
gives a conservative upper bound by replacing every nonzero covariance by its maximal
possible size.  The sharper object is the weighted dependence complexity
\(\Gamma_h(M)\), which keeps the covariance weights induced by the design contrast and
the potential-outcome table.

\subsection{Proof of Proposition~\ref{prop:weighted-ece-dependence}}
\paragraph{Restatement.}
Assume independent Bernoulli assignment.  Suppose the potential outcomes satisfy
neighborhood interference with known neighborhoods \(N_i\), and suppose
\(h_i(\Zvec)\) depends only on \(\Zvec_{\mathcal C_i}\), where
\(\mathcal C_i=\{i\}\cup N_i\).  Then a valid dependence graph for
\(\psi_i(\Zvec)=h_i(\Zvec)Y_i(\Zvec)\) is given by
\[
c_{ij}=\mathds{1}\{\mathcal C_i\cap\mathcal C_j\neq\emptyset\},
\qquad i\neq j.
\]
Moreover,
\[
\sup_{|Y|\le M}\operatorname{Var}(\hat\tau_h)
=
\frac{\Gamma_h(M)}{n^2}.
\]
If \(|h_i(\Zvec)|\le H_i\), then
\[
\operatorname{Var}(\hat\tau_h)
\le
\frac{M^2}{n^2}
\left\{
\sum_{i=1}^n H_i^2
+
\sum_{i=1}^n\sum_{j\neq i}c_{ij}H_iH_j
\right\}.
\]

\begin{proof}
By neighborhood interference, \(Y_i(\Zvec)\) is a function only of
\(\Zvec_{\mathcal C_i}\).  By assumption, \(h_i(\Zvec)\) is also a function only of
\(\Zvec_{\mathcal C_i}\).  Hence
\[
\psi_i(\Zvec)=h_i(\Zvec)Y_i(\Zvec)
\]
is a function only of \(\Zvec_{\mathcal C_i}\).  If
\(\mathcal C_i\cap\mathcal C_j=\emptyset\), then \(\Zvec_{\mathcal C_i}\) and
\(\Zvec_{\mathcal C_j}\) are independent under independent Bernoulli assignment.
Therefore \(\psi_i(\Zvec)\) and \(\psi_j(\Zvec)\) are independent.  The graph with
edge indicators \(c_{ij}=\mathds{1}\{\mathcal C_i\cap\mathcal C_j\neq\emptyset\}\)
is therefore a valid dependence graph.

For a fixed potential-outcome table \(Y\),
\[
\hat\tau_h=\frac1n\sum_{i=1}^n\psi_i(\Zvec),
\]
so
\[
\operatorname{Var}(\hat\tau_h)
=
\frac1{n^2}
\left[
\sum_{i=1}^n\operatorname{Var}\{\psi_i(\Zvec)\}
+
\sum_{i=1}^n\sum_{j\neq i}
\operatorname{Cov}\{\psi_i(\Zvec),\psi_j(\Zvec)\}
\right].
\]
Taking the supremum over all bounded potential-outcome tables gives
\[
\sup_{|Y|\le M}\operatorname{Var}(\hat\tau_h)
=
\frac{\Gamma_h(M)}{n^2}
\]
by Definition~\ref{def:weighted-ece-dependence}.

Finally, if \(|Y_i(\Zvec)|\le M\) and \(|h_i(\Zvec)|\le H_i\), then
\(|\psi_i(\Zvec)|\le MH_i\).  Thus
\[
\operatorname{Var}\{\psi_i(\Zvec)\}\le M^2H_i^2.
\]
If \(c_{ij}=0\), then \(\psi_i\) and \(\psi_j\) are independent and the covariance is
zero.  If \(c_{ij}=1\), then Cauchy--Schwarz gives
\[
|\operatorname{Cov}\{\psi_i(\Zvec),\psi_j(\Zvec)\}|
\le
\{\operatorname{Var}(\psi_i)\operatorname{Var}(\psi_j)\}^{1/2}
\le
M^2H_iH_j.
\]
Substituting these bounds into the variance expansion gives the stated inequality.
\end{proof}

\subsection{Proof of Proposition~\ref{prop:ede-variance-bound}}
\paragraph{Restatement.}
Assume the true potential outcomes satisfy neighborhood interference with neighborhoods
\(N_i\).  Assume Bernoulli assignment with treatment probabilities bounded away from zero
and one, let \(\Qpol_p\) denote this Bernoulli allocation policy, and assume
\(\sup_{\zvec}|Y_i(\zvec)|\le M\).  Then \(\hat\tau_{\mathrm{EDE}}^{HT}\) is
design-unbiased for \(\tau_{\mathrm{EDE}}(\Qpol_p)\), and
\[
\operatorname{Var}_{P_n}(\hat\tau_{\mathrm{EDE}}^{HT})
\le
C_p M^2 \frac{1+c_{\mathrm{avg}}}{n}.
\]
In particular, \(\hat\tau_{\mathrm{EDE}}^{HT}\) is design-MSE consistent if
\(c_{\mathrm{avg}}=o(n)\).

\begin{proof}
Write the estimator as \(n^{-1}\sum_i X_i\), where
\[
X_i=Y_i^{obs}\left\{Z_i/p_i-(1-Z_i)/(1-p_i)\right\}.
\]
First condition on \(Z_{-i}\).  Since the assignment is Bernoulli, \(Z_i\) is
independent of \(Z_{-i}\), and hence
\[
\mathbb E(X_i\mid Z_{-i})
=
Y_i(1,Z_{-i})-Y_i(0,Z_{-i}).
\]
Taking expectations over \(Z_{-i}\) gives
\[
\mathbb E(\hat\tau_{\mathrm{EDE}}^{HT})
=
\frac{1}{n}\sum_{i=1}^n
\left\{
\mathbb E[Y_i(1,Z_{-i})]
-
\mathbb E[Y_i(0,Z_{-i})]
\right\}
=
\tau_{\mathrm{EDE}}(\Qpol_p).
\]

The boundedness of the potential outcomes and the positivity of the treatment
probabilities imply \(|X_i|\le C_p M\).  If \(\mathcal C_i\cap\mathcal C_j=\emptyset\), then \(X_i\) and
\(X_j\) depend on disjoint sets of independent treatment assignments and are therefore
independent.  Thus only pairs with \(c_{ij}=1\) contribute to the covariance sum.  This
gives
\[
\operatorname{Var}(\hat\tau_{\mathrm{EDE}}^{HT})
\le
\frac{C_pM^2}{n^2}
\left(n+\sum_i\sum_{j\neq i}c_{ij}\right)
=
C_pM^2\frac{1+c_{\mathrm{avg}}}{n}.
\]
\end{proof}

\subsection{Proof of Corollary~\ref{cor:eie-variance-bound}}
\paragraph{Restatement.}
Assume the true potential outcomes satisfy neighborhood interference with known
neighborhoods \(N_i\).  Assume Bernoulli assignment with treatment probabilities
bounded away from zero and one, and assume \(\sup_{\zvec}|Y_i(\zvec)|\le M\).  Then
\(\hat\tau_{\mathrm{EIE}}^{G}\) is design-unbiased for \(\tau_{\mathrm{EIE}}(\pi)\),
and
\[
\operatorname{Var}(\hat\tau_{\mathrm{EIE}}^{G})
\le
\frac{C_pM^2}{n^2}
\left\{
\sum_i d_i^2
+
\sum_i\sum_{j\neq i}c_{ij}d_id_j
\right\}.
\]

\begin{proof}
The graph-aware EIE estimator is
\[
\hat\tau_{\mathrm{EIE}}^{G}
=
\frac1n\sum_{i=1}^nY_i^{obs}
\sum_{j\in N_i}
\left\{
\frac{Z_j}{p_j}
-
\frac{1-Z_j}{1-p_j}
\right\}.
\]
Its design unbiasedness follows by applying the ECE importance-weighting argument to
each local neighbor contrast and summing over \(j\in N_i\); equivalently, it follows
from the policy-derivative calculation in Proposition~\ref{prop:HuLiWager}.  To obtain
the variance bound, write
\[
h_i^{\mathrm{EIE},G}(\Zvec)
=
\sum_{j\in N_i}
\left\{
\frac{Z_j}{p_j}
-
\frac{1-Z_j}{1-p_j}
\right\}.
\]
If the treatment probabilities are bounded away from zero and one, then there is a
constant \(C_p\) such that
\[
|h_i^{\mathrm{EIE},G}(\Zvec)|\le C_p d_i.
\]
Moreover \(h_i^{\mathrm{EIE},G}(\Zvec)\) depends only on \(\Zvec_{N_i}\), and hence on
\(\Zvec_{\mathcal C_i}\).  Proposition~\ref{prop:weighted-ece-dependence}, with
\(H_i=C_pd_i\), gives
\[
\operatorname{Var}(\hat\tau_{\mathrm{EIE}}^{G})
\le
\frac{C_pM^2}{n^2}
\left\{
\sum_i d_i^2
+
\sum_i\sum_{j\neq i}c_{ij}d_id_j
\right\},
\]
after absorbing constants depending only on the treatment-probability bounds into
\(C_p\).
If \(d_i\le d_{\max}\), then
\[
\sum_i d_i^2+\sum_i\sum_{j\neq i}c_{ij}d_id_j
\le
d_{\max}^2\left\{n+\sum_i\sum_{j\neq i}c_{ij}\right\}
=
nd_{\max}^2(1+c_{\mathrm{avg}}),
\]
which gives the final display.
\end{proof}

\subsection{Heterogeneous ACE/ECE comparison under classical designs}
\label{app:ace-ece-separation}

The DTE/EDE comparison of Section~\ref{sec:ece-ht-estimators} extends to arbitrary
heterogeneous degree sequences, where it is governed by two graph functionals.  For a
fixed-exposure ACE, the relevant scale is the inverse propensity-score moment
\[
\mathcal A_n(d^+)
=
\frac{1}{n^2}\sum_{i=1}^n \pi_i(d_i^+)^{-1};
\]
for the direct-effect entry \((1,0_{N_i})\) under Bernoulli assignment,
\(\pi_i(1,0_{N_i})=p(1-p)^{d_i}\), so
\(\mathcal A_n(1,0)\asymp n^{-2}\sum_i \exp(\gamma d_i)\) is an exponential moment of
the degree sequence.  For the EDE under the same design, the relevant scale is the
average dependence degree, \(\mathcal B_n=c_{\mathrm{avg}}/n\).

The comparison is design-specific.  It applies to classical designs such as Bernoulli
assignment, and complete randomization with a fixed treatment fraction has the same
exponential propensity-score behavior for local fixed-exposure combinations, up to
finite-population corrections.  It does not rule out designs that are constructed to
increase the propensity scores of the ACE target combinations, such as the
independent-set designs.

\begin{table}[t]
\centering
\caption{Generic comparison of the ACE inverse-propensity scale and the EDE dependence
scale under Bernoulli assignment.  The ACE column corresponds to a direct-effect target
combination such as \((1,0_{N_i})\).  The EDE column gives the order of \(c_{\mathrm{avg}}\).}
\label{tab:ace-ede-graph-comparison}
\begin{tabular}{p{0.36\textwidth}p{0.28\textwidth}p{0.24\textwidth}}
\hline
Graph sequence & ACE scale \(\mathcal A_n\) & EDE dependence \(c_{\mathrm{avg}}\) \\
\hline
Bounded degree & \(O(n^{-1})\) & \(O(1)\) \\
Second-neighborhood regular, degree \(d\) & \(e^{\gamma d}/n\) & \(d^2\) \\
Sparse Erd\H{o}s--Renyi, mean degree \(\lambda\) & approximately \(e^{\gamma\lambda}/n\) & approximately \(\lambda+\lambda^2\) \\
Dense block model, block size \(s\) & approximately \(e^{\gamma s}/n\) & approximately \(s\) \\
Star graph & dominated by hub term \(e^{\gamma n}/n^2\) & \(n\) \\
Complete graph & \(e^{\gamma n}/n\) & \(n\) \\
\hline
\end{tabular}
\end{table}

The table is meant to show why a single average degree condition is not adequate for
heterogeneous graphs.  By Jensen's inequality,
\[
\frac{1}{n}\sum_{i=1}^n \exp(\gamma d_i)\ge \exp(\gamma\bar d),
\]
so large average degree is enough to make the ACE inverse-propensity scale large.
However, small average degree is not sufficient to make ACE estimation easy, because
hubs can dominate the exponential moment.  For EDE estimation, average degree is also
not the right quantity: a star graph has \(\bar d=O(1)\) but
\(c_{\mathrm{avg}}\asymp n\).

\begin{corollary}[Heterogeneous ACE/ECE separation under Bernoulli assignment]
\label{cor:ace-ece-exponential-separation}
Consider a sequence of undirected interference graphs with degrees \(d_i=|N_i|\), and
suppose treatment is assigned independently with a common probability \(p\) bounded
away from zero and one.  For the direct-effect ACE entry \((1,0_{N_i})\), there are
constants \(0<\gamma_1\le \gamma_2<\infty\), depending only on \(p\), such that
\[
\frac{1}{n^2}\sum_{i=1}^n \exp(\gamma_1 d_i)
\lesssim
\frac{1}{n^2}\sum_{i=1}^n \pi_i(1,0_{N_i})^{-1}
\lesssim
\frac{1}{n^2}\sum_{i=1}^n \exp(\gamma_2 d_i).
\]
The canonical ACE Horvitz--Thompson variance is bounded by
\[
\operatorname{Var}\{\hat\tau_{\mathrm{ACE}}^{HT}\}
\le
 \frac{C_pM^2}{n^2}
 \left\{
 \sum_{i=1}^n \exp(\gamma_2 d_i)
 +
 \sum_{i=1}^n\sum_{j\ne i}c_{ij}
 \exp\{\gamma_2(d_i+d_j)\}
 \right\},
\]
where \(c_{ij}\) is the interference-overlap graph for the ACE summands.  By
contrast, the canonical EDE Horvitz--Thompson estimator satisfies
\[
\operatorname{Var}\{\hat\tau_{\mathrm{EDE}}^{HT}\}
\le
C_pM^2\frac{1+c_{\mathrm{avg}}}{n}.
\]
Thus, under the same Bernoulli design and for arbitrary heterogeneous graph sequences,
the ACE benchmark involves exponential moments of the degree sequence and their
dependence-weighted analogues, whereas the EDE benchmark involves only the average
degree of the dependence graph.
\end{corollary}

The regular-graph comparison of Section~\ref{sec:ece-ht-estimators} is recovered as a
special case.  More generally, Table~\ref{tab:ace-ede-graph-comparison} should be read
through the two graph functionals appearing in the corollary: exponential degree
moments for ACE target combinations and average dependence degree for EDE
importance-weighted estimators.

\subsection{Proof of Corollary~\ref{cor:ace-ece-exponential-separation}}
\paragraph{Restatement.}
Under Bernoulli assignment with common treatment probability bounded away from zero and
one, the direct-effect ACE benchmark involves exponential moments of the degree
sequence and their dependence-weighted analogues, whereas the EDE benchmark involves
only the average dependence degree.

\begin{proof}
For the direct-effect entry \((1,0_{N_i})\) under Bernoulli assignment,
\[
\pi_i(1,0_{N_i})=p(1-p)^{d_i}.
\]
Since \(p\) and \(1-p\) are bounded away from zero and one, there are constants
\(0<\gamma_1\le\gamma_2<\infty\), depending only on \(p\), such that
\[
\exp(\gamma_1d_i)
\le
\pi_i(1,0_{N_i})^{-1}
\le
\exp(\gamma_2d_i)
\]
after absorbing fixed multiplicative constants into the comparison constants.  Summing
over \(i\) gives the inverse-propensity display.

For the ACE HT upper bound, Proposition~\ref{prop:generic-ace-ht} gives
\[
\operatorname{Var}\{\hat\tau_{\mathrm{ACE}}^{HT}\}
\le
 \frac{M^2}{n^2}
 \left\{
 \sum_i H_i+\sum_i\sum_{j\ne i}c_{ij}H_iH_j
 \right\}.
\]
Here the residual design-dependence term is zero because treatment is assigned
independently.
For the direct-effect contrast with no treated neighbors, both arm propensities are
bounded below by constants times \(\exp(-\gamma_2d_i)\), so
\(H_i\le C_p\exp(\gamma_2d_i)\).  Substitution gives the stated
dependence-weighted exponential bound.  The EDE bound is exactly
Proposition~\ref{prop:ede-variance-bound}.
\end{proof}

\subsection{Auxiliary calculation: DTE and TTE table rates}
\label{app:ace-table-calculations}
\begin{proposition}[DTE and TTE propensities under graph-unaware designs]
\label{prop:dte-tte-graph-unaware-propensities}
Let \(A_i\) denote the assignment coordinates that define the target
treatment--exposure combination for unit \(i\).  For arbitrary interference,
\(A_i=\{1,\ldots,n\}\), while under neighborhood interference
\(A_i=\mathcal C_i=\{i\}\cup N_i\).  Write \(a_i=|A_i|\).

\begin{enumerate}[label=(\roman*)]
\item Suppose the design assigns treatments independently, with
\(\epsilon\le P(Z_k=1)\le 1-\epsilon\) for all \(k\).  Then every fixed assignment
pattern on \(A_i\) has probability \(\exp\{-\Theta(a_i)\}\), with constants depending
only on \(\epsilon\).  Consequently, the DTE no-exposure conditions and the TTE
full-exposure/no-exposure conditions have inverse propensities
\(\exp\{\Theta(n)\}\) under arbitrary interference and \(\exp\{\Theta(d_i)\}\) under
neighborhood interference.

\item Suppose the design is exchangeable, so that conditional on \(T_n=\sum_k Z_k=m\)
all assignments with \(m\) treated units are equally likely.  Assume that, for some
\(\epsilon>0\), \(T_n/n\in[\epsilon,1-\epsilon]\) almost surely and that
\(a_i\le \epsilon n/2\).  Then every fixed assignment pattern on \(A_i\) has probability
\(\exp\{-\Theta(a_i)\}\), with constants depending only on \(\epsilon\).  Thus the same
\(\exp\{\Theta(d_i)\}\) inverse-propensity order holds for neighborhood DTE and TTE
conditions, including the no-exposure and full-exposure count/fraction exposure
representations.  Fixed-size complete randomization is a special case when the treatment
fraction is bounded away from zero and one.
\end{enumerate}

In both cases, the inverse-propensity lower bound in
Proposition~\ref{prop:ace-inverse-propensity} applies to any design-unbiased
non-parametric estimator of the corresponding DTE or TTE.  The exchangeable case covers
local treatment--exposure combinations.  It does not cover arbitrary-interference TTE/DTE
targets under fixed nontrivial treatment fractions, because such designs put zero
probability on the global all-treated or all-control assignments.
\end{proposition}

\begin{proof}[Proof of Proposition~\ref{prop:dte-tte-graph-unaware-propensities}]
For the independent-design claim, fix a set \(A\) with \(|A|=a\) and a pattern
\(z_A\in\{0,1\}^a\).  If \(\epsilon\le P(Z_k=1)\le 1-\epsilon\) for all \(k\), then
\[
\epsilon^a
\le
P(\Zvec_A=z_A)
\le
(1-\epsilon)^a.
\]
Thus every fixed pattern on \(A\) has probability \(\exp\{-\Theta(a)\}\).  For
arbitrary interference, the DTE and TTE target conditions impose fixed patterns on all
\(n\) assignments.  Under neighborhood interference, the DTE no-exposure conditions
\((1,0_{N_i})\) and \((0,0_{N_i})\), and the TTE full-exposure/no-exposure conditions
\((1,1_{N_i})\) and \((0,0_{N_i})\), impose fixed patterns on
\(\mathcal C_i=\{i\}\cup N_i\), whose size is \(d_i+1\).  The inverse propensities
therefore have the stated exponential orders.

For the exchangeable-design claim, condition on \(T_n=m\).  If a fixed pattern on
\(A\) contains \(t\) treated units, then
\[
P(\Zvec_A=z_A\mid T_n=m)
=
\frac{(m)_t(n-m)_{a-t}}{(n)_a},
\]
where \((x)_k=x(x-1)\cdots(x-k+1)\).  On the event
\(m/n\in[\epsilon,1-\epsilon]\) and for \(a\le \epsilon n/2\), each factor in this
product is bounded below by \(\epsilon/2\) and above by \(1-\epsilon\), after adjusting
constants depending only on \(\epsilon\).  Hence
\[
(\epsilon/2)^a
\le
P(\Zvec_A=z_A\mid T_n=m)
\le
(1-\epsilon)^a .
\]
The same bounds hold unconditionally because the support of \(T_n\) is contained in
\([\epsilon n,(1-\epsilon)n]\).  Applying this with \(A=\mathcal C_i\) gives the
neighborhood DTE and TTE orders.  Complete randomization with
\(m/n\in[\epsilon,1-\epsilon]\) is the special case in which \(T_n=m\) is fixed.  The
last statement follows immediately from Proposition~\ref{prop:ace-inverse-propensity}.
\end{proof}

\paragraph{Restatement.}
The orders in Tables~\ref{tab:dte-representation-rate} and
\ref{tab:tte-representation-rate} follow from the propensities of the target
treatment--exposure combinations under Bernoulli assignment.  Complete randomization
with a fixed treatment fraction has the same exponential order for the no-exposure and
full-exposure fixed entries, up to finite-population corrections.

\begin{proof}
Assume Bernoulli assignment with common treatment probability \(p\in(0,1)\) bounded
away from zero and one, and write \(d_i=|N_i|\).  For a fixed target contrast, the
HT weight scale \(H_i\) is the larger reciprocal propensity among the two target arms,
up to constants.  The uniform table scale is \(H=\max_i H_i\).  The lower-bound column
has the same propensity order by Proposition~\ref{prop:ace-inverse-propensity}, and
the HT variance-scale column follows by substituting \(H\) and the relevant dependence scale into
\(\mathcal V_n(H,c)=H/n+H^2c/n\).

For the direct-effect rows, arbitrary interference with no treated units except \(i\)
has target propensities
\[
p(1-p)^{n-1}
\qquad\text{and}\qquad
(1-p)^n,
\]
so \(H=e^{\Theta(n)}\).  Under neighborhood interference, the target conditions
\((1,0_{N_i})\) and \((0,0_{N_i})\) have propensities
\[
p(1-p)^{d_i}
\qquad\text{and}\qquad
(1-p)^{d_i+1},
\]
so \(H=e^{\Theta(d_{\max})}\).  The count and fractional no-exposure rows impose
\(S_i=0\), equivalently \(F_i=0\), and therefore have the same order.

For the binary any-treated direct-effect row, the DTE no-exposure condition is
\(U_i=0\), equivalently \(S_i=0\), so
\[
P\{Z_i=1,U_i=0\}=p(1-p)^{d_i},
\qquad
P\{Z_i=0,U_i=0\}=(1-p)^{d_i+1}.
\]
Thus \(H=e^{\Theta(d_{\max})}\).  For a typical count exposure \(k_i\asymp p d_i\),
\[
P\{S_i=k_i\}=\binom{d_i}{k_i}p^{k_i}(1-p)^{d_i-k_i}
=\Theta(d_i^{-1/2})
\]
uniformly for \(k_i\) in the central binomial range, by Stirling's approximation.
Multiplying by the constant arm probabilities \(p\) or \(1-p\) gives
\(H=\Theta(\sqrt{d_{\max}})\).

For the total-effect rows, arbitrary interference compares all treated with all
control, giving propensities \(p^n\) and \((1-p)^n\), hence \(H=e^{\Theta(n)}\).
Under neighborhood interference, the target conditions \((1,1_{N_i})\) and
\((0,0_{N_i})\) have propensities
\[
p^{d_i+1}
\qquad\text{and}\qquad
(1-p)^{d_i+1},
\]
so \(H=e^{\Theta(d_{\max})}\).  The count and fractional full-versus-none row imposes
\(S_i=d_i\) on the treated arm and \(S_i=0\) on the control arm, equivalently
\(F_i=1\) versus \(F_i=0\), and has the same order.  For the binary any-treated
version, the treated arm has probability \(p\{1-(1-p)^{d_i}\}\), but the control arm
with no exposure has probability \((1-p)^{d_i+1}\).  The no-exposure arm
therefore still gives \(H=e^{\Theta(d_{\max})}\).

Finally, under complete randomization with \(m\) treated units, the direct no-exposure
target propensities are
\[
\frac{\binom{n-d_i-1}{m-1}}{\binom{n}{m}},
\qquad
\frac{\binom{n-d_i-1}{m}}{\binom{n}{m}},
\]
whenever the binomial coefficients are well defined; uniformly over units this requires
\(1\le m\le n-d_{\max}-1\) for both direct-effect arms.  If \(m/n\to p\in(0,1)\),
these are the finite-population analogues of \(p(1-p)^{d_i}\) and
\((1-p)^{d_i+1}\); for example, when \(d_i=o(n)\) they are equal to the Bernoulli
propensities times \(\exp\{O(d_i^2/n)\}\).  For the full-exposure treated arm in the
TTE row, the corresponding propensity is
\[
\frac{\binom{n-d_i-1}{m-d_i-1}}{\binom{n}{m}},
\]
which is the finite-population analogue of \(p^{d_i+1}\).  Uniform positivity for the
full-versus-none TTE rows requires \(d_{\max}+1\le m\le n-d_{\max}-1\).  Thus complete
randomization has the same exponential dependence on \(d_i\) for local fixed
no-exposure and full-exposure entries, up to the usual finite-population correction,
but the variance bound also includes the residual design-dependence term
\(r_{\mathrm{avg}}\).  The arbitrary-interference rows are not covered by fixed
nontrivial complete randomization, since such a design does not assign positive
probability to both all-treated and all-control global regimes.
\end{proof}

\subsection{Auxiliary calculation: exposure compression and propensity scores}
\paragraph{Restatement.}
For an exposure mapping \(e=f_i(u)\),
\[
\pi_i^f(z,e)
=
\sum_{u\in f_i^{-1}(e)}\pi_i^{\mathrm{id}}(z,u),
\]
and hence
\[
\pi_i^f(z,e)\ge \pi_i^{\mathrm{id}}(z,u)
\qquad
\text{for every }u\in f_i^{-1}(e).
\]
Under balanced Bernoulli assignment,
\[
\pi_i^f(z,e)
=
2^{-(|N_i|+1)}\, |f_i^{-1}(e)|.
\]

\begin{proof}
The event \(\{Z_i=z,f_i(Z_{N_i})=e\}\) is the disjoint union of the events
\(\{Z_i=z,Z_{N_i}=u\}\) over \(u\in f_i^{-1}(e)\).  Summing their probabilities gives
the first equality and the inequality.  Under balanced Bernoulli assignment, each full
local configuration \((z,u)\) has probability \(2^{-(|N_i|+1)}\).
\end{proof}

\subsection{Auxiliary calculation: outcome-model span condition}
\paragraph{Restatement.}
Suppose an outcome model restricts \(\theta_i=X_i\gamma_i\).  Let \(O_i\) be the set of
local states with positive propensity score under the design.  A linear functional
\(\ell_i^\top\theta_i\) is determined by the potential outcomes in the observable
states \(O_i\) for every \(\gamma_i\) if and only if
\[
\ell_i^\top X_i
\in
\operatorname{rowspan}(X_{i,O_i}).
\]

\begin{proof}
The observable states determine the vector \(X_{i,O_i}\gamma_i\).  The functional
\(\ell_i^\top\theta_i=\ell_i^\top X_i\gamma_i\) is determined by these observable values
for every \(\gamma_i\) if and only if it is a linear function of
\(X_{i,O_i}\gamma_i\).  This is equivalent to the row vector \(\ell_i^\top X_i\) lying
in the row span of \(X_{i,O_i}\).
\end{proof}

\section{Proofs for Section~\ref{sec:designs}}
\label{app:design-proofs}

\subsection{Proof of Proposition~\ref{prop:ze-ht-unbiased}}
\begin{proof}
By stability, $\mathds{1}\{(Z_i,E_i)=d_i^+\}Y_i^{\mathrm{obs}}=\mathds{1}\{(Z_i,E_i)=d_i^+\}Y_i(d_i^+)$, and since the potential outcomes are fixed under the design,
\[
\mathbb E\!\left[\frac{\mathds{1}\{(Z_i,E_i)=d_i^+\}Y_i^{\mathrm{obs}}}{\pi_i(d_i^+)}\right]
=\frac{\pi_i(d_i^+)}{\pi_i(d_i^+)}\,Y_i(d_i^+)=Y_i(d_i^+).
\]
The same calculation gives $Y_i(d_i^-)$ for the second arm. Taking the difference and averaging over $i$ yields $\mathbb E[\hat\tau_{\mathrm{HT}}^{ZE}]=\tau$.
\end{proof}

\subsection{Proof of Lemma~\ref{lem:is1-propensity}}
\begin{proof}
The event $\{i\in\mathcal I\}$ equals $\{U_i<U_k\ \forall k\in N_i\}$, that is, $U_i$ is the strict minimum among the $d_i+1$ independent uniform draws in $\{i\}\cup N_i$. By exchangeability each draw is equally likely to be the minimum, so $\Pr(i\in\mathcal I)=1/(d_i+1)$. The coin $B_i$ is drawn independently of the draws $U$, so
\[
\rho_i^{(1)}=\Pr(i\in\mathcal I)\Pr(B_i=1)=\frac{q}{d_i+1},
\qquad
\rho_i^{(0)}=\Pr(i\in\mathcal I)\Pr(B_i=0)=\frac{1-q}{d_i+1}.
\]
On $F_i^{(1)}$, step (3) sets $Z_i=1$ and $Z_k=0$ for all $k\in N_i$, so $(Z_i,Z_{N_i})=(1,\mathbf 0)$; on $F_i^{(0)}$ it sets $Z_i=0$ and $Z_k=0$ for all $k\in N_i$, giving $(Z_i,Z_{N_i})=(0,\mathbf 0)$.
\end{proof}

\subsection{Proof of Proposition~\ref{prop:is1-variance}}
\begin{proof}
Write $\hat\tau_{\mathrm{IS}}^{D}=n^{-1}\sum_i W_i$ with $W_i=2(d_i+1)\mathds{1}\{i\in\mathcal I\}\{Y_i(1,\mathbf 0_{N_i})B_i-Y_i(0,\mathbf 0_{N_i})(1-B_i)\}$. Conditional on $\mathcal I$, the coins $B_i$ are independent, and $\mathbb E[W_i\mid\mathcal I]=\mathds{1}\{i\in\mathcal I\}\,\delta_i/p_i$; taking expectations gives $\mathbb E[W_i]=\delta_i$. For $i\ne j$, independence of $B_i,B_j$ given $\mathcal I$ yields
\[
\mathbb E[W_iW_j]=\frac{\Pr(i,j\in\mathcal I)}{p_ip_j}\,\delta_i\delta_j,
\qquad
\operatorname{Cov}(W_i,W_j)=\delta_i\delta_j\left(\frac{\Pr(i,j\in\mathcal I)}{p_ip_j}-1\right).
\]
If the closed neighborhoods of $i$ and $j$ are disjoint, the events $\{i\in\mathcal I\}$ and $\{j\in\mathcal I\}$ depend on disjoint sets of draws, so $\Pr(i,j\in\mathcal I)=p_ip_j$ and the covariance vanishes; this occurs whenever $\mathrm{dist}_G(i,j)>2$. Summing the diagonal variances and these covariances gives the stated exact identity.

For the pairwise ratio, if $i\sim j$ then $i$ and $j$ cannot both be closed-neighborhood minima, so $\Pr(i,j\in\mathcal I)=0$. If $i\not\sim j$ share $s_{ij}=|N_i\cap N_j|$ neighbors, set $a=d_i+1$, $b=d_j+1$, $s=s_{ij}$. Conditioning on $U_i=x$ and $U_j=y$, both are minima iff the $a-1-s$ private neighbors of $i$ exceed $x$, the $b-1-s$ private neighbors of $j$ exceed $y$, and the $s$ shared neighbors exceed $\max(x,y)$:
\[
\Pr(i,j\in\mathcal I)=\int_0^1\!\!\int_0^1 (1-x)^{a-1-s}(1-y)^{b-1-s}(1-\max\{x,y\})^{s}\,dx\,dy=\frac{a+b}{ab(a+b-s)}.
\]
Since $p_ip_j=1/(ab)$, the ratio is $(a+b)/(a+b-s)=(d_i+d_j+2)/(d_i+d_j+2-s_{ij})$, and the excess over $1$ is $s_{ij}/(d_i+d_j+2-s_{ij})$.

For the bound, $\mathbb E[W_i^2]=2(d_i+1)\{Y_i(1,\mathbf 0_{N_i})^2+Y_i(0,\mathbf 0_{N_i})^2\}\le 4M^2(d_i+1)$, so $\operatorname{Var}(W_i)\le 4M^2(d_i+1)$. Taking absolute values in the covariance sum and using $|\delta_i\delta_j|\le 4M^2$ with the pairwise ratios (contribution $1$ for each of the $2|E|$ ordered adjacent pairs, and $s_{ij}/(d_i+d_j+2-s_{ij})$ for each non-adjacent overlapping pair) gives the displayed inequality.
\end{proof}

\subsection{Proof of Corollary~\ref{cor:is1-rate}}
\begin{proof}
On a $d$-regular graph, $\sum_i(d_i+1)=n(d+1)$ and $2|E|=nd$. Each adjacent pair contributes at most $1$, and each non-adjacent overlapping pair contributes $s_{ij}/(2d+2-s_{ij})\le 1$ since $s_{ij}\le d-1$, while each ordered wedge $\sum_k d_k(d_k-1)=nd(d-1)$ counts at most one non-adjacent overlapping pair. The number of non-adjacent overlapping pairs is therefore $O(nd^2)$, and the weighted sum of the pairwise ratios is $O(nd^2)$; this yields the cruder rate $O(M^2d^2/n)$ after dividing by $n^2$. If the graph is locally tree-like, $s_{ij}=O(1)$, the pairwise ratio satisfies $s_{ij}/(2d+2-s_{ij})=O(1/d)$ (since the denominator is at least $d+3$), and the weighted sum over the $O(nd^2)$ overlapping pairs is $O(nd)$; dividing by $n^2$ gives $O(M^2 d/n)$. For a general graph, drop the denominators using $d_i+d_j+2-s_{ij}\ge 1$ and count ordered wedges: $\sum_{i\ne j:\ i\not\sim j}s_{ij}\le\sum_k d_k(d_k-1)$. Adding $2|E|=\sum_i d_i$ and the diagonal $\sum_i(d_i+1)=n+\sum_i d_i$ bounds the bracket by $n+\sum_i d_i+\sum_i d_i^2$, giving the stated $O(M^2(1+\bar d+\overline{d^2})/n)$.
\end{proof}

\subsection{Proof of Lemma~\ref{lem:is2-propensity}}
\begin{proof}
The event $\{i\in\mathcal I_2\}$ is $\{U_i<U_k\ \forall k\in B_i^{(2)}\setminus\{i\}\}$, i.e.\ $U_i$ is the strict minimum among the $b_i$ independent draws in $B_i^{(2)}$; by exchangeability this has probability $1/b_i$. On $\{i\in\mathcal I_2,R_i=1\}$, step (3) sets the entire closed neighborhood to $1$, giving $(Z_i,Z_{N_i})=(1,\mathbf 1)$, and $R_i$ is independent of the draws, so $\rho_i^{(1)}=1/(2b_i)$; the control side is symmetric. The lower bound $d_i+1\le b_i$ holds because the closed neighborhood is contained in the radius-two ball. The upper bound follows by counting vertices reached from $i$ in zero, one, or two steps: at most $1+d_i+\sum_{j\in N_i}(d_j-1)\le 1+d_i d_{\max}$.
\end{proof}

\subsection{Proof of Proposition~\ref{prop:is2-variance}}
\begin{proof}
Unbiasedness follows directly from the focal reveal events. On $\{i\in\mathcal I_2,R_i=1\}$ the observed outcome equals $Y_i(1,e_i^{\mathbf 1})$, and on $\{i\in\mathcal I_2,R_i=0\}$ it equals $Y_i(0,0)$; each event has probability $1/(2b_i)$. Writing $\hat\tau_{\mathrm{IS}}^{T}=n^{-1}\sum_i V_i$, the argument of Proposition~\ref{prop:is1-variance} applies with $p_i$ replaced by $q_i=1/b_i$ and $\delta_i$ by $\Delta_i$: conditioning on $\mathcal I_2$ gives $\mathbb E[V_i\mid\mathcal I_2]=\mathds{1}\{i\in\mathcal I_2\}\Delta_i/q_i$ and, for $i\ne j$, $\operatorname{Cov}(V_i,V_j)=\Delta_i\Delta_j(\Pr(i,j\in\mathcal I_2)/(q_iq_j)-1)$. If $B_i^{(2)}\cap B_j^{(2)}=\emptyset$ the covariance vanishes. If $\mathrm{dist}_G(i,j)\le2$, then $i$ and $j$ cannot both be strict minima in their radius-two balls, so $\Pr(i,j\in\mathcal I_2)=0$. If $\mathrm{dist}_G(i,j)\in\{3,4\}$, the same integral as before with ball sizes $(b_i,b_j)$ and overlap $s=s_{ij}^{(2)}$ gives $\Pr(i,j\in\mathcal I_2)/(q_iq_j)=(b_i+b_j)/(b_i+b_j-s)$. Finally $\mathbb E[V_i^2]=2b_i\{Y_i(1,e_i^{\mathbf 1})^2+Y_i(0,0)^2\}\le 4M^2 b_i$, and taking absolute values in the covariance sum gives the bound: distance-one and distance-two pairs contribute at most one each, and distance-three and distance-four overlapping pairs contribute $s_{ij}^{(2)}/(b_i+b_j-s_{ij}^{(2)})$.
\end{proof}

\subsection{Proof of Corollary~\ref{cor:is2-rate}}
\begin{proof}
On a $d$-regular locally tree-like graph, $b_i=1+d+d(d-1)=d^2+1$, so the diagonal term is $4M^2 n^{-2}\sum_i b_i=O(M^2d^2/n)$. The number of ordered pairs at distance at most two is $O(nd^2)$, so the distance-two exclusion term is also $O(M^2d^2/n)$. For the positive radius-two overlap term, each unit has at most $d(d-1)^2$ distance-three partners, each with overlap $s=2$ and denominator of order $d^2$, contributing $O(nd)$; each unit has $O(d^4)$ distance-four partners, each with overlap $s=1$ and denominator of order $d^2$, contributing $O(nd^2)$, which is the same order as the diagonal and exclusion terms. Thus the total is $O(M^2 d^2/n)$. For a general graph, the diagonal and distance-two exclusion terms are bounded by $O(\sum_i b_i)$, and wedge-counting in the radius-two overlap graph bounds the positive overlap sum by $\sum_k b_k(b_k-1)$, giving $O(M^2\overline{b^2}/n)$.
\end{proof}

\subsection{Proof of Proposition~\ref{prop:is-seeding}}
\begin{proof}
\textbf{Unbiasedness.} On $F_i^{(1)}=\{Z_i=0,U_i=1\}$ the observed outcome equals $Y_i(0,1)$, and on $F_i^{(0)}=\{Z_i=0,U_i=0\}$ it equals $Y_i(0,0)$. Since $\rho_i^{(b)}>0$ for every unit of degree at least one,
\[
\mathbb E\!\Big[\frac{\mathds 1\{F_i^{(1)}\}Y_i^{\mathrm{obs}}}{\rho_i^{(1)}}\Big]=Y_i(0,1),
\qquad
\mathbb E\!\Big[\frac{\mathds 1\{F_i^{(0)}\}Y_i^{\mathrm{obs}}}{\rho_i^{(0)}}\Big]=Y_i(0,0),
\]
so $\mathbb E[\hat\tau_{\mathrm{seed}}]=n^{-1}\sum_i\{Y_i(0,1)-Y_i(0,0)\}=\tau_{\mathrm{IE}}(0;0,1)$.

\textbf{Focal probabilities.} Because $\mathcal I$ is an independent set, no ego has an ego neighbor. Consequently the event $\{U_i=1\}$---at least one treated neighbor---forces $i\notin\mathcal I$ and hence $Z_i=0$. Writing $K_i=|\mathcal I\cap N_i|$, the number of treated ego neighbors is a binomial thinning of $K_i$ by the independent coins $B_j$, so $\Pr(U_i=1)=\mathbb E[1-(1-q)^{K_i}]$, which gives $\rho_i^{(1)}$. Further, $Z_i=0$ holds when either $i\notin\mathcal I$ or $i\in\mathcal I$ with $B_i=0$, so
\[
\Pr(Z_i=0)=1-q\,\Pr(i\in\mathcal I)=1-\frac{q}{d_i+1},
\]
and $\rho_i^{(0)}=\Pr(Z_i=0)-\rho_i^{(1)}=\mathbb E[(1-q)^{K_i}]-\frac{q}{d_i+1}$, as claimed in \eqref{eq:seeding-focal}.

\textbf{Poisson limit on locally tree-like $d$-regular graphs.} Suppose the induced subgraph on the radius-two ball of $i$ is a tree, as on locally tree-like $d$-regular graphs. Given $U_i=u$, the ego events of the $d$ neighbors of $i$ are determined by disjoint sets of draws and are therefore conditionally independent, each with common probability
\[
p(u)=\Pr\Big(U_j<u,\ U_j<U_\ell\ \forall\,\ell\in N_j\setminus\{i\}\Big)
=\int_0^u (1-x)^{d-1}\,dx
=\frac{1-(1-u)^d}{d}.
\]
Hence $K_i\mid U_i=u\sim\mathrm{Binomial}(d,p(u))$ and, for $s\in[0,1]$,
\[
\mathbb E[s^{K_i}]
=\int_0^1 \Big(1-(1-s)\,p(u)\Big)^d du
=\int_0^1 \Big(1-\frac{1-s}{d}\big(1-(1-u)^d\big)\Big)^d du
\xrightarrow{d\to\infty} e^{-(1-s)},
\]
by dominated convergence, since $d\,p(u)=1-(1-u)^d\to 1$ for $u\in(0,1)$. Thus $K_i\Rightarrow\mathrm{Poisson}(1)$. Taking $s=1-q$ gives $\mathbb E[(1-q)^{K_i}]\to e^{-q}$; hence $\rho_i^{(1)}\to 1-e^{-q}$, and since $q/(d+1)\to 0$, $\rho_i^{(0)}\to e^{-q}$.

\textbf{Star center.} For the center $c$ of a star, the leaves are degree-one units: given $U_c=u$, each leaf is an ego exactly when its single draw falls below $u$, so the leaf ego events are independent Bernoulli($u$) and
\[
\mathbb E[(1-q)^{K_c}]
=\int_0^1 (1-qu)^d\,du
=\frac{1-(1-q)^{d+1}}{q(d+1)}
=O\!\Big(\frac{1}{qd}\Big),
\]
so $\rho_c^{(0)}$ has the same polynomial order and $\rho_c^{(1)}\to 1$. In general $K_i\le d_i$, so $\mathbb E[(1-q)^{K_i}]\ge (1-q)^{d_i}$ and $\rho_i^{(0)}\ge(1-q)^{d_i}-q/(d_i+1)$: when the ego-neighborhood count $K_i$ concentrates on values of order $d_i$ the unexposed arm is no larger than the classical $(1-q)^{d_i}$ scale. The constant-order guarantee is therefore tied to units whose ego-neighborhood is $O(1)$ in expectation.

\textbf{Variance.} Write $\hat\tau_{\mathrm{seed}}=n^{-1}\sum_i W_i$. Since $|Y_i|\le M$,
\[
\mathbb E[W_i^2]
=\mathbb E\Big[\frac{\mathds 1\{F_i^{(1)}\}(Y_i^{\mathrm{obs}})^2}{(\rho_i^{(1)})^2}+\frac{\mathds 1\{F_i^{(0)}\}(Y_i^{\mathrm{obs}})^2}{(\rho_i^{(0)})^2}\Big]
\le M^2\Big(\frac1{\rho_i^{(1)}}+\frac1{\rho_i^{(0)}}\Big),
\]
so the diagonal term is at most $M^2 n^{-2}\sum_i(1/\rho_i^{(1)}+1/\rho_i^{(0)})=O(M^2/n)$ on locally tree-like $d$-regular graphs. The focal event $F_i^{(b)}$ is a function of the draws $U$ in the radius-two ball of $i$ and of the coins of the egos in $N_i$, so for $\mathrm{dist}_G(i,j)\ge 5$ the focal events of $i$ and $j$ are independent and $\operatorname{Cov}(W_i,W_j)=0$. The remaining covariance contributions come from the $O(nd^4)$ pairs at distance at most four and obey the same radius-two overlap anatomy as in the proof of Proposition~\ref{prop:is1-variance}; combined with the diagonal term this is exactly the structure of the generic ACE variance bound of Proposition~\ref{prop:generic-ace-ht}.
\end{proof}

\vskip 0.2in
\bibliography{references,interference, references_pnas,mechvspolocy}

\begin{thebibliography}{42}
\providecommand{\natexlab}[1]{#1}
\providecommand{\url}[1]{\texttt{#1}}
\expandafter\ifx\csname urlstyle\endcsname\relax
  \providecommand{\doi}[1]{doi: #1}\else
  \providecommand{\doi}{doi: \begingroup \urlstyle{rm}\Url}\fi

\bibitem[Adhikari and Zheleva(2025)]{adhikari2025}
Shishir Adhikari and Elena Zheleva.
\newblock Inferring individual direct causal effects under heterogeneous peer
  influence.
\newblock \emph{Machine Learning}, 114\penalty0 (4):\penalty0 113, 2025.

\bibitem[Aral and Walker(2011)]{aral2011}
Sinan Aral and Dylan Walker.
\newblock Creating social contagion through viral product design: A randomized
  trial of peer influence in networks.
\newblock \emph{Management Science}, 57\penalty0 (9):\penalty0 1623--1639,
  2011.

\bibitem[Aral and Walker(2012)]{aralwalker2012identifying}
Sinan Aral and Dylan Walker.
\newblock Identifying influential and susceptible members of social networks.
\newblock \emph{Science}, 337\penalty0 (6092):\penalty0 337--341, 2012.

\bibitem[Aral et~al.(2013)Aral, Muchnik, and Sundararajan]{aral2013engineering}
Sinan Aral, Lev Muchnik, and Arun Sundararajan.
\newblock Engineering social contagions: Optimal network seeding in the
  presence of homophily.
\newblock \emph{Network Science}, 1\penalty0 (2):\penalty0 125--153, 2013.

\bibitem[Aronow(2012)]{aronow2012general}
Peter~M Aronow.
\newblock A general method for detecting interference between units in
  randomized experiments.
\newblock \emph{Sociological Methods \& Research}, 41\penalty0 (1):\penalty0
  3--16, 2012.

\bibitem[Aronow and Samii(2017)]{aronow2017estimating}
Peter~M. Aronow and Cyrus Samii.
\newblock Estimating average causal effects under general interference.
\newblock \emph{The Annals of Applied Statistics}, 11\penalty0 (4):\penalty0
  1912--1947, 2017.

\bibitem[Athey et~al.(2018)Athey, Eckles, and Imbens]{athey2018exact}
Susan Athey, Dean Eckles, and Guido~W Imbens.
\newblock Exact p-values for network interference.
\newblock \emph{Journal of the American Statistical Association}, 113\penalty0
  (521):\penalty0 230--240, 2018.

\bibitem[Baird et~al.(2018)Baird, Bohren, McIntosh, and
  {\"O}zler]{baird2018optimal}
Sarah Baird, J~Aislinn Bohren, Craig McIntosh, and Berk {\"O}zler.
\newblock Optimal design of experiments in the presence of interference.
\newblock \emph{Review of Economics and Statistics}, 100\penalty0 (5):\penalty0
  844--860, 2018.

\bibitem[Basse and Feller(2018)]{basse2018design}
Guillaume~W. Basse and Avi Feller.
\newblock Design of experiments in the presence of interference.
\newblock \emph{Annual Review of Statistics and Its Application}, 5:\penalty0
  45--69, 2018.
\newblock \doi{10.1146/annurev-statistics-031017-100449}.

\bibitem[Basse et~al.(2019)Basse, Feller, and Toulis]{basse2019randomization}
Guillaume~W Basse, Avi Feller, and Panos Toulis.
\newblock Randomization tests of causal effects under interference.
\newblock \emph{Biometrika}, 106\penalty0 (2):\penalty0 487--494, 2019.

\bibitem[Bhattacharya et~al.(2020)Bhattacharya, Malinsky, and
  Shpitser]{bhattacharya2020causal}
Rohit Bhattacharya, Daniel Malinsky, and Ilya Shpitser.
\newblock Causal inference under interference and network uncertainty.
\newblock In \emph{Uncertainty in Artificial Intelligence}, pages 1028--1038.
  PMLR, 2020.

\bibitem[Bond et~al.(2012)Bond, Fariss, Jones, Kramer, Marlow, Settle, and
  Fowler]{bond201261}
Robert~M Bond, Christopher~J Fariss, Jason~J Jones, Adam~DI Kramer, Cameron
  Marlow, Jaime~E Settle, and James~H Fowler.
\newblock A 61-million-person experiment in social influence and political
  mobilization.
\newblock \emph{Nature}, 489\penalty0 (7415):\penalty0 295--298, 2012.

\bibitem[Bramoull{\'e} et~al.(2009)Bramoull{\'e}, Djebbari, and
  Fortin]{bramoulle2009identification}
Yann Bramoull{\'e}, Habiba Djebbari, and Bernard Fortin.
\newblock Identification of peer effects through social networks.
\newblock \emph{Journal of Econometrics}, 150\penalty0 (1):\penalty0 41--55,
  2009.

\bibitem[Cai et~al.(2024)Cai, Zhang, and Airoldi]{cai2024independent_set}
Chencheng Cai, Xu~Zhang, and Edoardo~M. Airoldi.
\newblock Independent-set design of experiments for estimating treatment and
  spillover effects under network interference.
\newblock In \emph{International Conference on Learning Representations
  (ICLR)}, 2024.

\bibitem[Centola(2010)]{centola2010}
Damon Centola.
\newblock The spread of behavior in an online social network experiment.
\newblock \emph{Science}, 329\penalty0 (5996):\penalty0 1194--1197, 2010.

\bibitem[De~Giorgi et~al.(2010)De~Giorgi, Pellizzari, and
  Redaelli]{degiorgi2010identification}
Giacomo De~Giorgi, Michele Pellizzari, and Silvia Redaelli.
\newblock Identification of social interactions through identifying
  restrictions.
\newblock \emph{American Economic Journal: Applied Economics}, 2\penalty0
  (2):\penalty0 275--291, 2010.
\newblock \doi{10.1257/app.2.2.275}.

\bibitem[Giffin et~al.(2023)Giffin, Reich, Yang, and
  Rappold]{giffin2023generalized}
Andrew Giffin, BJ~Reich, Shu Yang, and AG~Rappold.
\newblock Generalized propensity score approach to causal inference with
  spatial interference.
\newblock \emph{Biometrics}, 79\penalty0 (3):\penalty0 2220--2231, 2023.

\bibitem[Graham(2008)]{graham2008identifying}
Bryan~S Graham.
\newblock Identifying social interactions through conditional variance
  restrictions.
\newblock \emph{Econometrica}, 76\penalty0 (3):\penalty0 643--660, 2008.

\bibitem[Granovetter(1978)]{granovetter1978}
Mark Granovetter.
\newblock Threshold models of collective behavior.
\newblock \emph{American Journal of Sociology}, 83\penalty0 (6):\penalty0
  1420--1443, 1978.

\bibitem[Hu et~al.(2022)Hu, Li, and Wager]{huLiWagerBiometrika}
Yuchen Hu, Shuangning Li, and Stefan Wager.
\newblock Average direct and indirect causal effects under interference.
\newblock \emph{Biometrika}, 109\penalty0 (4):\penalty0 1165--1172, 2022.

\bibitem[Hudgens and Halloran(2008)]{hudgens2012toward}
Michael~G Hudgens and M~Elizabeth Halloran.
\newblock Toward causal inference with interference.
\newblock \emph{Journal of the American Statistical Association}, \penalty0
  (482):\penalty0 832--842, 2008.

\bibitem[Karwa and Airoldi(2018)]{karwa2018systematic}
Vishesh Karwa and Edoardo~M Airoldi.
\newblock A systematic investigation of classical causal inference strategies
  under mis-specification due to network interference.
\newblock \emph{arXiv preprint arXiv:1810.08259}, 2018.

\bibitem[Leskovec et~al.(2007)Leskovec, Adamic, and
  Huberman]{leskovec2007dynamics}
Jure Leskovec, Lada~A Adamic, and Bernardo~A Huberman.
\newblock The dynamics of viral marketing.
\newblock \emph{ACM Transactions on the Web (TWEB)}, 1\penalty0 (1):\penalty0
  5--es, 2007.
\newblock \doi{10.1145/1232722.1232727}.

\bibitem[Li and Wager(2022)]{liWagerAnnals}
Shuangning Li and Stefan Wager.
\newblock Random graph asymptotics for treatment effect estimation under
  network interference.
\newblock \emph{The Annals of Statistics}, 50\penalty0 (4):\penalty0
  2334--2358, 2022.

\bibitem[Manski(1993)]{manski1993identification}
Charles~F Manski.
\newblock Identification of endogenous social effects: The reflection problem.
\newblock \emph{The Review of Economic Studies}, 60\penalty0 (3):\penalty0
  531--542, 1993.

\bibitem[Manski(2013)]{manski2013identification}
Charles~F Manski.
\newblock Identification of treatment response with social interactions.
\newblock \emph{The Econometrics Journal}, 16\penalty0 (1):\penalty0 S1--S23,
  2013.

\bibitem[Ohnishi et~al.(2025)Ohnishi, Karmakar, and
  Sabbaghi]{ohnishi2025degree}
Yuki Ohnishi, Bikram Karmakar, and Arman Sabbaghi.
\newblock Degree of interference: A general framework for causal inference
  under interference.
\newblock \emph{Journal of Machine Learning Research}, 26\penalty0
  (120):\penalty0 1--37, 2025.

\bibitem[Paluck et~al.(2016)Paluck, Shepherd, and Aronow]{paluck2016changing}
Elizabeth~Levy Paluck, Hana Shepherd, and Peter~M. Aronow.
\newblock Changing climates of conflict: A social network experiment in 56
  schools.
\newblock \emph{Proceedings of the National Academy of Sciences}, 113\penalty0
  (3):\penalty0 566--571, 2016.

\bibitem[Patacchini et~al.(2017)Patacchini, Rainone, and
  Zenou]{patacchini2017heterogeneous}
Eleonora Patacchini, Edoardo Rainone, and Yves Zenou.
\newblock Heterogeneous peer effects in education.
\newblock \emph{Journal of Economic Behavior \& Organization}, 134:\penalty0
  190--227, 2017.

\bibitem[Pouget-Abadie et~al.(2019)Pouget-Abadie, Saint-Jacques, Saveski, Duan,
  Ghosh, Xu, and Airoldi]{pouget2019testing}
Jean Pouget-Abadie, Guillaume Saint-Jacques, Martin Saveski, Weitao Duan,
  Souvik Ghosh, Ya~Xu, and Edoardo~M Airoldi.
\newblock Testing for arbitrary interference on experimentation platforms.
\newblock \emph{Biometrika}, 106\penalty0 (4):\penalty0 929--940, 2019.

\bibitem[Puelz et~al.(2022)Puelz, Basse, Feller, and Toulis]{puelz2022graph}
David Puelz, Guillaume Basse, Avi Feller, and Panos Toulis.
\newblock A graph-theoretic approach to randomization tests of causal effects
  under general interference.
\newblock \emph{Journal of the Royal Statistical Society Series B: Statistical
  Methodology}, 84\penalty0 (1):\penalty0 174--204, 2022.

\bibitem[Rigdon and Hudgens(2015)]{rigdon2015exact}
Joseph Rigdon and Michael~G Hudgens.
\newblock Exact confidence intervals in the presence of interference.
\newblock \emph{Statistics \& probability letters}, 105:\penalty0 130--135,
  2015.

\bibitem[Rosenbaum(2007)]{rosenbaum2007interference}
Paul~R Rosenbaum.
\newblock Interference between units in randomized experiments.
\newblock \emph{Journal of the American Statistical Association}, 102\penalty0
  (477), 2007.

\bibitem[S{\"a}vje(2024)]{savje2024causal}
Fredrik S{\"a}vje.
\newblock Causal inference with misspecified exposure mappings: separating
  definitions and assumptions.
\newblock \emph{Biometrika}, 111\penalty0 (1):\penalty0 1--15, 2024.

\bibitem[S{\"a}vje et~al.(2017)S{\"a}vje, Aronow, and
  Hudgens]{savje2017average}
Fredrik S{\"a}vje, Peter~M Aronow, and Michael~G Hudgens.
\newblock Average treatment effects in the presence of unknown interference.
\newblock \emph{arXiv preprint arXiv:1711.06399}, 2017.

\bibitem[Sobel(2006)]{sobel2006randomized}
Michael~E Sobel.
\newblock What do randomized studies of housing mobility demonstrate? causal
  inference in the face of interference.
\newblock \emph{Journal of the American Statistical Association}, 101\penalty0
  (476):\penalty0 1398--1407, 2006.

\bibitem[Ugander et~al.(2013)Ugander, Karrer, Backstrom, and
  Kleinberg]{ugander2013graph}
Johan Ugander, Brian Karrer, Lars Backstrom, and Jon Kleinberg.
\newblock Graph cluster randomization: Network exposure to multiple universes.
\newblock In \emph{Proceedings of the 19th ACM SIGKDD international conference
  on Knowledge discovery and data mining}, pages 329--337. ACM, 2013.

\bibitem[Viviano(2026)]{viviano2024design}
Davide Viviano.
\newblock Experimental design under network interference.
\newblock \emph{Journal of Political Economy: Microeconomics}, 2026.
\newblock First version 2020; arXiv:2003.08421.

\bibitem[Weinstein and Nevo(2026)]{weinstein2026causal}
Bar Weinstein and Daniel Nevo.
\newblock Causal inference with misspecified network interference structure.
\newblock \emph{Biometrics}, 82\penalty0 (1):\penalty0 ujag023, 2026.

\bibitem[Yu et~al.(2022)Yu, Airoldi, Borgs, and Chayes]{yu2022pnasTTE}
Christina~Lee Yu, Edoardo~M Airoldi, Christian Borgs, and Jennifer~T Chayes.
\newblock Estimating the total treatment effect in randomized experiments with
  unknown network structure.
\newblock \emph{Proceedings of the National Academy of Sciences}, 119\penalty0
  (44):\penalty0 e2208975119, 2022.

\bibitem[Zigler and Papadogeorgou(2021)]{zigler2021bipartite}
Corwin~M Zigler and Georgia Papadogeorgou.
\newblock Bipartite causal inference with interference.
\newblock \emph{Statistical Science}, 36\penalty0 (1):\penalty0 109--123, 2021.

\bibitem[Zigler et~al.(2012)]{zigler2012}
Corwin~M Zigler et~al.
\newblock Estimating causal effects of air quality regulations.
\newblock \emph{Biostatistics}, 13\penalty0 (2):\penalty0 289--302, 2012.

\end{thebibliography}

\end{document}